\documentclass{emulateapj}

\shorttitle{Stellar orbits in the Galactic Center}
\shortauthors{Gillessen et al.}

\begin{document}

\title{Monitoring stellar orbits around the Massive Black Hole in the Galactic Center}

\author{S.~Gillessen\altaffilmark{1}, F.~Eisenhauer\altaffilmark{1}, S.~Trippe\altaffilmark{1}, T.~Alexander\altaffilmark{3}, R.~Genzel\altaffilmark{1,2}, F.~Martins\altaffilmark{4}, T.~Ott\altaffilmark{1}}
\altaffiltext{1}{Max-Planck-Institut f\"ur Extraterrestrische Physik, 85748 Garching, Germany}
\altaffiltext{2}{Physics Department, University of California, Berkeley, CA 94720, USA}
\altaffiltext{3}{Faculty of Physics, Weizmann Institute of Science, POB 26, Rehovot 76100, Israel; William Z. and Eda Bess Novick Career Development Chair}
\altaffiltext{4}{GRAAL-CNRS, Universit\'e Montpellier II, Place Eug\`ene Bataillon, 34095 Montpellier, France}

\begin{abstract}
We present the results of 16 years of monitoring stellar orbits around 
the massive black hole in center of the Milky Way using high
resolution near-infrared techniques. This work refines our previous 
analysis mainly by greatly improving the definition of the coordinate
system, which reaches a long-term astrometric accuracy of $\approx 300\,\mu$as, 
and by investigating in detail the individual systematic error
contributions. The combination of a long time baseline and the
excellent astrometric accuracy of adaptive optics data allow us to
determine orbits of 28 stars, including the star S2, which has
completed a full revolution since our monitoring began. Our main
results are: all stellar orbits are fit extremely well by a single
point mass potential to within the astrometric uncertainties, which
are now $\approx 6\times$ better than in previous studies. The central object
mass is $(4.31 \pm 0.06|_\mathrm{stat}\pm0.36|_{R_0})\times 10^6 M_\odot$ 
where the fractional statistical error of $1.5\%$ is nearly independent from $R_0$
and the main uncertainty is due to the uncertainty in $R_0$.
Our current best estimate for the distance to the
Galactic Center is $R_0=8.33 \pm 0.35$ kpc. The dominant errors in
this value is systematic. The mass scales with distance as
$(3.95 \pm 0.06)\times 10^6(R_0/8\, {\rm kpc})^{2.19}
M_\odot$. The orientations of orbital angular
momenta for stars in the central arcsecond are random. We identify six
of the stars with orbital solutions as late type stars, and six early-type stars as
members of the clockwise rotating disk system, as was previously
proposed. We constrain the extended dark mass enclosed between the
pericenter and apocenter of S2 at less than $0.066$, at the $99\%$ confidence level, of the
mass of Sgr~A*. This is two orders of magnitudes larger than
what one would expect from other theoretical and observational
estimates.
\end{abstract}


\section{Introduction}
\begin{figure*}[htbp]
\begin{center}
\plotone{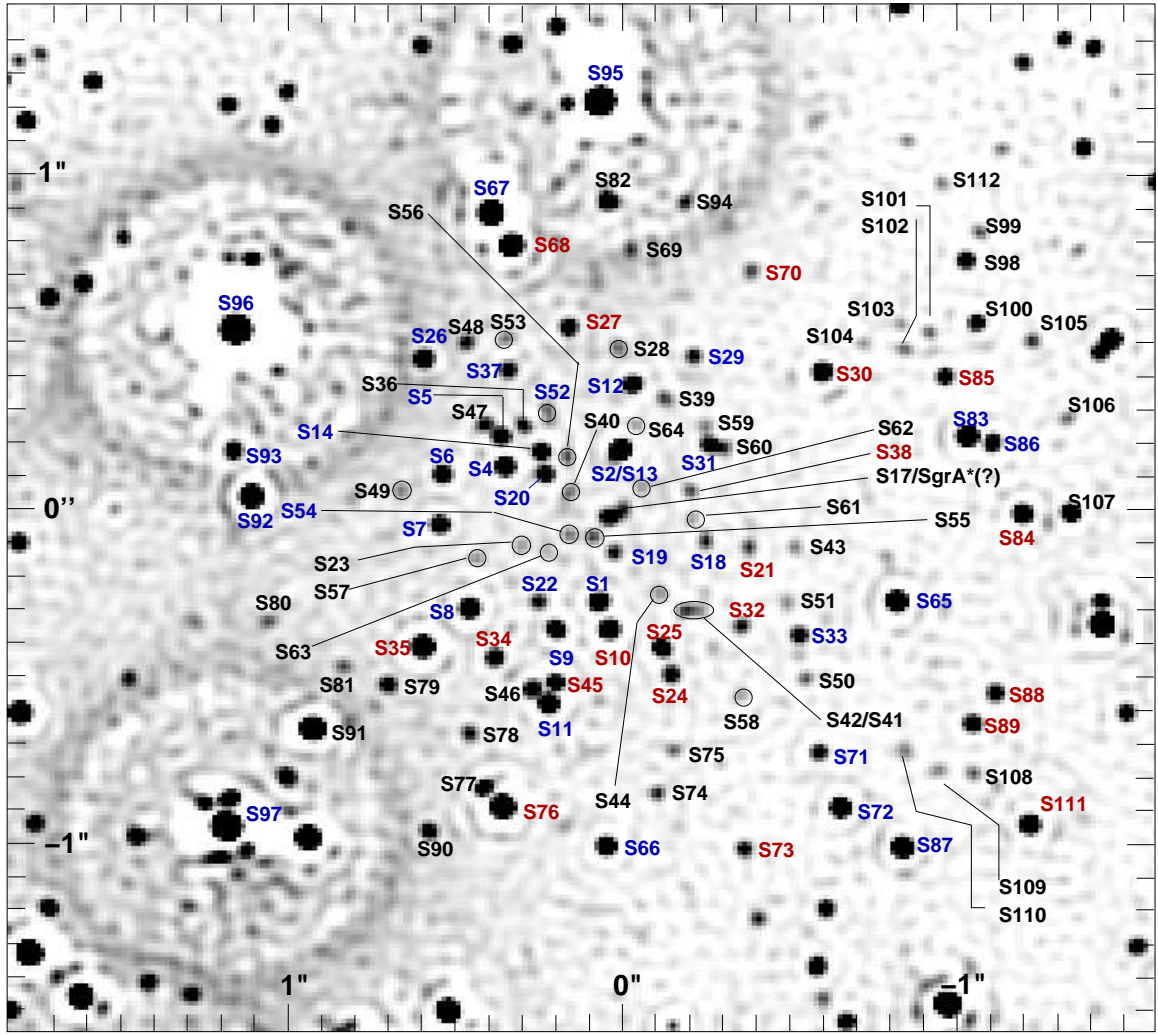}
\caption{Finding chart of the S-star cluster.  This figure is based on a
natural guide star adaptive optics image obtained as part of this study, using NACO at UT4 (Yepun) of the VLT on July 20, 2007 in the H-band. The original
image with a FWHM of $\approx 75\,$mas was deconvolved with the Lucy-Richardson algorithm and beam restored with a Gaussian beam with FWHM$\,=2\,$pix=$26.5\,$mas.
Stars as faint as $m_H=19.2$ (corresponding roughly to $m_K=17.7$) are detected at the $5\sigma$ level. Only stars that are unambiguously identified in several images have designated
names, ranging from S1 to  S112. Blue labels indicate early-type stars, red labels late-type stars. Stars with unknown spectral type are labelled in black. At the position
of Sgr~A* some light is seen, which could be either due to Sgr~A* itself or due to a faint, so far unrecognized star being confused with Sgr~A*. }
\label{findchart}
\end{center}
\end{figure*}

Observations of Keplerian stellar orbits in the Galactic Center (GC) 
that revolve in the
gravitational potential created by a highly concentrated mass of
roughly $4\times10^6\,M_\odot$ \citep{sch02,eis05,ghe03,ghe05}
currently constitute the best proof for the existence of an astrophysical massive black hole.
In this experiment the stars in the innermost arcsecond (the so-called S-stars) of 
our galaxy are used as test particles to probe the potential in which they move. Unlike gas, 
the motion of stars is determined solely by gravitational forces. 
Since the beginning of the observations in 1992 one of the stars, called S2, 
has now completed one full orbit. Its orbit \citep{sch02,ghe03}
has a period of 15 years. Since 2002 the
number of reasonably well-determined orbits has grown from one to 28; 
in total we currently monitor 109 stars, see Figure~\ref{findchart}\footnote{This work is based on observations
collected between 1992 and 2008 at the European Southern Observatory, both on Paranal and La Silla, Chile.}.

Due to the
high interstellar extinction of $\approx30$ magnitudes in the optical
towards the GC the measurements have to be
performed in the near infrared (NIR), where the extinction amounts to only
$\approx3$ magnitudes. The first
positions of S-stars were obtained in 1992 by Speckle imaging
at ESO's NTT in La Silla, a 4m-telescope, and in 1995 at the Keck telescope, a 10m-telescope.
Since 1999 (Keck: \cite{ghe01}) and 2002 (VLT: \cite{sch02}) the combination of 8m/10m-class telescopes and
adaptive optics (AO) has been routinely used for deep (H$\,\approx 19$) diffraction limited (FWHM 40-100 mas) imaging and spectroscopy.

The GC is a uniquely accessible laboratory for exploring the interactions between a massive black hole (MBH) 
and its stellar environment. By tracking the orbits of stars close to the MBH one can gather information on the
gravitational potential in which they move. Of prime interest is the value of $R_0$, the distance to the GC, as it
is one of the fundamental quantities in models for our Galaxy. Equally interesting is the nature of the mass responsible
for the strong gravitational forces observed. While the measured mass makes a compelling case for a MBH, the exact
form of the potential encodes answers to many interesting questions. Clearly, testing general relativity for such a heavy object is among
the goals; the first step would be to detect the Schwarzschild precession of the pericenters of some orbits. A measurable deviation from
a point mass potential would give access to a possible cluster of dark objects around the MBH, testing many theoretical ideas, such as 
mass segregation or the concept of a loss-cone. Another focus of interest are the properties of the stellar orbits. 
The distributions of the orbital elements may have conserved valuable information about the formation scenario of the respective stars. 
This addresses for example the so-called 'paradox of youth' for the stars in the central arcsecond \citep{ghe03} or the
puzzling existence of a large number of O-stars and Wolf-Rayet stars in the GC \citep{pau06}.

This paper is the continuation of our long-term work on stellar motions in the vicinity of Sgr~A*. 
We reanalyzed all data available to our team from 16 years. The basic steps of the analysis are:
\begin{itemize}
\item Obtain high quality, astrometrically unbiased maps of the S-stars. Obtain high quality spectra for these stars.
\item Extract pixel positions from the maps and radial velocities from the spectra.
\item Transform the pixel positions to a common astrometric coordinate system; transform the radial velocities to the local standard of rest (LSR). For the astrometric data several steps are needed:
\begin{itemize}
\item Relate the fainter S-stars positions to those of the brighter S-stars (Speckle data only).
\item Relate the S-stars positions to a set of selected reference stars.
\item Relate the reference stars to a set of SiO maser stars, of which the positions relative to Sgr~A* are known with good accuracy from radio (VLA) observations \citep{rei07}.
\end{itemize}
\item Fit the data with a model for the potential and gather in that way orbital parameters as well as information about the potential.
\end{itemize}
We organize this paper according to these steps. 

\section{Data base}
The present work relies on data obtained over many years with different
instruments. In this section we briefly describe the different data sets.
\subsection{SHARP}
\label{sharp}
The first high resolution imaging data of the GC region were obtained in 1992 with the SHARP camera built at MPE \citep{hof92,eck94}. 
SHARP was used by MPE scientists until 2002 at ESO's 3.5m NTT in Chile. The data led to the detection of 
high proper motions close to Sgr~A* \citep{eck96}. 
The camera was operating in speckle mode with exposure times of $0.3\,$s, $0.5\,$s and $1.0\,$s, which was the optimum compromise between sufficient signal to noise and 
fast sampling of the atmospheric turbulence. The data are described in \cite{sch03}, a summary is given in table~\ref{summ_sstars}. We used the simple 
shift-and-add (ssa) technique \citep{chr91} in order to obtain deep diffraction limited images from the raw 
frames. Compared to our previous analysis \citep{sch03,eis05} we did not base the astrometry on images combined from multiple pointing positions. Due to the 
camera's image distortions one should not trust the larger scale astrometry of such multi-pointing images since 
coadding different pointings in the presence of static image distortions will lead to discontinuities in the effective distortion map of the combined image.
These in turn would not be described well by the polynomial relations we use to map pixel positions onto the astrometric coordinate system, resulting in
astrometric biases \citep{rei03}. 
Single pointing combinations are astrometrically unbiased - although not necessarily linear. They do not show discontinuities and represent smooth coordinate grids. Hence, we co-added the frames pointingwise, yielding typically four coadded maps per observing epoch \citep{ott02}. In order to get deeper in the central arcsecond, we also co-added all frames into one single map per epoch. Of the latter map the astrometry however can only be trusted for a region as small as the central arcsecond (the region which was present
in all pointing positions), leading to an additional step of cross-calibration between the deep map and the four single-pointing maps per epoch. The ssa maps had a typical Strehl ratio of 30\%. 
We further cleaned them
using careful deconvolution and beam restoration, following the strategy outlined in \cite{sch03}. In order to assess the errors introduced by the deconvolution we used two different deconvolution methods:
The Lucy-Richardson algorithm \citep{luc74} and an iterative blind deconvolution process \citep{jef93}, yielding two different (although not independent) maps. 
\subsection{NACO}
\label{naco}
The first AO imaging data available to us of the GC region was obtained in 2002 with the Naos-Conica (NACO) system mounted at the fourth unit telescope Yepun of the VLT \citep{len98, rou98}. Compared to the SHARP data the NACO data are superior due to the larger telescope aperture ($8.0\,$m versus $3.5\,$m) and the higher Strehl ratios (typically 40\% for NACO) of the AO which is equipped with an IR wavefront sensor, allowing the use of the nearby K=$6.5\,$mag star IRS7 as AO guide star. Furthermore, the sampling is increased compared to the Speckle data. For NACO we have typically ten epochs per year, compared to one per year for SHARP.
We obtained images both in the $27\,$mas/pix and the $13\,$mas/pix image scales.
\begin{itemize}
\item In order to measure the positions of the SiO maser stars IRS9, IRS10EE, IRS12N, IRS15NE, IRS17, IRS19NW, IRS28 and SiO-15 \citep{rei07}, we used the $27\,$mas/pix image scale both in H- and K-band in all years since 2002. The data are described in \cite{rei07,tri08} and summarized in table~\ref{summ_maser}. 
The typical single detector integration time was two seconds, such that the bright IR sources present in the $r\approx20''$ field covered did not get saturated. 
Mostly, we used a dither pattern of four positions that guaranteed that the central ten arcseconds are imaged in each pointing position.
The number of useful maser positions per image varied between six and eight. IRS19NW was not in the images in 2002, 2003 and 2006; SiO-15 was not
covered in 2003. Due to their brightness IRS17 and IRS9 were in the non-linear regime of detector in the observations from June 12, 2004 and thus excluded for that epoch. 
Since the NACO camera when operated in the $27\,$mas/pix mode exhibits notable geometric image distortions we constructed de-distorted mosaics from the individual images by applying a distortion correction, 
involving rebinning of the measured flux distribution to a new pixel grid.
The procedure is described in detail in \cite{tri08} and relies on comparing distances between stars present in the different pointings. The distortion model
used is $\vec{p}=\vec{p}\,' (1 - \beta\, \vec{p}\,'^{\,2})$ with $\beta\approx3\times10^{-9}$ where $\vec{p}$ and $\vec{p}\,'$ denote true and distorted pixel positions with respect
to some origin in the image that also is determined from the data. 
See also fig.~\ref{distortion27}.
We did not apply deconvolution techniques on these images. 
\item The positions of the S-stars were determined mostly from images obtained in the $13\,$mas/pix image scale. (Only when no image in the $13\,$mas/pix scale was available sufficiently close in time, we used also images obtained in the $27\,$mas/pix scale.) A typical data set contains two hours of data. The single detector integration time was mostly 
around 15 seconds, and the field of view was moved after a few integrations successively to four positions such that the central four arcseconds are present in all frames.
The data are summarized in table~\ref{summ_sstars} and a complete list of the data sets used is given in the table in appendix~\ref{nacocom}.
The reduction followed the usual steps of sky subtraction and flat-fielding. Manually selected high quality frames were combined to a single ssa map per epoch since the 
optical distortions are small enough to be neglected in the $13\,$mas/pix scale \citep{tri08} for the frame combination. A distortion model of the same type as for the
$27\,$mas/pix scale images was constructed for each epoch; however the best-fitting model parameters varied more than expected between
the different epochs. We concluded that we were not able to solve for the distortion parameters with our observations. Hence, we did not apply distortion models
to the $13\,$mas/pix data but used higher order transformations when relating pixel positions to astrometric positions (see fig.~\ref{distortion13}).
In order to separate sources we moderately deconvolved the central five arcseconds of these maps with the Lucy-Richardson algorithm. The latter 
used a point spread function constructed from the map itself obtained by applying the starfinder code \citep{dio00}.
In order to estimate the deconvolution error we divided each $13\,$mas/pix data set into two and obtained two coadded maps, each with half of the integration. Both maps were then deconvolved the same way as the full coadd.
\end{itemize}

\begin{table}
\caption{\label{summ_sstars} Summary of the yearly number of epochs for which we obtained S-star images and the 
yearly mean number of S-star positions determined per epoch.}
{\scriptsize 
\begin{center}
\begin{tabular}{lccc}
Year & Instrument & \# epochs & \# $\frac{\mathrm{S-star\, positions}}{\mathrm{epoch}}$\\
\hline
1992 & SHARP & 1 & 33\\
1994 & SHARP & 1 & 41\\
1995 & SHARP & 1 & 38\\
1996 & SHARP & 2 & 38.5\\
1997 & SHARP & 1 & 38\\
1998 & SHARP & 1 & 33\\
1999 & SHARP & 1 & 39\\
2000 & SHARP & 1 & 38\\
2000 & GEMINI & 1 & 31\\
2001 & SHARP & 1 & 39\\
2002 & SHARP & 1 & 21\\
2002 & NACO & 8 & 81.9\\
2003 & NACO & 12 & 83.8\\
2004 & NACO & 11 & 75.3\\
2005 & NACO & 5 & 86\\
2006 & NACO & 13 & 72.8\\ 
2007 & NACO & 13 & 101.2\\
2008 & NACO & 5 & 104.8
\end{tabular}
\end{center}
}
\end{table}

\begin{table}
\caption{\label{summ_maser} Summary of the number of available maser star mosaic images, number of maser stars present in each frame and the respective FWHM of the point spread function in the images.}
{\scriptsize 
\begin{center}
\begin{tabular}{lccc}
Date & \# mosaics & FWHM [mas] & \# maser stars\\
\hline
May 2002 & 1 &70 & 7\\
May 2003 & 1&74  & 6\\
June 2004 & 3&70, 81, 86 & 6, 6, 8\\
May 2005 & 1&88  & 8\\
April 2006 & 2&100, 100  & 7, 7\\
March 2007 & 2&80, 80 & 8, 8 \\
March 2008 & 1& 84 & 8
\end{tabular}
\end{center}
}
\end{table}

\subsection{SINFONI}
Spectroscopy enables one to determine radial velocities of stars if
the positions of known atomic or molecular lines can be measured in the stellar
spectra. The GC is best exploited with integral field spectroscopy as
one is interested in the radial velocities of all stars for which one can hope to determine orbits,
i.e. all stars in the central arcsecond. In the
NIR the K-band ($2.0-2.4\,\mu$m) is best suited since it contains the hydrogen line Bracket-$\gamma$
at $2.16612\,\mu$m. This line is present in absorption
for B type stars, the most common spectral type for the S-stars \citep{eis05}. 
For late-type stars the CO band heads 
at $2.2935\,\mu$m, $2.3227\,\mu$m, $2.3535\,\mu$m and $2.3829\,\mu$m
are also covered by the K-band.

Since July 2004 we regularly monitored the GC with the 
AO assisted field spectrometer SINFONI \citep{eis03a, bon04}. The
instrument is mounted at the Cassegrain focus of ESO's UT-4 (Yepun) and
offers several operation modes concerning pixel scale and
wavelength coverage. For the GC we operated SINFONI mostly in the AO scale,
mapping 0.8''$\times$0.8'' onto 64$\times$32 spatial pixels. 
We used the K-band grating and the combined H+K grating of SINFONI, with spectral
resolutions of 4000 and 1500 respectively. For most of the data sets, the single
exposure time per frame was $10\,$minutes; a few data sets also used
$5\,$minute exposures. We chose various mosaicking patterns inside the central
arcsecond for the different runs;
mainly with the aim to have a good compromise between monitoring the activity of Sgr~A* and
building up integration on the S-stars. For stars at somewhat larger radii ($r>1"$) where confusion 
is less severe we also used
data originally obtained for other scientific programs in the $100\,$mas/pix scale offering a field of view of 
3.2"$\times$3.2". 

The SINFONI AO works in the optical. Since the GC region is 
heavily extincted, one has to use a guide 
star relatively far away from Sgr~A*. It is located 
10.8'' East and 18.8'' North of Sgr~A* and has a magnitude
of $m_R=14.65$. As a result the performance of the AO strongly depends
on the seeing conditions. Therefore the quality of our SINFONI data
is variable over the data set.
For a typical run, one can detect Bracket-$\gamma$ absorption of early-type stars as faint 
as $m_\mathrm{K}=15.5$ and the CO band heads of late-type stars
up to $m_\mathrm{K}=16.0$. A summary of our data is given in Table~\ref{summ_sinf}.

We applied the standard data reduction for SINFONI data, including detector calibrations (such as bad pixel corrections, flat-fielding and distortion corrections) and cube reconstruction. The wavelength scale was
calibrated by means of emission line lamps and finetuned on the atmospheric OH lines. The remaining uncertainty of the wavelength scale corresponds to typically $\lesssim10\,$km/s. We did not trust the SINFONI cubes for their astrometric precision, they were used only for their spectral dimension. Nevertheless it is easy to identify stars in the cubes.

\begin{table}
\caption{Summary of SINFONI data used for this work. The exposure time is the effective shutter-open time on S2, for other stars the actual exposure time might be different since the observations were
mosaicing around Sgr~A*. The FWHM was determined from a median image of the respective cube on the unconfused star S8.
\label{summ_sinf}}
{\scriptsize 
\begin{center}
\begin{tabular}{l|cccc}
Date& Band &$t_\mathrm{exp}$ on S2 & FWHM &\# S-stars\\
&&[min]&[mas]&with velocities\\
\hline
July 14 2004 &H+k& 40 & 79 & 7 \\
July 17 2004 &K& 110 & 93 & 25 \\
August 18/19 2004 &K& 80 & 88 & 23 \\
February 26 2005 &K& 20 & 108 & 4\\
March 18 2005 &K& 10 & 150 & 4\\
March 19 2005 &K& 40 & 69 & 16\\
June 15 2005 &K& 200 & 113 & 8\\
June 17 2005 &K& 440 & 88 & 25\\
Aug 28 - Sep 5 2005&K&10&$>$250&5\\
October 2-6 2005 &H+K& 120& 74 & 22 \\
March 16 2006 &H+K& 110 & 76 & 27 \\
April 21 2006 &H+K&10 &100 & 6\\
August 16/17 2006 &H+K& 100 & 88 & 18 \\ 
March 26 2007&H+K& 20 & 86 & 10\\
July 18-23 2007 &H+K& 133 & 78 & 15 \\
September 3/4 2007 &H+K& 70 & 81 & 15\\
April 4-9 2008 &H+K& 200 &65& 40\\
June 4 2008&H+K&10&84&3
\end{tabular}
\end{center}
}
\end{table}
\clearpage
\subsection{Other}
Beyond the data sets described so far, we included a few more data points which we
describe briefly in this section.
\begin{itemize}
\item {\bf Positions from public Gemini data for 2000:}
In addition to our observations we included images from
the Galactic Center Demonstration Science Data Set obtained
in 2000 with the 8-m-telescope Gemini North on Mauna Kea,
Hawaii, using the AO system Hokupa'a in combination with
the NIR camera Quirc. These images were processed by the
Gemini team and released to be used freely. We treated this data in the same way as the SHARP data.
\item {\bf Published radial velocities of S2 in 2002:} 
The first radial measurements of S2 were obtained by \cite{ghe03}. We included
the two published radial velocities since they extend the sampled time range by
one year and clearly contribute significantly in fixing the epoch of pericenter passage $t_\mathrm{P}$ for S2.
\item {\bf Radial velocities from longsplit spectroscopy with NACO in 2003:} 
We used NACO in its spectroscopic mode to measure the radial velocity of S2
in 2003. The data are described in \cite{eis03b}.
\item {\bf Radial velocities from integral field spectroscopy with SPIFFI in 2003:} 
SPIFFI is the integral field spectrometer inside SINFONI. We used it without
AO in 2003 as guest instrument at ESO-VLT UT-4 (Yepun) under superb atmospheric conditions and obtained cubes from which radial velocities for 18 stars (namely S1, S2, S4, S8, S10, S12, S17, S19, S25, S27, S30, S35, S65, S67, S72 S76, S83, S95, S96).
The data are described in \cite{eis03b}.
\end{itemize}

\section{Analysis of astrometric data}
\begin{figure}[htbp]
\begin{center}
\plotone{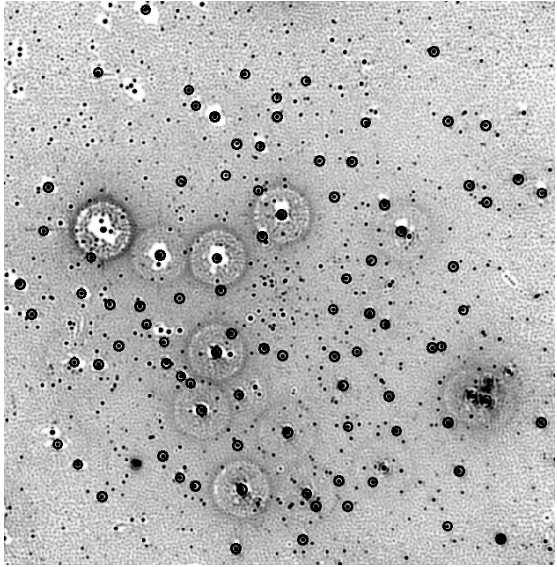}
\caption{The open symbols mark the sample of 91 reference stars which are used to define the astrometric frame for the S-stars. The underlying
image was obtained on April 3, 2007 in H-band, deconvolved and beam-restored with a beam of $2\,$pix. North is up, East is left. The field is
9.3"$\times\,$9.3".}
\label{refstars}
\end{center}
\end{figure}

This section describes in some technical detail the astrometric calibration of our data.
The first step is to measure the positions of stars on the astrometric maps. Next,
these positions of stars on the detector have to be transformed into a common astrometric reference frame. This procedure ultimately relies on measurements of eight SiO maser stars of which positions can be determined both in the radio and in NIR images. However, a direct comparison of the central arcsecond and the maser stars on one and the same image is impractical for two reasons: a) The exposure times necessary to obtain sufficiently deep images for the S-stars saturates the detector at the positions of the maser stars. b) The field of view of the $13\,$mas/pix pixel scale is too small to show enough maser stars. Therefore we need to crosscalibrate the S-stars images with the maser star images. This is done by a set of selected reference stars (fig.~\ref{refstars}), which are present both in the S-star images and the maser star images. For the SHARP data, even an additional step of cross-calibration is taken. We selected reference stars with
$1"\le r\le4"$ that are brighter than $m_\mathrm{K}\approx14.5$ and apparently unconfused, yielding a sample size of 91 stars.
\subsection{Extraction of pixel positions}
All pixel positions were obtained by two-dimensional Gaussian fits in the images. The fits yielded both the positions and estimates for the statistical error of the positions (section \ref{staterrpixpos}). For each epoch for which we have useful S-stars data we extracted pixel positions for the S-stars and for the reference stars. 
\subsubsection{SHARP}
Only star images that are not visually distorted (e.g. due to a confusion event) were used from the SHARP data.  
\begin{itemize}
\item {\bf Reference stars}:  We obtained the reference stars' positions from the four single-pointing maps from each epoch. Due to the limited field of view, in each frame only a subset of the reference stars is present. 
\item {\bf Brighter S-stars}: For the brighter S-stars (e.g. S2, S1, S8, S10, S30, S35) typically all four different pointing positions could be used. The astrometric position of each star was
determined from the corresponding four pixel positions using the astrometric average position (see section~\ref{staterrpixpos}).
\item {\bf Fainter S-stars}: In order to detect faint S-stars we used the fifth coadded map which can be trusted astrometrically only for the innermost arcsecond. The limiting magnitude for a non-confused source was typically $m_\mathrm{K}\approx15.8$. We determined the pixel positions 
of the weaker S-stars as well as the ones of the brighter S-stars. The latter served as reference for relating 
the fainter stars to the astrometric coordinate system (see section~\ref{staterrpixpos}).
\end{itemize}
Since we had two different deconvolutions at hand, we extracted pixel positions from both sets of images. Thus, up to eight (= two deconvolutions $\times$ four pointings) pixel positions were obtained per star and epoch.
\subsubsection{NACO}
For the NACO data, we used both the $27\,$mas/pix data and the $13\,$mas/pix data.
\begin{itemize}
\item {\bf SiO maser stars}: Positions for the SiO maser stars were obtained by Gaussian fits to the stars' images in the $27\,$mas/pix mosaics. The SiO maser stars were unconfused in all mosaics. 
\item {\bf Reference stars}: The positions of the reference stars were measured both on the $27\,$mas/pix mosaics as well as on the $13\,$mas/pix maps (Table~\ref{summ_sstars}), since they serve as cross-calibration between the two sets. They were selected to be unconfused, thus essentially it was possible to use all reference stars visible on any given frame.
\item {\bf S-stars}: For isolated S-stars, the positions were obtained from a simple Gaussian fit to the manually identified stars in the maps.
Due to the higher sampling rate with NACO confusion events can be tracked much better in the AO data than in the SHARP data. Therefore it was reasonable to also measure positions when stars are partly overlapping. In such a case, a simultaneous, multiple Gaussian fit to the individual peaks was used, resulting
of course in larger statistical uncertainties of the obtained positions.
\end{itemize}

\subsection{Relating the reference stars to the SiO maser stars}
\label{lawofmotionrefstars}
The goal of this step is to obtain linear models for the motions of the reference stars, i.e. to express their velocities and positions with linear functions $x(t), y(t)$ in terms of astrometric coordinates. These models then define a common reference frame that is calibrated in position and velocity such that radio Sgr~A* should be at rest at the origin of the system. Such a coordinate system allows one to test if the center of mass obtained from orbital fitting coincides with the compact radio source\footnote{Systematic problems of the coordinate system could be absorbed into the orbital fitting by allowing the center of mass to have an offset from 0/0 and a non-zero velocity, at the cost of not being able to test the coincidence of the center of mass with radio Sgr~A*.}. 

In the following we present two ways to obtain the desired calibration. They differ in the way in which the positions and velocities for the reference stars are determined: either all maser star images are tied to the respective maser positions (multi-epoch cross calibration); or only one maser star image is used to tie to the radio maser positions, and the other maser star images are matched to that by an additional step of cross calibration that only involves infrared data (single epoch cross calibration). It turns out that both ways have their specific advantages and disadvantages in terms of position and velocity calibration of the resulting coordinate systems. We finally constructed a third coordinate system combining the advantages and rejecting the disadvantages.
\subsubsection{Multi-epoch cross calibration with all maser star images}
\label{masersys}
Using the results from \cite{rei07} we calculated the expected radio maser positions for the given observation epochs. The different maser images contained between seven and nine SiO masers of which we used six to eight since we excluded IRS7 due to its brightness of $m_\mathrm{K}\approx6.5$. By allowing for a linear transformation of type $\vec{x} = \vec{x_0}+M.\vec{p}$ between the astrometric positions $\vec{x}$ and the pixel positions $\vec{p}$ in the respective image we determined a transformation by which any detector position can be converted into astrometric coordinates. Note that the use of a linear transformation is justified since the IR images were distortion-corrected mosaics. 
The rms of the 1D-residuals of the SiO masers (thus applying the transformation to the SiO masers' pixel positions and comparing the result with the expected
radio positions) was $2.28\,$mas. Correspondingly we expect that from our 11 images a coordinate system can be defined to at most an 1D-accuracy of $2.28/\sqrt{11}\,$mas$\approx0.7\,$mas
if the measurement errors from the 11 images are uncorrelated.
The transformation was applied to the sample of reference stars in each image. We then fitted the resulting astrometric positions of the reference stars as a function of time with linear functions. From these linear fits we obtained residuals, allowing obvious outliers to be identified and rejecting them. 
We excluded stars that had a residual different from the median residual by twice that value, if the deviation was larger than $2\,$mas. That excluded between 0 and at most 10
of the 91 reference stars for the various mosaics, the reason for these outliers being confusion that in some mosaics affects the fainter reference stars
due to the varying image quality. After this moderate data cleaning we repeated the fitting. The mean rms of the 1D-residuals per image 
had then a value of $1.45\,$mas. 

The next step of refinement was to compare all measured positions in one mosaic with the positions expected from the fits, effectively checking how well a given image fits to the other 10 images. A visual inspection of maps of residual vectors showed that the residuals are not randomly distributed but unveiled some systematic shift and rotation for each image. Since each image is compared with 10 other images, any systematic problem in the given image is
most likely to come from that image and not from a combined effect of the others. Indeed, the interpretation of the observed systematic effect is straightforward, it means that each individual mosaic is not registered perfectly with respect to the sample average, i.e.
the transformation for the respective image is slightly wrong. This systematic error is 
naturally explained by measurement errors of the positions of the SiO maser stars in the respective image. Such an error translates into an error of the parameters of the linear transformation used to tie the astrometric frame to the pixel positions in the mosaic and shows up as a systematic effect in the residuals of the independent set of reference stars.
Thus, we were able to determine better transformation parameters by adding  to the
original linear transformation the linear transformation that minimizes the residuals of the reference star sample, yielding a corrected linear transformation. We applied it to the data and obtained the final linear motion models for the reference stars. The
rms of the 1D-residuals now was $0.55\,$mas. This step changed the position of the origin by $(\Delta \alpha,\Delta \delta) =(-0.01,\,0.05)\,$mas and the velocity 
of the system by less than $4\,\mu$as/yr, these quantities being the mean differences of the respective quantities for the reference stars before and after the refinement. 
Hence, the refinement effectively did not change the coordinate system calibration. We call the coordinate system so defined the `maser system' in the following.

The position of the origin of the maser system and its velocity are uncertain due to two effects: a) the non-zero errors of the SiO maser stars' radio positions and velocities and b) the IR positions of the SiO maser stars show some residuals to the best fitting linear motion, indicative of residual image distortions and of measurement errors in the pixel positions in the IR images. The propagation of the statistical errors into the definition of the coordinate system was addressed using a Monte-Carlo technique. We varied the input to the transformations according to the measured errors and residuals. We created $10^5$ realizations of transformations, assuming a Gaussian distribution of the simulated values around the original values. The standard deviation of the positions obtained for Sgr~A* estimates the positional uncertainty of the maser system under the assumption of uncorrelated measurement errors. We obtained 
$(\Delta \alpha,\Delta \delta) = (0.46, 0.77)\,$mas. Similarly, the standard deviation of the velocities obtained for Sgr~A* estimates the uncertainty of the maser system's velocity under the same assumption. We obtained  $(\Delta v_\alpha,\Delta v_\delta) = (0.29, 0.55)\,$mas/yr.

However, in our data the assumption that the errors from the 11 maser images are uncorrelated is not fulfilled. We rather observe a typical residual per SiO maser star for all epochs
when comparing the transformed, measured positions with the predicted radio positions. Possible reasons are: first, the linear motion models obtained for the SiO masers 
could be inaccurate due to some unknown some unknown systematic problem of the radio positions. Secondly, the radio positions could not be applicable to the IR positions, for instance
if the maser emission would originate from far away of the stellar surface. Thirdly, the correlation could arise due to some unaccounted systematics in the
infrared frames, such as uncorrected distortion. We fitted the residuals of each star with linear functions and obtained in that way estimates for the mean position and
mean velocity uncertainty for each star. Then we calculated the mean deviation (over the SiO maser stars which are $<15$'' away from Sgr~A*) of these linear motion model parameters as 
estimates for the positional and velocity uncertainty of the maser system given the correlations in our data.
With our initial transformation we obtained
$(\Delta \alpha,\Delta \delta) = (0.92\pm0.42,\, 2.22\pm0.43)\,$mas and $(\Delta v_\alpha,\Delta v_\delta) = (0.41\pm0.24,\, 0.29\pm0.24)\,$mas/yr. After
the refinement we got $(\Delta \alpha,\Delta \delta) = (0.95\pm0.73,\, 2.35\pm0.58)\,$mas and $(\Delta v_\alpha,\Delta v_\delta) = (0.38\pm0.41,\, 0.28\pm0.33)\,$mas/yr.
Finally we conservatively adopt for the uncertainties of the maser system  $(\Delta \alpha,\Delta \delta) = (1.0,\, 2.5)\,$mas and 
$\Delta v_\alpha=\Delta v_\delta = 0.5\,$mas/yr. The positional uncertainty is considerably larger than what one would have obtained for uncorrelated residuals.

While the maser system is a direct crosscalibration of maser and reference stars, the resulting velocities of the reference stars are directly sensitive to errors in the velocities of the SiO maser stars, both in the radio data and the NIR $27\,$mas/pix mosaics.
\subsubsection{Single-epoch cross calibration with one maser star image}
\label{clustersys}
The sensitivity of the reference star velocities to errors in the SiO maser velocities can be avoided by an additional step of cross-calibration: We can measure the positions of the reference stars in all maser images with respect to a much larger sample of stars in these images. This cluster of stars is assumed to be non-rotating and not moving with respect to Sgr~A*. The cluster is tied to the astrometric frame for just one epoch, as given by the radio positions of the SiO masers, which can be calculated for the chosen epoch from \cite{rei07}. For all other epochs it is assumed that the mean cluster is stationary in time. Hence, the velocity calibration relies on the statistical argument that for a sufficiently large sample of cluster stars the mean velocity of the cluster is expected to become very small. For a typical velocity of $v$ for a cluster star and $N$ stars, the error of this mean should be of order $v/\sqrt{N}$.

Effectively the few maser stars are only used once in this scheme. Any error in their radio positions, radio velocities or NIR detector positions will therefore translate into a positional offset, but not into a systematic velocity of the coordinate system. The latter is instead connected to the validity of the assumption that the cluster mean is stationary. In order to ensure the best estimate of the velocity calibration we adopted the following procedure:
\begin{enumerate}
\item We selected the maser mosaic from 12 May 2005, which was chosen since it is of good quality and roughly corresponds to the middle of the range in time covered with NACO.
Building upon the work done by \cite{tri08}
we selected an ensemble of stars in that mosaic of which the positions can be measured with high reliability: Take all stars that have a peak flux of more than 25 counts which at the given noise level of 1.9 counts selects high-significance stars. In a second step many stars get excluded again: All stars with more than 700 counts (they 
could be saturated in other frames with longer single detector integration times) and all stars that have a potential source (peak with 5 counts) within 10 pixels.
Furthermore a Gaussian fit was required to yield a FWHM$<0.05\,$pix and the fitted position must coincide with the position obtained using DAOPHOT FIND.
This yielded a sample of 433 stars.
We determined the astrometric positions at the given epoch of the 433 stars by means of a linear transformation that was determined from the eight maser stars.
\item We then determined preliminary astrometric positions for a much larger sample of 6037 stars in all 11 mosaics, by tying their
pixel positions at all 11 epochs to the astrometric positions of the 433 stars at the reference 
epoch with a linear transformation. Note that not taking into account
proper motions in that step makes the velocity calibration independent from 
the radio measurements. The error due to the omission of the proper motions 
is minimized by using 433 stars instead of few masers for the cross-calibration.
\item We fitted linear motion models to all 6037 stars. After the fit we determined the residuals of all star positions in all mosaics. By inspecting the residuals of any mosaic as function of position, we were able to map the residual image distortions in the given mosaic. These residual image distortions can arise due to imperfect registration of the individual exposures to the respective mosaic, or due to an error in the distortion correction applied to the individual frames.
\item For each star in each mosaic we determined an estimate of the residual image distortion by calculating the mean of the residuals of the stars in the vicinity ($r<2''$) of the chosen star. The radius was chosen such that a suitable number of stars was present in the area from which the correction was determined and the area was
sufficiently local. A value of $r<2"$ was a good compromise, yielding 30-50 stars typically. 
The estimate for the residual image distortion was then subtracted from the given star; the typical values
applied were $\Delta x = 0.66 \pm 0.20\,$mas and $\Delta y = 0.62 \pm 0.24\,$mas.
\item In a second fit we used the corrected astrometric positions in order to obtain updated linear motions models for the 6037 stars. 
\item Then we defined the final cluster: it consists of all stars which were present in all 11 mosaics, with radii between 2'' and 15'' and which are not known early-type stars. These criteria yielded a cluster sample size of 2147 stars.
\item We determined the cluster mean velocity, yielding $(-0.04,\,0.00)\,$mas/yr, and subtracted that value from all velocities of the 6037 stars. This then is the final, velocity-calibrated list of linear motions from which the reference star sample is extracted. The mean radius of the 2147 stars of the cluster sample is 9.89'', the root mean square (rms) speed of the stars in the sample is $157\,$km/s$\,\approx4.15\,$mas/yr (for $R_0=8\,$kpc).
\end{enumerate}

We call the coordinate system defined in this way the `cluster system'. Since we expect the mean of the cluster to
show a net motion of order $157/\sqrt{2147}\,\mathrm{km/s}=3.4\,\mathrm{km/s}=0.09\,$mas/yr, we estimate the 
uncertainty of the velocity calibration to be of the same size. We checked this number more thoroughly by means of a Monte Carlo
simulation: We divided the cluster into nine radial bins with boundaries $[2,5,7,8.5,10,11,12,13,14,15]\,$mas (selected
such that in each bin roughly the same number of stars is present). For each bin we have determined the rms velocity
yielding ${3.7, 3.3, 3.0, 2.9, 2.8, 2.7, 2.6, 2.6, 2.6}\,$mas/yr respectively. We then simulated clusters in proper motion space,
for each bin the Gaussian width of the velocity distribution was set to the respective rms velocity and the number of stars was matched to the
real numbers in each bin. For each simulated cluster we were able to obtain in that way a mean velocity; simulating 10000 clusters
allowed us then to estimate the uncertainty of the mean cluster velocity. We obtained $0.06\,$mas/yr; even a bit better than the
simple estimate. Hence, if the assumption of isotropy is correct, the cluster system should allow for a better calibration of the reference star velocities
than with the maser system. The assumption could be wrong, for example if a net streaming motion were present in the GC cluster.

The statistical positional uncertainty of the origin of the cluster system was estimated by the same means as for the maser system. We obtained $(\Delta \alpha,\Delta \delta) = (0.85, 1.51)\,$mas. In addition to these uncertainties, the residuals of the SiO masers also need to be considered, for the epoch at hand the mean deviation
is $(\Delta \alpha,\Delta \delta) = (1.87, 3.12)\,$mas. The uncertainties here are greater than the respective numbers for the maser system due to the fact that the position of Sgr~A* in the cluster system is effectively measured only on one frame while in the maser system it is measured in several and the residuals are not fully correlated. 

\subsubsection{The final, combined coordinate system}
\label{combinedsys}
The maser system has a smaller systematic error in its position calibration, while the cluster system is superior with respect to the velocity calibration. Hence, by combining the two we were able to construct a system that 
combines both advantages. The idea simply is to correct either the velocity calibration of the maser system such that it agrees with the one from the cluster system or to correct the origin of the cluster system such that it coincides with the origin of the maser system (taking into account that the systems refer to two different epochs). Note that this implicitly uses
the fact that the second, refining transformation of the maser system did not change its calibration properties.

We used the sample of reference stars to compare the two systems. 
The mean positional offset between the two lists of positions for the epoch of the cluster system was 
\begin{eqnarray}
\vec{p}_\mathrm{CSys} - \vec{p}_\mathrm{MSys} 
& =& \left( \begin{array}{c}-1.87\\+1.87\end{array}\right)
\pm \left( \begin{array}{c}0.04\\0.04\end{array}\right)\,\mathrm{mas}\,\,.
\label{eqn11} 
\end{eqnarray}
Here, `Csys' denotes the cluster system, `MSys' the maser system. The errors are the standard deviation of the
sample of differences.
We also calculated the differences of the reference star velocities, as given by the two linear motion models obtained for each reference star. We obtained
\begin{eqnarray}
\vec{v}_{CSys} - \vec{v}_{MSys}
&=& \left( \begin{array}{c}-0.60\\+0.56\end{array}\right)
\pm \left( \begin{array}{c}0.08\\0.06\end{array}\right)\,\frac{\mathrm{mas}}{\mathrm{yr}}\,\,,
\label{eqn10} 
\end{eqnarray}
where again the errors are the standard deviation of the
sample of differences.

This means that the two coordinate systems differ significantly in position and velocity calibration in a systematic way. It should be noted that
only the difference between the two coordinate systems is that well defined; for the question how well each of the coordinate systems
relates to Sgr~A*, the larger, systematic errors of sections~\ref{masersys} and~\ref{clustersys} need to be considered.
It is exactly the fact the difference between the coordinate systems is well defined that allowed us to 
combine the two coordinate systems and to gain accuracy in the combined system that way.
Also note that the size of the offsets occurring here are consistent with the combined uncertainties of the two
coordinate system; much larger offsets would have meant that the coordinate systems would be inconsistent
with each other.

Finally we chose the method which corrects the cluster system by a positional offset. The
positional difference from equation~(\ref{eqn11})
was subtracted from all positions of the cluster stars (and thus also from the reference stars that are a subset of the cluster). 
This combined coordinate system has the same prior as the cluster system, namely that the cluster is at rest with respect to Sgr~A*.
The linear motion models so obtained were then used for the further analysis.

\subsection{Relating the S-stars to the reference stars}
We constructed the transformation from pixel positions on the detector to astrometric positions by means of the reference stars. For each given image, we calculated the expected astrometric positions of the reference
stars using the linear motions models as obtained in section~\ref{combinedsys}. Given the pixel positions of the reference stars in the respective image, we related the two sets of positions by means of a cubic transformation (20 parameters) of type
\begin{eqnarray}
x_\mathrm{sky}& =& p_0 + p_1 x + p_2 y + p_3 x^2 + p_4 x y + p_5 y^2 + \nonumber \\
&&p_6 x^3 + p_7 x^2 y + p_8 x y^2 + p_9 y^3 \nonumber \\ 
y_\mathrm{sky}& =& q_0 + q_1 x + q_2 y + q_3 x^2 + q_4 x y + q_5 y^2 + \nonumber \\
&&q_6 x^3 + q_7 x^2 y + q_8 x y^2 + q_9 y^3 \,\,.
\label{trafoCubic}
\end{eqnarray}
Once the transformations are known, it is straight-forward to apply them to the pixel positions of the S-stars.

The parameters $p_i, q_i$ were found by demanding that the transformation should map the two lists of positions optimally in a $\chi^2$ sense. Since the problem is linear, the parameter set can be found with a pseudo-inverse matrix (we always used at least 50 stars, thus 100 coordinates, for 20 parameters). The procedure also allows for an outlier rejection. For this purpose we applied the transformation to reference stars themselves and calculated the residuals to the expected astrometric positions. By only keeping reference star positions which are not more off than $15\,$mas from the expected position we cleaned our sample. This excluded in total 19 of the 7189 reference star
positions. For the cleaned set we redetermined the linear motion model for each star under the side condition that the refinement would not change the mean position or the mean velocity of the sample of reference stars, thus avoiding a change of the origin of the coordinate system and a change of its velocity. 
Compared to previous work the number of reference stars used is roughly a factor eight larger. This reduced the statistical uncertainty of this calibration step to a very small level\footnote{Actually some of the reference stars relatively close to Sgr~A* were also considered as S-stars for which we tried to determine orbits. Indeed, four of those stars showed
significant accelerations. However, we did not exclude them from the sample of reference stars. Therefore, an additional, obvious step of refinement would be 
to allow for quadratic motion models for the reference stars.}.

For the SHARP data we had to use some additional steps for relating the S-star positions to the reference stars, since
for a given epoch we used two deconvolutions for which we had four single-pointing frames and one combined map respectively. 
We used the pixel positions of the reference stars in the two times four single-pointing images together with 
the predicted astrometric positions of the reference stars to set up eight transformations of the type given in equation~\ref{trafoCubic}. Not all reference stars
are present in all pointings, but in all cases their number exceeded 50, such that the transformation parameters were well determined. With these 
transformations we calculated the astrometric positions of the brighter S-stars detected in the eight frames and used the average astrometric position in the end.
The standard deviation of the eight astrometric positions was included
in the error estimate. For the fainter S-stars we used the coadded maps.
For the two coadded maps (two deconvolutions) per epoch we set up two times four full first order transformations relating pixel positions of the brighter S-stars in each coadded map to the respective pixel positions in the four single-pointing frames. With these transformations we determined the
pixel positions of the fainter S-stars which they would have had in the single-pointing frames. These fictitious pixel positions were then transformed
with the cubic transformation of the respective single-pointing frame into astrometric positions. The average of the latter was used in the end,
the standard deviation was included in the error estimate. 

\subsection{Estimation of astrometric errors}
\label{astroerr}
The goal of this section is to understand the errors of the astrometric data. This includes both statistical and systematic error terms. The statistical error is due to the uncertainty of the measured pixel positions. Among the systematic error terms are the influence of the coordinate system, residual image distortions, transformation errors and unrecognized confusion.
\subsubsection{Offset and velocity of the coordinate system}
The accuracy in 2D-position ($\Delta x$, $\Delta y$) and 2D-velocity ($\Delta v_x$, $\Delta v_y$) of the combined coordinate system is given by the numbers in sections~\ref{masersys} and~\ref{clustersys}.
In the third dimension, we don't use any priors for $\Delta z$, since we wish to determine $R_0$ from our data. 

For $\Delta v_z$ we use the prior that Sgr~A* is not moving radially, based both
on theoretical arguments and on radio and NIR measurements. Even if Sgr~A* is dynamically relaxed in the stellar cluster surrounding it, some random Brownian motion
due to the interaction with the surrounding stars is expected. \cite{mer07} calculated this number and concluded that the motion should be $\approx0.2\,$km/s.
This is consistent with the findings of \cite{rei04} who show that Sgr~A* has  a proper motion of $v_l = 18 \pm 7\,$km/s in galactic longitude and $v_b = -0.4 \pm 0.9\,$km/s 
in galactic latitude (assuming
$R_0=8\,$kpc). The significance of the fact that $v_l \ne0$ is disputed, and furthermore it is not clear, whether it is truly due to a peculiar motion of Sgr~A*
or due to a difference between the global and local measures of the angular rotation rate of the Milky Way \citep{rei04}. Clearly, the motion of Sgr~A* perpendicular to
the galactic plane is very small as expected. In the third dimension, the velocity of Sgr~A* can only be determined indirectly by radial velocity measurements of the stellar
cluster surrounding it. Using a sample of 85 late-ytpe stars \cite{fig03} found that the mean radial velocity of the cluster is consistent with 0: $v_z = -10 \pm 11\,$km/s. \cite{tri08}
used a larger sample of 664 late-type stars and found consistently $v_z = 4.6 \pm 4.0\,$km/s. Compared to that, the uncertainty $\Delta U\approx 0.5\,$km/s in the definition of the local standard of rest is much smaller \citep{deh98}. We conclude that all measurements are consistent with Sgr~A* being at rest at the dynamical center of the Milky Way and we assume a prior of $v_z = 0 \pm 5\,$km/s for our coordinate system.

Summarizing, our combined coordinate system should be accurate to the numbers listed here, of which finally used the conservatively rounded values.:
\begin{eqnarray}
\Delta x &=& 0.95 \,\mathrm{mas} \approx 1.0 \,\mathrm{mas}  \nonumber \\
\Delta y &=& 2.35 \,\mathrm{mas}  \approx 2.5\,\mathrm{mas} \nonumber \\
\Delta v_x &=&0.06\, \mathrm{mas/yr}  \approx 0.1 \,\mathrm{mas/yr} \nonumber \\
\Delta v_y &=&0.06\,\mathrm{mas/yr} \approx 0.1 \,\mathrm{mas/yr}  \nonumber \\
\Delta v_z &=&5\,\mathrm{km/s}\,\, .
\label{priorsUsed}
\end{eqnarray}

\subsubsection{Rotation and pumping of the coordinate system}
Potentially, there are two more degrees of freedom, which could affect the reliability of the chosen coordinate system, namely rotation and pumping. An artificial rotation can be introduced if the selected stars by chance preferentially move on tangential tracks with a preferred sense of rotation. Similar, artificial pumping can occur: suppose that by chance all selected stars move on perfect radial trajectories and that stars further out move faster than stars closer to Sgr~A*.
Such a pattern, which would be somewhat similar to the Hubble flow of galaxies, would yield under the set of transformations a time-dependent plate scale and otherwise stationary stars.
Both effects can affect the selection of the reference star sample and (less important) the selection of cluster stars.

The chosen coordinate system relies on the assumption that the cluster does not show any net motion (see section~\ref{clustersys}), net rotation or net pumping. The selection of a finite number of cluster stars however limits the
accuracy with which these conditions can be satisfied. Given 2147 stars with a RMS velocity of $\approx157\,$km/s and a typical distance of 10'' we expect that any selection leads to a pumping or rotation effect of the order of $9\,\mu$as/yr/''. 

Due to the errors in the SiO maser positions, the maser system can show artificial pumping or rotation. Similar to what was done in sections~\ref{masersys} and \ref{clustersys}
we simulated in a Monte Carlo fashion the error propagation. From $10^5$ realizations of the transformations, assuming the observed errors of the
SiO maser positions in the NIR and radio, we created perturbed sets of reference stars. The standard deviation of the pumping and rotation motion ($v_r/r$ and $v_t/r$ respectively) over these sets
then estimate the stability of the maser system. We obtained 
\begin{eqnarray}
\label{eq4}
v_r/r|_\mathrm{MSys} &=& 37\, \mu\mathrm{as/yr} /'' \,\,,\nonumber \\
v_t/r|_\mathrm{MSys} &=& 33 \,\mu\mathrm{as/yr} /'' \,\,.
\end{eqnarray}

The cluster system (and therefore also the combined system) can be checked against the maser system. By calculating the difference in velocity for each reference star and subtracting from those the difference of the two coordinate system velocities we obtained a vector field of residual velocities, which is well described by:
\begin{eqnarray}
\label{eq5}
v_r/r|_\mathrm{CSy}-v_r/r|_\mathrm{MSy} &=& (32 \pm 2|_\mathrm{stat} \pm 9|_\mathrm{sys})  \,\mu\mathrm{as/yr} /''\nonumber\\
v_t/r|_\mathrm{CSy}-v_t/r|_\mathrm{MSy}& = &(6 \pm 2|_\mathrm{stat} \pm 9|_\mathrm{sys})\, \mu\mathrm{as/yr} /'' 
\end{eqnarray}

The combined size of the effects from equations~\ref{eq4} and \ref{eq5}
estimate the error made when using the assumption that the combined coordinate system is non-rotating and non-pumping. 
At 1'' these effects can sum up over 15 years to at most $0.7\,$mas, while for the maximum projected distance of S2 ($\approx0.2$") the resulting positional errors are even a factor 5 smaller. 
We therefore neglected these effects in the following.
\subsubsection{Statistical errors of the pixel positions}
\label{staterrpixpos}
This paragraph deals with the uncertainties of the stellar positions on a given image; the unit of this error term as measured is therefore pixels. The error which is most easily accessible is the formal fit error of the Gaussian fit to a source. However, in deconvolved and beam-restored images it might be a bad estimator for the positional uncertainties. Therefore we compared additionally different deconvolutions of the same image for each epoch in order to get a more robust estimate.

\begin{figure}[htbp]
\begin{center}
\plotone{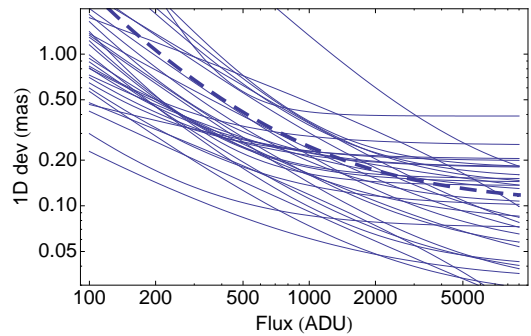}
\caption{The statistical errors of the pixel positions for the NACO K-band data as a function of arbitrary detector units of flux. The thin lines show the respective error model for each epoch; the thick dashed line is the mean for the data.
The mean has a floor at $99\,\mu$as, the median (not shown) at $84\,\mu$as.}
\label{statpixerrors}
\end{center}
\end{figure}

For the SHARP data we used up to eight (= two deconvolutions $\times$ four pointings) pixel positions. The standard deviation of the astrometric positions was 
included in the error estimate for the statistical position error. For stars which were present only in one frame, the typical error of the
epoch was used instead. 

\begin{figure}[htbp]
\begin{center}
\plotone{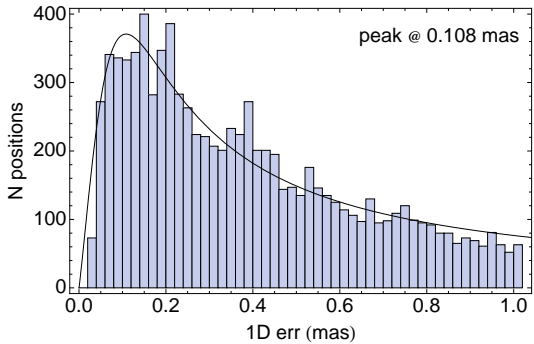}
\caption{The measured distribution of the statistical errors of the pixel positions for the NACO  data. The characteristic statistical error (defined as the peak of the distribution) is $108\,\mu$as, the
systematic error terms have to be added to this to come to a fair estimate of the true uncertainty.}
\label{measuredNACOerr}
\end{center}
\end{figure}

For NACO we split up each data set into two parts and deconvolved both co-added images with the same point spread function as the co-added image of the complete data set  (see Section \ref{naco}). We determined the pixel positions of the reference and S-stars in the two deconvolved frames and applied a pure shift between
the two lists of pixel positions such that the average pixel position is the same for both.
The remaining difference between respective positions of one star estimates the statistical uncertainty for that star. The error estimates obtained this way were a strong function of the stellar brightness. Therefore we described the error estimates as a function of flux for each epoch (see Figure~\ref{statpixerrors}) using a simple empirical model of the form $a x^{-n} +b$. The mean floor $\bar{b}$ over all data sets is $99\mu$as, while for lower fluxes the error increases up to $2\,$mas. We used the empirical description of each image to assign an error to all stellar positions obtained from that frame. Finally we checked whether the formal fit error of the positions was greater than the estimate from the empirical error model. In such a case we used the formal fit error instead. Figure~\ref{measuredNACOerr} shows the final distribution of statistical errors for the NACO data. It is effectively the mean error model folded with the brightness distribution of the S-stars.

For the SHARP data we obtained a broad distribution of the statistical pixel position errors with no clear maximum and a tail to $2\,$mas. The median error is $360\,\mu$as, the mean error $760\,\mu$as in the SHARP data.
\subsubsection{Residual image distortions}
\begin{figure}[htbp]
\begin{center}
\plotone{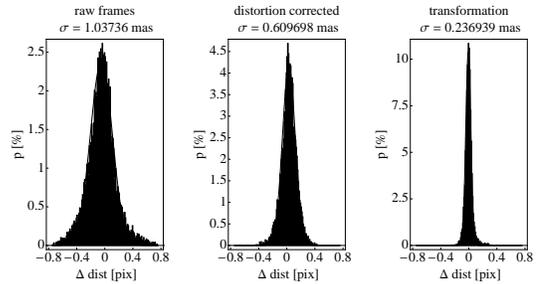}
\caption{Determination of residual image distortions for the NACO H-band data from September 8, 2007, $13\,$mas/pix. The histograms show the differences of detector distances for a set of bona fide stars as measured in the four pointing positions with a dither offset of 7''. Left: Using the raw frames. Middle: After application of a distortion model, Right: After transforming the raw positions with a cubic transformation onto a common grid. The corresponding 1D coordinate errors are determined from Gaussian fits to the distributions and are quoted at the top of each panel.}
\label{distortion13}
\end{center}
\end{figure}

A main source of error at the sub-milliarsecond level is image distortions. We estimated this error term by comparing distances of stars in different pointing positions
with a dither offset of $7"$ (see Figure~\ref{distortion13}). 
If we had used only the raw positions and linear transformations, the resulting mean 1D position error would be as large as $1\,$mas for the $13\,$mas/pix NACO data. By applying a distortion model (see section~\ref{naco}) plus a linear transformation this error can be reduced to $600\,\mu$as. Allowing for a cubic transformation onto a common grid yields an error of $240\,\mu$as only. This justifies our choice to use a high order transformation rather than to de-distort the $13\,$mas/pix NACO images. The numbers obtained in this way are actually the combined error of the statistical and transformation uncertainties with the residual image distortions. Subtracting the former we conclude that residual image distortions have a contribution of $210\,\mu$as to the error budget of each individual astrometric data point. We thus added this value in squares to all other error terms, effectively
acting as a lower bound for the astrometric errors. 

\begin{figure}[htbp]
\begin{center}
\plotone{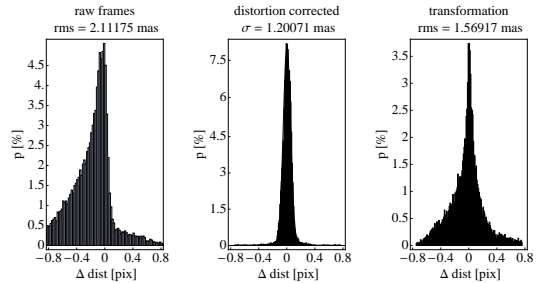}
\caption{Determination of residual image distortions for the NACO K-band data from March 16, 2007, $27\,$mas/pix. The histograms show the differences of detector distances for a set of bona fide stars as measured in the four pointing positions with a dither offset of 14''. Left: Using the raw frames. Middle: After application of a distortion model, Right: After transforming the raw positions with a cubic transformation onto a common grid. The corresponding 1D coordinate errors are quoted at the top of each panel, in the middle panel the value is determined from a Gaussian fit, for the other two the rms is quoted due to the non-Gaussianity of the distributions.}
\label{distortion27}
\end{center}
\end{figure}

We applied the same analysis to the $27\,$mas/pix NACO data which had a dither offset of $14"$ (see Figure~\ref{distortion27}). The raw differences showed a skewed distribution, indicating the presence of image distortions. The rms of this distribution is $2.1\,$mas. After applying the distortion model the typical residual error is reduced to $1.3\,$mas and the distribution is a nice Gaussian. Interestingly, mapping the positions with a cubic transformation onto each other does less well here. The distribution becomes less skewed, however it is still non-Gaussian and the rms is $1.6\,$mas. This justifies a-posteriori the use of the distortion correction for the $27\,$mas/pix NACO data when determining the motion models for the reference stars. 
For the SHARP data we obtained a characteristic error due to residual image distortions of $0.8\,$mas and a median of $1.2\,$mas.

\subsubsection{Transformation errors}
It is important to notice that any error in deriving the motions for the reference stars only translates into a global uncertainty of the coordinate system (which could show up as an offset of the center of mass from 0/0 or a net motion of the coordinate system). It will however not affect the accuracy of individual data points in this system. Only the selection of reference stars and transformation errors contribute to the errors of the individual data points. We estimated them by performing all coordinate transformations not only once but also with subsets of the available reference stars. The standard deviation of the sample of obtained astrometric positions was then included in the astrometric error estimate. The typical uncertainty introduced by the transformations was quite small, namely $23\,\mu$as for the NACO data. 
This is consistent with the fact that $\approx100$ stars have been used of which each can be determined with an accuracy of $\approx200\,\mu$as. For the SHARP data we found a value of $100\,\mu$as, again consistent with the characteristic single position error of $\approx1\,$mas.
\subsubsection{Differential effects in the field of view}
At the sub-mas level, there is a multitude of differential effects over the field of view that can influence astrometric positions. The most prominent ones are relativistic light deflection in the gravitational field of the sun, light aberration due to Earth's motion or refraction in the atmosphere. Since our analysis is based on relative astrometry, the absolute magnitudes of the effects do not matter. Only the differential effects over the field of view can contribute to the positional uncertainties.
\begin{itemize} 
\item Over a field of view of 20'' the differential effects of aberration can be described by a global change of image scale \citep{lin06}. Since we fit the image scale for each epoch separately, the differential aberration is absorbed into the linear terms of the transformation and thus is not affecting the astrometry. The size of the effect for a small field of view with a diameter $f$ amounts to $f \times v/c \times \cos \Psi$ where $\Psi$ is the angle between the observation direction and the apex point. For $f\approx10''$ and $v\approx30\,$km/s
this yields $\approx1\,$mas at most.
\item The light deflection can be approximated by $4\,\mathrm{mas}\times \cot \Psi/2$ where $\Psi$ is the angle between observation direction and Sun \citep{lin06}. The differential effect over 20'' will not exceed $100\,\mu$as as long as $\Psi>3.6^\circ$, which is guaranteed for all our data.
\item From the usual refraction formula $R=44'' \, \tan z$ (for a standard pressure of $740\,$mbar at Paranal) we find a differential effect of $4-8\,$mas over 20'' or $2-4\,$mas over the field in which we selected the reference stars. The effect will be a change in one direction (towards zenith) of the image scale. Since the effect is at most quadratic over the field, it will be absorbed completely into the first and second order terms of the transformations. Note that it is crucial to allow also for skew terms, i.e. it is not sufficient to use a shift, rotation and scale factor only in the linear terms, but the off-diagonal terms in the transformation matrix are also required.
\end{itemize}
\subsubsection{Unrecognized confusion}
One important contribution to the position errors is the fact that stars can be confused and that sometimes the confusion is not recognized. This problem is more severe for the SHARP data than for the NACO data due to the lower resolution. Of course we excluded positions for which we know that they are confused. However, unrecognized confusion cannot be dealt with by principle. We therefore simply accept that these events happen. This means in turn, that we expect to find a reduced $\chi^2>1$ when trying to describe the motions with smooth functions. In addition we note that for a sufficiently large amount of data points unrecognized confusion events should only lower the precision but not the accuracy since no global bias is expected. Still, if a confusion event happens during an unfortunate part of the orbit (for example at an end point or during pericenter passage)
a bias in the results of an orbit fit can be introduced. 
\subsubsection{Gravitational lensing}
Gravitational lensing might affect the measured positions. A quantitative analysis shows that the effects are very small except in unusual, exceptional geometric
configurations. For a star at a distance $z\ll R_0$ sufficiently far behind Sgr~A* the angle of deflection as measured from Earth is
\begin{equation}
\theta=\frac{z}{R_0} \frac{4 G M_\mathrm{MBH}}{c^2\,b},
\end{equation}
where $b$ is the impact parameter. For the GC, this evaluates to $\theta\approx20\,\mu\mathrm{as} \times z/b$, indicative of a very small astrometric effect unless $z/b\gg1$
This rough estimate is consistent with the rigorous treatment of the problem from \cite{nus04}, who show that in order to achieve a displacement of $1\,$mas, a star at $z\approx1000\,$AU needs to have $b\approx2\,$mas$\,\approx16\,$AU. In our data set, none of the stars get close to the regime that gravitational lensing
actually becomes important. Therefore, we neglected the effect.
\subsubsection{Comparison of error estimates with noise}
We were able to check how well our error estimates agree with the intrinsic noise of the data. For this purpose we fitted all measured positions of the reference stars with simple quadratic functions. After exclusion of $3\sigma$-outliers, we have calculated the reduced $\chi^2$ for each reference star. The mean reduced $\chi^2$ for the NACO data is $2.0\pm0.7$, while for the SHARP data we obtained values between 0.5 and 2.0 with a mean of 1.0.

\begin{figure}[htbp]
\begin{center}
\plotone{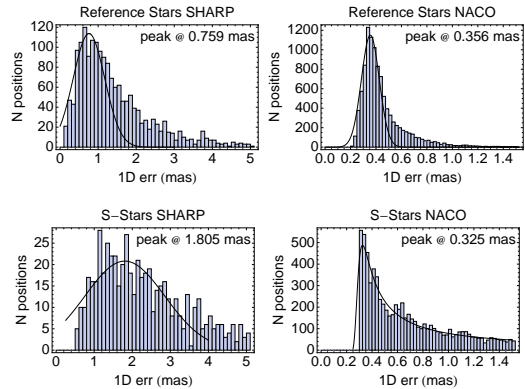}
\caption{Final distribution of total astrometric errors for our data. Left column: SHARP data, right column: NACO data. Top row: reference stars, bottom row: S-stars. The curves show empirical fits to the histograms in order to determine the respective characteristic error as the peak of the distribution.}
\label{finalerrors}
\end{center}
\end{figure}

Since our data set consists of two subsets (SHARP and NACO), each covering roughly the same amount of time, the relative weight of the two subsets matters. Given that we seem to underestimate the errors for NACO a bit, while the SHARP errors seem consistent with the noise in the data, we decided to apply a global rescaling factor of $r=1.42$ to all NACO data points. This procedure adjusts the relative weight between the two subsets.
Still we expect a reduced $\chi^2>1$ when performing orbit fits due to unrecognized confusion events.

In Figure~\ref{finalerrors} we show the final error distributions (after rescaling all NACO errors with the global factor) for the S-stars and reference stars, both for the NACO and SHARP data. The characteristic error for a reference star in the NACO data is $360\,\mu$as, in the SHARP data it amounts to $760\,\mu$as. For the S-stars, the histogram of the NACO errors has a peak also around $325\,\mu$as and a tail towards larger errors, essentially telling us that for bright S-stars the astrometry is as good as one could hope for (since it is equally good as for the reference stars). The tail is due to the fact that many of the S-stars are faint (hence the statistical error is severe) and probably also unrecognized confusion events affect the statistical error since confusion can alter the shapes of the images of faint stars. In the SHARP data, the typical S-stars error is $2\,$mas and the lower end of the distribution at $\approx1\,$mas is consistent with what could be expected from the reference stars.
\subsection{S2 in 2002}
\label{s22002}
\begin{figure}[htbp]
\begin{center}
\plotone{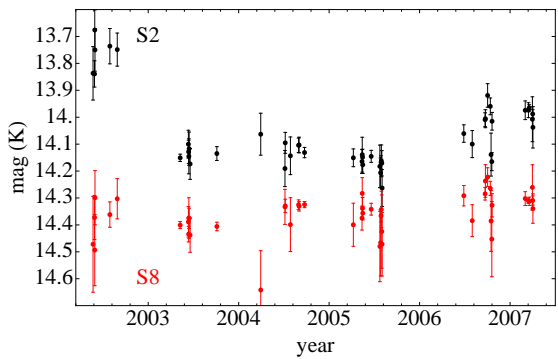}
\caption{The K-band magnitude of S2 as function of time in the NACO data, determined by means of PSF photomery (black data). For comparison the star S8 is shown (red data).}
\label{s2brightness}
\end{center}
\end{figure}

Our data set covers the pericenter passages of several stars. Particularly important to our analysis is the one of the star S2. The star is one of the brightest in the sample and we observed a full orbit (see Figure~\ref{s2orbit}). In 2002 S2 passed its pericenter, thus changing quickly in velocity throughout a period of a few months. These data are particularly useful for constraining the potential of the MBH. However, as we will now discuss, the photometry of the star near pericenter-passage is puzzling and may indicate that the positional information is affected by a possible confusion event with another star. Figure~\ref{s2brightness} shows a K-band PSF-photometrically determined light curve for the star \citep{ran07}. It is clear that S2 was brighter in 2002 than in the following years. There are several reasons why a star could change its apparent brightness. 

\begin{enumerate}
\item In 2002, S2 was positionally nearly coincident with Sgr~A* and thus confused with the NIR counterpart of the MBH. Typically,
Sgr~A* is fainter than $m_\mathrm{K}=17$ and thus the extra-light from Sgr~A* in quiescence is not sufficient to explain the observed
increase in brightness of S2. However, Sgr~A* is known to exhibit flares that can reach a brightness level that could
account for the observed increase in brightness \citep{gen03a,tri07}. In that case we would expect to see intra-night variability of S2 in the 2002 data.
Assuming conservatively that we can determine the relative flux of S2 to $\Delta m_\mathrm{K}=0.1$ in each frame and given
the brightness of S2 ($m_\mathrm{K}\approx14$) we estimate that we would have noticed any variations in Sgr~A* that exceed $m_\mathrm{K}\approx16.5$.
Since we did not observe any intra-night variability we exclude that flares from Sgr~A* significantly contributed to the increased brightness of S2 in 2002.
\item Intrinsic variability of S2 might explain the observed light curve. However, it is unlikely to be the correct explanation, since it would be a big coincidence that the brightening happens during the pericenter passage. Also an eclipsing binary seems unlikely given the slow variation.
\item The star could change its properties during the pericenter passage. While tidal heating \citep{ale05} cannot plausibly change the temperature of a star within a few months, the interaction of S2 with some ambient medium does not seem ruled out. Such an encounter would primarily change the surface temperature of the star and therefore would act nearly instantaneously. Effectively the light curve would then be a direct trace of the density of the surrounding gas encountered along the orbital path of S2. However, energetically, this scenario seems unlikely: Given the maximum velocity of S2 at pericenter ($v\approx8000\,$km/s), the radius of the star ($r=11R_\odot$, \cite{mar08}) and assuming that the kinetic energy of the gas that hits the geometric cross section of the star is converted to radiation, one can estimate the number density $n$ necessary to produce the observed brightness increase of $\Delta m_K\approx 0.5$. We obtained $n\approx10^{11}\mathrm{cm}^{-3}$, which is unrealistically high, and so
we do not favor this scenario.
\item \cite{loe04} proposed that the stellar winds of early-type stars passing their pericenters close to the MBH could
alter the accretion flow onto Sgr~A*. Such an event would produce a change in the brightness of Sgr~A* on the timescale
of months, compatible with Figure~\ref{s2brightness}. However, \cite{mar08} showed that the mass loss rate of S2 is too low for this mechanism to work.
\item The extinction could be locally smaller than the average value. For instance, Sgr~A* could remove dust in
the interstellar medium in its vicinity. This hypothesis can be tested in the future by observing other S-stars passing close to Sgr~A* during the pericenters
of their orbits.
\item The brightness of S2 could be affected by dust in the accretion flow
onto the MBH. The dust
would be heated by S2 and account for the excess brightness, a proposal that was used by
\cite{gen03b} to explain the MIR excess of S2/Sgr~A*.
\item The star could be confused with another star. If S2 had been located very close to another star in projection, the true nature of this encounter could remain undiscovered, but the observed brightness of S2 would be increased.
\end{enumerate}
Of the three viable explanations (5 to 7), the first would not lead to astrometric biases, the others however would displace S2 artificially. Given the importance of the 2002 data, we decided not to discard it completely but to estimate the astrometric error assuming a confusion event, given the measured increase in brightness.

\begin{figure}[htbp]
\begin{center}
\plotone{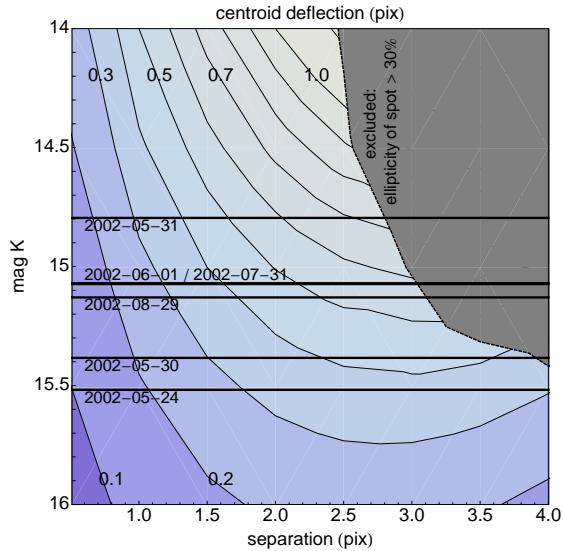}
\caption{Simulation of a confusion event. The contour lines show by which amount a $m_\mathrm{K}=14$ source is displaced if it is confused with a second source that has certain magnitude and that is located in a given distance. The units are pixels, the simulation assumed simple Gaussian point spread functions that are sampled as it is the case for the NACO detector in K-band. The area to the top right can be excluded since a relatively bright source a few pixels apart from the primary would produce an elongated shaped image (which is not observed for S2 in 2002). The line denotes the limit at which the major axis is 30\% larger than the minor axis. The horizontal lines indicate the brightnesses that a secondary source would have needed to push the S2 brightness up by the observed amount for the observed magnitudes at the dates indicated. For each date a mean deflection can be read from this plot. That value is used as astrometric error for S2 at the given date.}
\label{s2deflection}
\end{center}
\end{figure}

For this purpose we simulated confusion events. We assumed simple Gaussian point spread functions and sampled them as they are sampled by the 13$\,$mas/pix scale of the NACO camera in K-band. By polluting a primary source with a fainter secondary source we generated a confused stellar image. This was then fit by a two-dimensional Gaussian and the displacement from the position of the primary source was determined. We varied brightness ratio and distance between the two sources systematically, yielding a displacement map (Figure~\ref{s2deflection}). This map allows the determination of the possible range of displacements if the brightness of the secondary source is known. The range can be constrained further, since a bright secondary source in a few pixels distance will lead to very eccentric images that would be easily detected in the data. We excluded all points that would lead to a stellar image of which the major axis is more than 30\% larger than the minor axis. 
Thus, from the measured S2 fluxes, the known, unconfused brightness of S2 and the roundness of the S2 images, we were able to constrain the astrometric bias due to confusion. 
For each date we looked up in figure~\ref{s2deflection} the possible range of astrometric displacements given the observed brightness of S2, essentially determining the
profile along a horizontal line in the plot. The mean of this distribution was then considered as an additional 2D error to be added to the respective astrometric errors for that date.
The such obtained error terms ranged between $2.37\,$mas and $3.76\,$mas  

\begin{figure}[htbp]
\begin{center}
\plotone{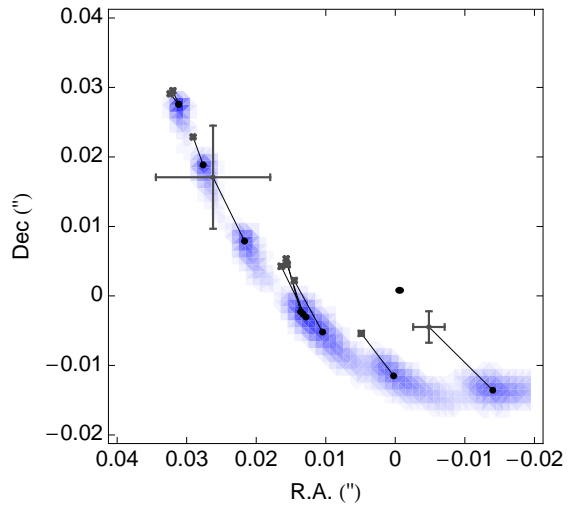}
\caption{The 2002 data of S2. The grey symbols show the measured positions, the errors are as obtained
from the standard analysis and are not yet enlarged by the procedure described in section~\ref{s22002}. The black dots are the positions predicted for the observation dates
using an orbit fit obtained from all data other than 2002. The blue shaded areas indicate the uncertainties in the predicted 
positions resulting from the uncertainties of the orbital elements and of the potential, taking into account parameter correlations. The little ellipse
close to the origin denotes the position of the fitted mass and the uncertainty in it. This plot shows that the S2 positions
are dragged for most of the data by $\approx10\,$mas to the NE; they are not biased towards Sgr~A*.}
\label{s2residua}
\end{center}
\end{figure}

We checked whether the residuals of the 2002 data, relative to an orbit fit to the data other than 2002, show some systematic trend (figure~\ref{s2residua}) and found that indeed all points appear to be shifted systematically by $10\,$mas$\,\approx1\,$pix towards the NE. Still, this is hard to interpret. In particular, S2 does not appear systematically displaced towards Sgr~A*. Extrapolating backwards the track of S19 that was observed from 2003 on shows that it also was located close to the S2 positions in 2002. Again, there is no indication that S2 would be displaced towards the extrapolated positions of S19. Also, S19 with $m_\mathrm{K}\approx16.0$
is too faint to account for the observed increase in brightness of S2. Any other star that potentially was close to S2 in 2002 (candidates are S23, S38, S40, S56) is even fainter. 
From figure~\ref{s2deflection} one can see that a star with $m_\mathrm{K}\approx 14.4 - 14.0$ in a distance of $2-2.5$ pixels would be required to account for the
observed shift. Furthermore, that secondary source would have to move for a few months and for $\approx40\,$mas nearly parallel to S2.
It is extremely unlikely that we have missed such an event.

From this analysis, it is clear that the weight of the 2002 data will influence the resulting orbit fits, since these points will systematically change the orbit figure at its pericenter. At the same
time we have no plausible explanation for the increase in brightness and the systematic residuals in the 2002 data; in particular a confusion event seems unlikely. Thus, it
is clear that using the 2002 data will affect the results, but we cannot decide whether it biases towards the correct solution or away from it.
Therefore
we use in the following two options: a) we include the 2002 data with the increased error bars; b) we completely disregard the 2002 data of S2.

\section{Analysis of spectroscopic data}
Most of the radial velocities were obtained with SINFONI. For the few non-SINFONI data we used the already published values \citep{ghe03, eis03b}.  

From the SINFONI cubes we determined spectra by manually selecting on- and off-pixels for each S-star and calculating the mean of the on-pixels minus the mean of the off-pixels. The spectra were then used to determine the radial velocities of the respective stars at the given epoch. We only used spectra in which we were able to visually identify the stellar absorption lines without doubt. The most prominent features are the Br-$\gamma$ line for early-type stars and the CO band heads for late-type stars. 

Both line profiles are non-trivial, possibly biasing the result when using a simple Gaussian profile to fit the line. The bias can be avoided by crosscorrelating the spectra with a template and determining the maximum of the crosscorrelation.  

For the CO band heads we used a template spectrum from \cite{kle86}. We used the well-established tool `fxcor' which is part of NOAO-package in iraf. We identified the following stars as late-type stars: S10, S17, S21, S24, S25, S27, S30, S32, S34, S35, S38, S45, S68, S70, S73, S76, S84, S85, S88, S89, S111.

\begin{figure}[htbp]
\begin{center}
\plotone{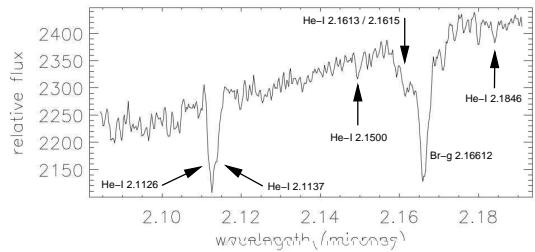}
\caption{The combined S2 spectrum from the 2004 - 2006 SINFONI data, used as velocity template.}
\label{s2spec}
\end{center}
\end{figure}

Also for the early-type stars one might be worried that radial velocity measurements are biased due to
a complex line profile. In particular, Br-$\gamma$ might be affected by nearby He lines. We tested this for the bright star S2, by generating a template from our 2004 - 2006 data\footnote{The combined S2 spectrum created in this context was also the basis for the work of \cite{mar08}.}: we estimated for all S2-spectra the velocities by simple Gaussian fits to the Br-$\gamma$ line. We then Doppler-shifted all spectra to the 0-velocity (using the iraf task 'dopcor') and coadded them (using the iraf task 'scombine').
This resulted in a first template for S2. With this template we crosscorrelated all individual S2-spectra in the wavelength range $2.08-2.20\,\mu$m (using
the iraf task 'fxcor') and obtained better estimates for the velocities. With these new velocities we reassembled the template spectrum. We stopped after this first iteration since the velocity differences had already converged to a mean deviation of $0.2\,$km/s with a standard deviation of $2\,$km/s. This template spectrum is shown in Figure~\ref{s2spec}. We used it to determine the final S2-velocities. Comparing the results to the initial estimates of the velocities showed that the Gaussian fits were not notably biased.
The mean velocity difference was $8\,$km/s with a standard deviation of $27\,$km/s.
Therefore we simply used the Gaussian fits to the Br-$\gamma$ line for the other early-type stars. We identified the following stars as early-type:
S1, S2, S4, S5, S6, S7, S8, S9, S11, S12, S13, S14, S18, S19, S20, S22, S26, S31, S33, S37, S52, S54, S65, S66, S67, S71, S72, S83, S86, S87, S92, S93, S95, S96, S97.

Before the measured velocities can be used in a fit they have to get referred to a common reference frame. The most suitable choice is the LSR. We used standard tools to determine the corrections which for our data only depend on the observing date and the source location. The observatory's position on Earth does not matter at the level of $15\,$km/s accuracy, since it leads to a correction $<0.5\,$km/s.
\subsection{Radial velocity errors}
\label{rverrs}
All radial velocities crucially depend on an exact wavelength calibration. 
The errors in the radial velocities were estimated from the following terms: 
\begin{itemize}
\item {\bf The formal fit error}. For radial velocities which were obtained from a cross correlation with a template
spectrum, the formal fit error is given by the fit error of the peak in the cross correlation, which is calculated 
routinely with the cross correlation routine. For the data for which we fitted a simple line profile to the spectrum the formal fit error is also an output of the fit routine. The magnitude of this error depends on the spectral type and the SNR in the spectrum. For a bright late-type star, e.g. S35 with $m_\mathrm{K}\approx13.3$, the formal fit error can be as small as $10\,$km/s, for a bright early-type star, e.g. S2 with $m_\mathrm{K}\approx14.0$, a typical value is $30\,$km/s. 
\item {\bf Accuracy of wavelength calibration for Br-$\gamma$}. We used the non-sky-subtracted data cubes in order to determine the positions of atmospheric OH-lines. Comparing those to the nominal positions allowed us to estimate the accuracy of the wavelength calibration in the range of Br-$\gamma$ and the He-lines around $2.11\,\mu$m. The rms of the OH-line positions around their nominal positions yielded
errors in the order of $2-3\,$km/s.  
\item {\bf Accuracy of wavelength calibration for CO band heads}. Since there are no OH emission lines at wavelengths longer than $2.25\,\mu$m, we used atmospheric absorption features in the non-atmosphere-divided spectra of the respective standard stars in order to asses the accuracy of the wavelength calibration at the wavelengths of the CO band heads. This was possible since our standard stars were early-type stars (spectral type around B5) that do not show
spectral features at the region of interest. 
We divided the region from $2.25\,\mu$m to $2.40\,\mu$m into short windows of $\Delta \lambda = 0.05\,\mu$m and cross correlated each with a respective theoretical spectrum of the atmosphere. The typical resulting deviation was measured to be $10\,$km/s. The accuracy of the procedure is limited however by the accuracy by which the individual deviations can be measured, which yielded a value of $10\,$km/s, too. So probably the calibration is even more accurate than $10\,$km/s and consistent with what is found for the accuracy of the calibration for the shorter wavelengths.
\item {\bf Uncertainty of the underlying spectrum}. The GC region is highly confused. Therefore we did not use an automated procedure to extract the spectra from the data cubes but selected the respective signal and off pixels manually.
Since there is no clear prescription for what the optimum way for that procedure is, we extracted each spectrum several times. This allowed us to estimate the error due to the selection of signal and off pixels. While for bright stars ($m_K\approx 14)$ this error term is below $10\,$km/s, it becomes dominant for fainter stars. For an early-type star of $m_K\approx 15.5$ a value of $100\,$km/s is common.  
\end{itemize}
Since the wavelength calibration is determined independently for all data sets, these errors will average out with an increasing database.

\section{Orbital fitting}
\label{orbitalfitting}
The aim of the orbital fitting is to infer the orbits of the individual stars as well as
information on the gravitational potential. A Keplerian orbit can be described by the 
six parameters semi major axis $a$, eccentricity $e$, inclination $i$, angle of the line of nodes
$\Omega$, angle from ascending node to pericenter $\omega$ 
and the time of the pericenter
passage $t_\mathrm{P}$. If the orbit is only approximately Keplerian, these parameters
should be interpreted as the osculating orbital parameters.
The parameters describing a simple point mass potential are the 
distance to the GC, $R_0$,
the mass of the central object, $M_\mathrm{MBH}$, its position and velocity. Note that
the potential might also be more complicated, for example due to an extended mass component
or due to the corrections arising from the Schwarzschild metric.
These parameters can be inferred from our data by orbital fitting. 

After 16 years of high-precision astrometry of the innermost stars in our galaxy and a 
few years of Doppler-based radial velocity measurements the accuracy of the available data has 
reached a level at which one might hope to detect deviations from the Keplerian orbits on which 
the stars apparently move due to the existence of the MBH at the dynamical center 
of the Milky Way. Such deviations may be due to relativistic effects or are the effect of an extended mass 
component possibly residing in the vicinity of the MBH. Both cases are scientifically highly interesting.
In order to analyze these effects we implemented a general orbital fitting routine 
that permits the fitting of orbits in an arbitrary potential and that can take into account also relativistic effects.

For a $1/r$ potential it is well-known that the solutions of the equations of motion of test particles 
are (Kepler) ellipses. Assuming such a potential, orbits can be fitted 
by adjusting the orbital elements, since there is a straightforward prescription for
the calculation of the position and velocity vectors at any given time from the
orbital elements. 
However, a more general approach is needed if an arbitrary potential determines 
the dynamics. Then the trajectory has to be determined numerically. 
The problem can be described by the initial conditions of each test 
particle plus the parameters describing the potential. For each set of 
parameters a $\chi^2$ with respect to the measured data can be calculated. 
One seeks the parameter values which minimize the $\chi^2$. This is a 
computationally demanding problem as at each step of the high-dimensional 
minimization the equations of motion are solved numerically. We chose 
the high-level tool {\it Mathematica} \citep{wol05} for the implementation and tested it thoroughly, e.g. by comparing results with results obtained from the former routine that explicitly uses ellipse-shaped orbits and that was used for the work of \cite{eis05}. Some features of the new routine are:
\begin{itemize}
\item The NIR flares of SgrA* are believed to appear at the position of the center of mass for the orbits \citep{gen03a}. When a flare occurs it therefore is reasonable to take the measured position of the flare into account and to identify it with the center of mass. This can be achieved by letting this measurement contribute to the $\chi^2$ of the fit. In total we measured 22 times a position of Sgr~A* (at various brightness levels, typically at $m_\mathrm{K}\approx$15). Note that with such a fit, while possibly constraining the potential parameters better, one gives up the possibility of testing whether the center of mass and the NIR counterpart of Sgr~A* coincide.
\item We implemented four relativistic effects:\\ 
a) the geometric retardation due to the finite speed of light, also called the Roemer effect. This involves numerically solving the retardation equation $t_\mathrm{obs}=t_\mathrm{em}-z(t_\mathrm{em})/c$, where z is the coordinate along the line of sight, in order to
know the position and velocity of the star at the time of emission.\\ 
b) the relativistic Doppler formula, giving rise to the so-called transverse Doppler effect, affecting only the radial velocities.\\ 
c) the gravitational redshift due to the potential of the central point mass, altering the conversion of line positions to radial velocities. \cite{zuc06} show that effects a) - c) might become visible in the radial velocity measurements during a close periastron passage of a star.\\ 
d) the first general relativistic correction to the Newtonian potential as given by the Schwarzschild metric: $V(r)=-G M_\mathrm{MBH}/r + G M_\mathrm{MBH} l^2/c^2 r^3$ where $l$ is the orbital angular momentum of the star.\\ 
Within the fitting routine all four effects can be turned on or off, or the strength of the effect can be used as a fit parameter where 0 means the effect is not present and 1 corresponds to the case in which the effect is as strong as expected from the theory. 
\item We allow for additional mass components in the potential, described by an arbitrary number of additional parameters, all of which can be either treated as fixed or as free fit parameters. The additional mass components can be given either as a term in the potential or as a function describing the density as function of the spatial coordinates. In the latter case the routine determines the potential from the mass distribution by solving the Poisson equation $\nabla^2 V(r) = 4 \pi G \rho(r)$.  Here, one encounters either a case in which a closed solution for $V(r)$ can be found or it might happen that for each set of parameters for which $\chi^2$ is calculated during the fit the Poisson equation has to be solved numerically. 
\item For some of the parameters of the problem there could exist independent measurements which one might want to take into account during the fit.
An example is the position of the central mass. We used radio measurements of Sgr~A* to determine the coordinate system and thus we expect the central
point mass to reside in the origin of the chosen coordinate system.    
We therefore implemented the use of priors for any of the parameters, which can be done straightforwardly by including them into the calculation of $\chi^2$.
\item Instead of fitting the semi-major axis, we fit the periastron distance $p$. This has the advantage that we can allow values of $e<0$, effectively exchanging the role of major and minor semi axis. By using $p$ the parameter space is compact and the fitting routine can smoothly pass $e=0$.
\end{itemize}
We followed the usual approach when calculating the statistical fit errors \citep{pre92}. For the given best fit solution at a certain set of values $\{p_i\}$ for the parameters we determine the Hessian matrix from the curvature of the $\chi^2$-surface: $\partial^2 \chi^2/\partial p_i \partial p_j$. The formal fit errors are the diagonal elements of the inverse of that matrix. Note that still these are only formal, statistical fit errors. Possible systematic errors come in addition to them. Parameter correlations are taken into account by the matrix inversion. All orbital elements for a given star are correlated with each other and with the potential parameters. However, the other matrix elements describing correlations between orbital elements of different stars can be set to 0. This reflects the test particle approach in which one star can only influence the fit result for another star via its influence on the potential.
We explicitly use the test particle approach also when calculating $\chi^2$ for more than one star. It allows one to
use several CPUs in parallel since the contributions to $\chi^2$ from the individual stars are independent.

\section{Results}
\label{results}
In order to predict the motion of a star in a given gravitational potential one has to know six phase space coordinates, e.g. its position and velocity at a given time. Since the radial position is not measurable for any of the S-stars and only for a few the radial velocity is measured, one needs additional dynamical quantities. As such one can use accelerations, either in the proper motion or in the radial velocity. Also higher order derivatives (e.g. $d a/dt$) of the astrometric data can be used as additional dynamical measurables. 
If more than six dynamical quantities are measures, the star can be used to retrieve information about the potential.

This section is organized as follows: First, we check by polynomial fits (going up to third order), for which stars we can expect to find orbital solutions and which stars can contribute in the determination of the potential. Then we determine the potential, yielding also the orbits of the stars used in this step. Finally, we determine the orbits of the remaining stars in the given potential.

\subsection{Polynomial fits}
\label{polyfits}

\begin{figure*}[htbp]
\begin{center}
\epsscale{1}
\plotone{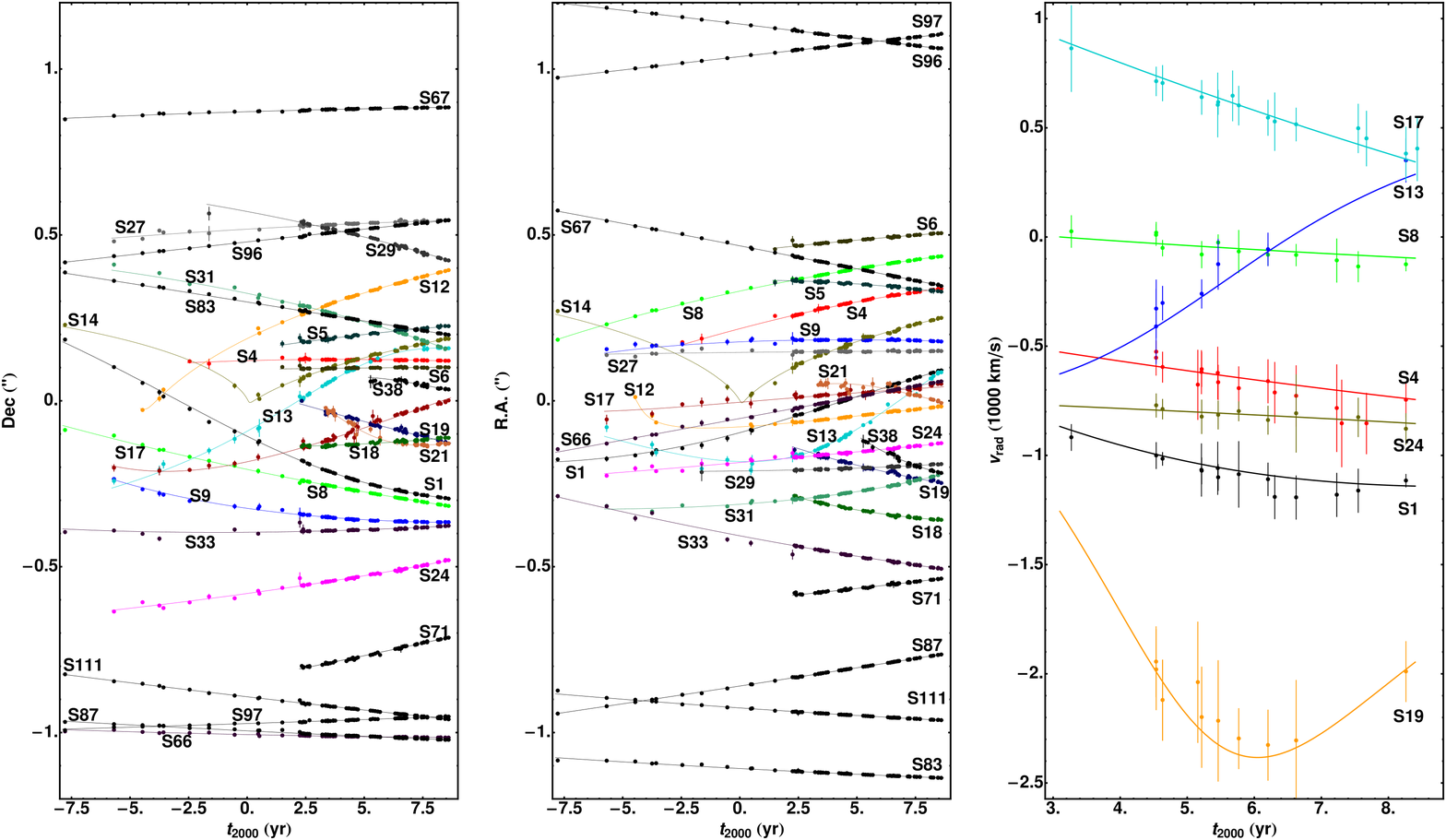}
\epsscale{1.0}
\caption{The orbital data for the S-stars other than S2, the data of which is given in figure~\ref{s2orbit}. Left: The measured declinations as function of time for the stars for which we were able to determine orbits together with the orbital solution. Middle: The same plot for right ascenscion. Right: The measured radial velocity for those stars for which we
were able to measure changes in the radial velocity together with the orbital solutions. The radial velocities for the other stars are given in table~\ref{polytab}.} 
\label{sstardata}
\end{center}
\end{figure*}

For stars for which a significant part of the orbit is sampled, the astrometric data cannot be described by polynomial fits anymore. Most prominently,
in our data set this is S2 of which our astrometric measurements cover more than one complete revolution. 
For all other stars we report the polynomial fits to the astrometric data in the table in appendix~\ref{polytab}.  
We also give there polynomial fits to the radial velocity data of those stars
for which we were able to determine orbits. The order of the polynomials in all cases was
chosen such that the highest order term still differed significantly (at the 5-$\sigma$ level) from 0. Significances were calculated after rescaling the errors 
such that the reduced $\chi^2$ of the respective fit was 1, which is a conservative approach.

Astrometrically, we found significant $d a/dt$ (requiring at least a 5-$\sigma$ level) values for the stars S1, S4, S12, S13, S14, S17 and S31. Significant astrometric accelerations
(at the 5-$\sigma$ level or above) were found in addition for S5, S6, S8, S9, S18, S19, S21, S23, S24, S27, S28, S29, S33, S38, S39, S40, S48, S58, S66, S67, S71, S83, S87 and S111, where we checked that the acceleration vector actually points towards Sgr~A*.

We measured changes in the radial velocity for S1, S2, S4, S8, S13, S17, S19 and S24 (all $>5$-$\sigma$, except S24 with $4.8\sigma$). Additionally, we were able to determine radial velocities for S5, S6, S7, S9, S10, S11, S12, S14, S18, S20, S21, S22, S25, S26, S27, S29, S30, S31, S32, S33, S34, S35, S37, S38, S45, S52, S54, S65, S66, S67, S70, S71, S72, S73, S76, S83, S84, S85, S86, S87, S88, S89, S92, S93, S95, S96, S97 and S111.

Summarizing, we expect
\begin{itemize}
\item that the S2 data will dominate the problem of determining the gravitational potential;
\item that S1, S4, S8, S12, S13, S14, S17, S19, S24 and S31 can be used additionally to constrain the
potential further;
\item that we can find orbits in addition for S5, S6, S9, S18, S21, S27, S29, S33, S38, S66, S67, S71, S83, S87 and S111. 
\end{itemize}
The data for the stars for which we found orbital solutions is presented in figures~\ref{sstardata} and~\ref{s2orbit}, see also table~\ref{polytab}.

\subsection{Mass of and distance to Sgr~A*}
Here and in the following we report always
the fit results including the (downweighted) 2002 data of S2 and excluding it. 
The coordinate system priors were used as given in equation~\ref{priorsUsed}.
The fit errors reported are rescaled such that the reduced $\chi^2=1$. Note that these errors include the
formal fit errors, taking into account parameter correlations between the 
parameters reported here and the respective orbital elements determined simultaneously. The
systematic uncertainty due to the coordinate system is included here as well, since these parameters
were varied during the fits, too. 
The importance of this was pointed out also by \cite{nik08}.
 
\subsubsection{$R_0$ and mass from S2 data only}
\begin{figure}[htbp]
\begin{center}
\epsscale{1.0}
\plotone{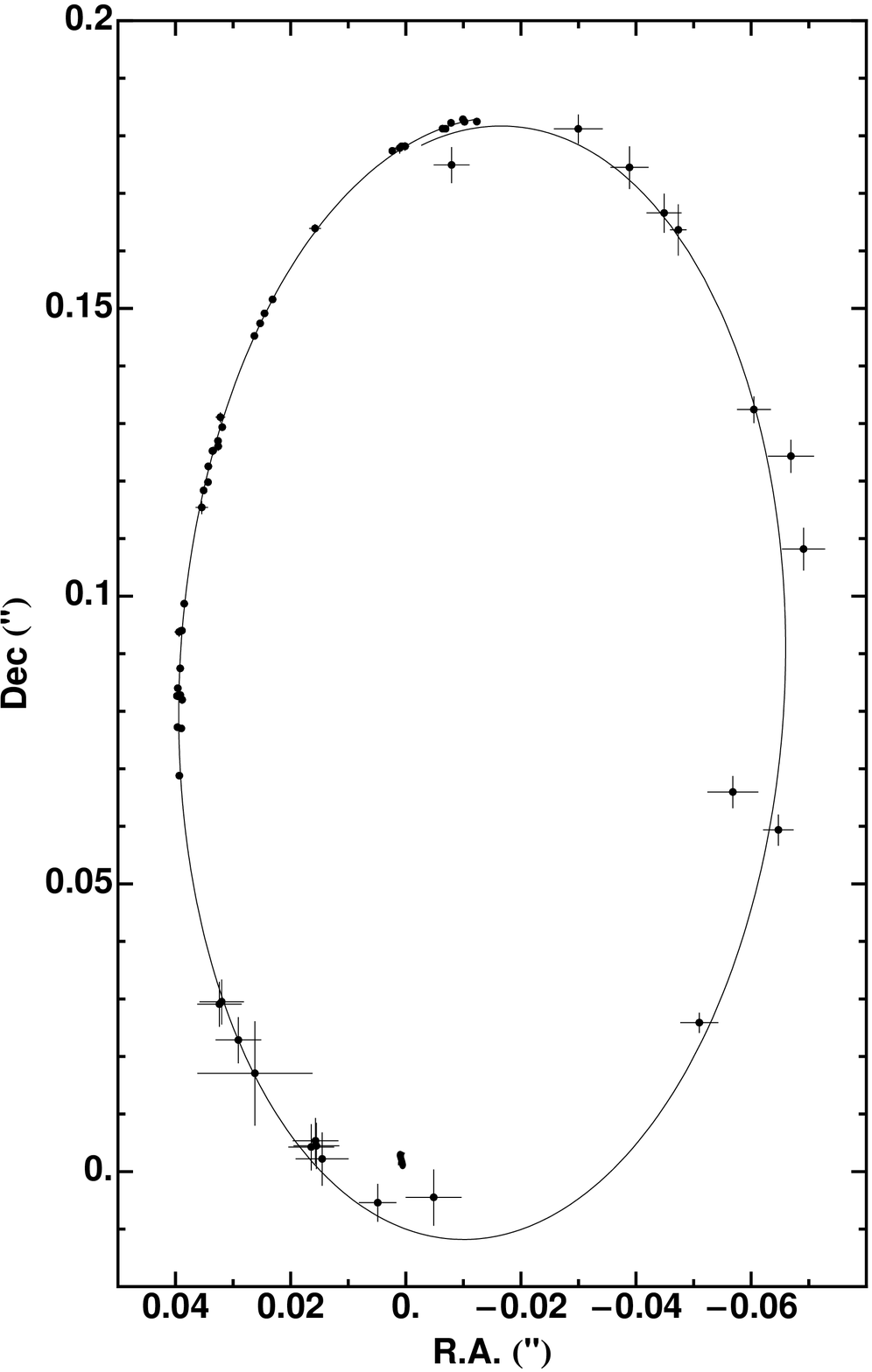}
\plotone{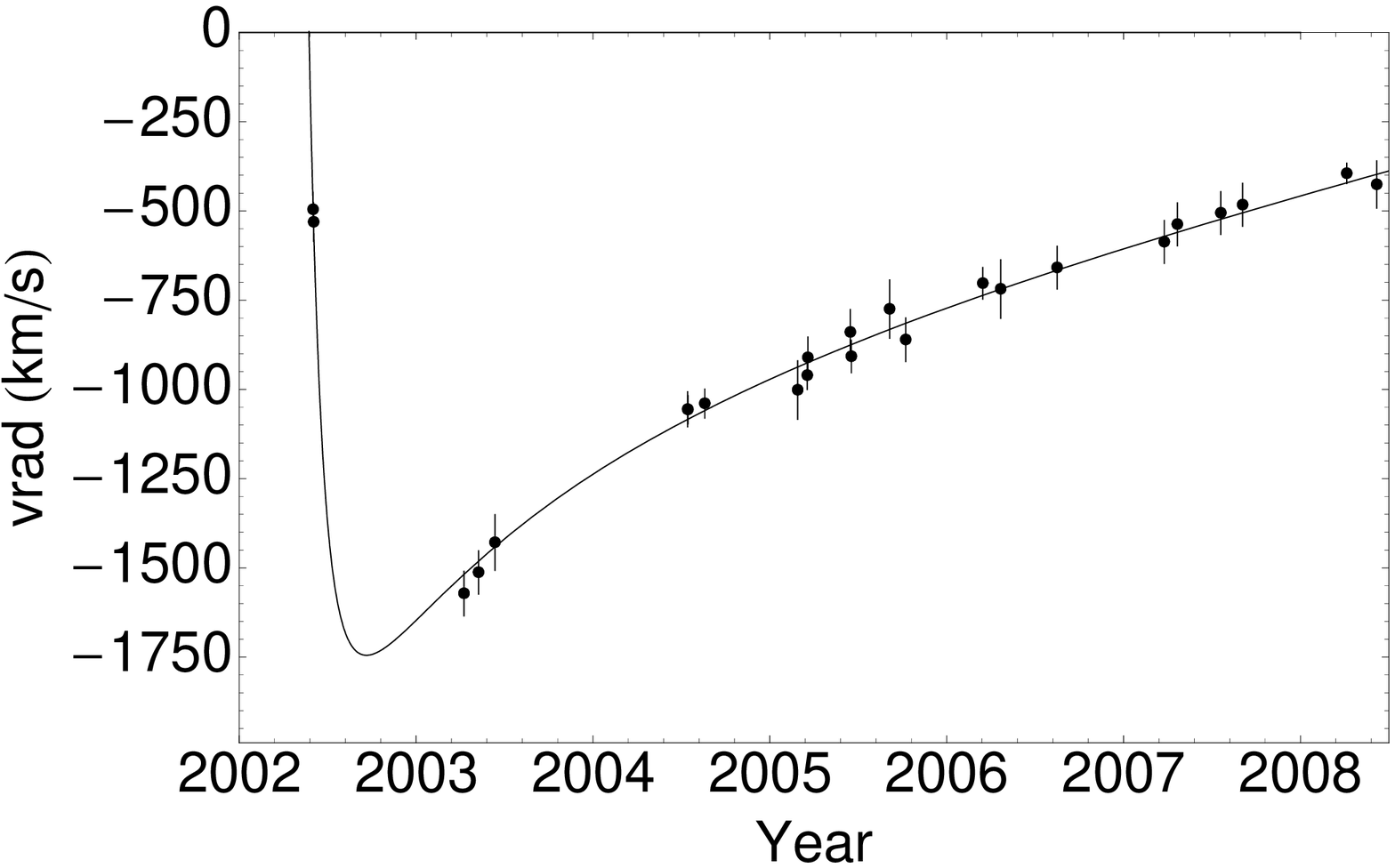}
\epsscale{1.0}
\caption{Top: The S2 orbital data plotted in the combined coordinate system and
fitted with a Keplerian model in which the velocity of the central point mass and its position were free fit parameters. 
The non-zero velocity of the central point mass is the reason why the orbit figure does not close exactly in the overlap region 1992/2008 close to apocenter.  
The fitted position of the central point mass is indicated by the elongated dot inside the orbit near the origin; its shape is determined from the uncertainty in the position and the fitted velocity, which leads to the elongation.
Bottom: The measured radial velocities of S2 and the radial velocity as calculated from the orbit fit. }
\label{s2orbit}
\end{center}
\end{figure}

\begin{figure}[htbp]
\begin{center}
\plotone{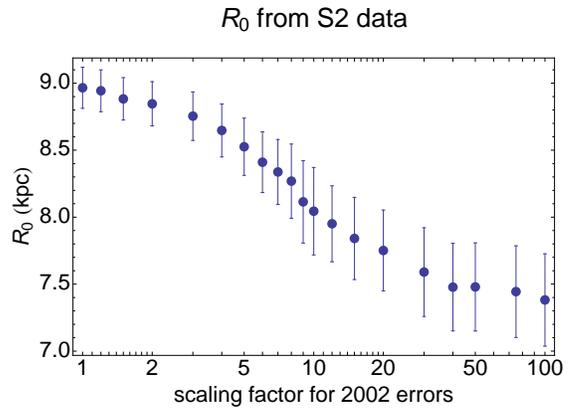}
\caption{Fitted value of $R_0$ for various scaling factors of the S2 2002 data, using a fit with the coordinate system priors. 
The factor by which the 2002 astrometric errors of the S2 data is scaled up strongly influences the distance. The mean factor determined in Figure~\ref{s2deflection} is $\approx7$,
corresponding to $R_0\approx8.1\,$kpc.}
\label{r0Sys}
\end{center}
\end{figure}

First, we used the S2 data only to determine a Keplerian gravitational potential (see figure~\ref{s2orbit}). Using the priors
as obtained in equation~\ref{priorsUsed}, the fits yield the numbers in the first and second row of table~\ref{summ_fit}.
The two values for $R_0$ differ by more than what the errors suggest; indicating that the 2002 data influences $R_0$. This confirms
the presumption from section~\ref{s22002}. We exploited this further in figure~\ref{r0Sys}. Assigning the 2002 data higher weights (smaller errors) pushes the distance estimate up, smaller weights lower it. 
 
\begin{figure}[htbp]
\begin{center}
\plotone{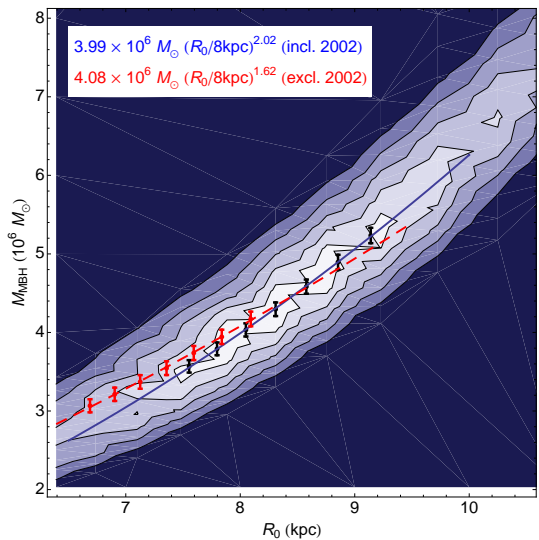}
\caption{Contour plot of $\chi^2$ as function of $R_0$ and central point mass. The two parameters are strongly correlated.
The contours are generated from the S2 data including the 2002 data; fitting at each point all other parameters both of the potential
and the orbital elements. The black dots indicate the position and errors of the best fit values of
the mass for the respective distance; the blue line is a power law fit to these points; the corresponding function is given in the upper row of the text box. 
The central point is chosen at the best fitting distance. The red points and the red dashed line are the respective data and fit for the S2 data excluding the 2002 data;
the fit is reported in the lower row of the text box.
The contour levels are drawn at confidence levels corresponding to $1\sigma,\,3\sigma,\,5\sigma,\,7\sigma,\,9\sigma$.}
\label{md}
\end{center}
\end{figure}

Mass and distance are strongly correlated parameters,
see Figure~\ref{md}. The scaling of mass with $R_0$ in our data set is a power law with
$M_\mathrm{MBH}\sim R_0^{\,\,2}$. For a purely astrometric data set one would have an exponent of 3 and a complete degeneracy; the fact that the exponent is $<3$ and that the degeneracy is not complete is due to the influence of the radial velocity information in our data set and due to the use of priors. The degeneracy can be understood qualitatively. Changing $R_0$ effectively changes the conversion from measured angles (in mas) to physical lengths (in pc), i.e. changing $R_0$ changes the semi major axis. Since the orbital period is well determined in our data, the mass has to change in order to fulfill Kepler's third law.

The strong dependency means that the uncertainties for mass and distance are coupled.  Fixing the distance yields a very small fractional error on the mass of $\Delta M_\mathrm{MBH} \approx 0.02 M_\mathrm{MBH}$. This shows that the error of the fitted mass is completely dominated by the uncertainty in the distance. Once the distance is known, the mass immediately follows from the scaling relation
\begin{eqnarray}
M_\mathrm{MBH} &=& (3.99 \pm 0.07|_\mathrm{stat} \pm 0.32|_\mathrm{R_0})\nonumber \\ 
&& \times 10^6\,M_\odot \left( \frac{R_0}{8\,\mathrm{kpc}} \right)^{2.02}\;(\mathrm{incl. 2002})\,,\nonumber \\
M_\mathrm{MBH} &=& (4.08 \pm 0.09|_\mathrm{stat} \pm 0.39|_\mathrm{R_0})\nonumber\\
&&\times 10^6\,M_\odot \left( \frac{R_0}{8\,\mathrm{kpc}} \right)^{1.62}\;(\mathrm{excl. 2002})\,,
\end{eqnarray}
where the error due to $R_0$ corresponds to the fit error reported in table~\ref{summ_fit}.

\subsubsection{Position of the central point mass}
\label{posMBH}
By construction the position of the radio source Sgr~A* in our coordinate system is located at the origin. Since it is clear that Sgr~A* is the MBH candidate 
we used this fact when applying the priors of equation~\ref{priorsUsed}. However, our data actually
allows us to test this hypothesis. By leaving the position and proper motion of the mass completely free, we can check how well the position of the mass
coincides with Sgr~A*.
Using the S2 data only, no 2D priors but the prior in $v_z$ from equation~\ref{priorsUsed} we obtained the numbers presented in the third and fourth row of table~\ref{summ_fit}.
We note that the mass is located within $\approx2\,$mas at the expected position. 
The current accuracy by which this statement holds is an improvement of a factor $\approx 2$ over the
work from \cite{sch03}.

We also report the S2-only fits when not using any coordinate system priors at all (rows 5 and 6 in table~\ref{summ_fit}). This enlarges the errors on $R_0$ and $M_\mathrm{MBH}$
substantially, the fit values however are not significantly different from the respective fits in which the $v_z$-prior was applied. Not applying the $v_z$-prior also
shows a large uncertainty on $v_z$ of $\approx 50\,$km/s; this parameter also is degenerate with $R_0$.

~\\
From the numbers it seems that the fit excluding the 2002 data agrees better with the expectations for the coordinate system (equation~\ref{priorsUsed}) than
the fit including it. The latter is marginally consistent with the priors, while the former is fully consistent. This means that the 2002 data not only affects
$R_0$ (which we want to measure and thus cannot use to judge the result) but also the position and velocity of the mass for which we have an independent measurement via
the coordinate system definition. This argument points towards rejecting the 2002 data. 
\clearpage
\subsubsection{Position of the IR counterpart of Sgr~A*}
At 22 epochs we have identified a source in the NACO data between 2003 and 2008 that might be associated with Sgr~A*. In some cases, e.g. when a bright flare occurred, the 
identification seems unproblematic. In other cases, one cannot be sure that the emission is not due to an unrecognized star at or very close to the position of Sgr~A*; 
an example is Figure~\ref{findchart}. Due to this probably very frequent confusion we expect that the measured positions are very noisy and we decided not 
to include them into the orbital fits. However, we checked whether the measured positions are compatible with the orbital fits.
Fitting a linear motion model to the Sgr~A* data we obtained
\begin{eqnarray}
\alpha [\mathrm{mas}]&=& (1.2 \pm 0.8)  + (0.15 \pm 0.46) \times (t[\mathrm{yr}] - 2005.91) \nonumber \\
\delta [\mathrm{mas}]&=& (2.7 \pm 0.7)  - (0.73 \pm 0.39) \times (t[\mathrm{yr}] - 2005.91)
\end{eqnarray}
The errors here are rescaled for a reduced $\chi^2$ of 1. The velocity errors are approximately a factor $5$ larger than the priors from equation~\ref{priorsUsed}, justifying
our choice not to incorporate this data into the orbital fits. Given the uncertainties, the position of the IR counterpart of Sgr~A* is consistent with
the position of the central point mass. Interestingly, that data seems to prefer a position of Sgr~A* marginally North of the expected position, which
is also the case for the orbit fits which include the 2002 data of S2. This weakens again the conclusion from section~\ref{posMBH} that the 2002
data should be rejected.

\subsubsection{$R_0$ and mass from a combined orbit fit}
\label{combifit}
Given the large uncertainties due to the 2002 data of S2, we decided to obtain more information about the potential by using a combined
orbit fit and the coordinate system priors. For comparison, we also excluded S2 completely. We used the stars S1, S2, S8, S12, S13, S14. We selected 
these stars from the sample that can contribute to the potential (section~\ref{polyfits}) since for them a large fraction of the respective orbit is covered.
We did not select S4 and S17 as they suffered confusion in the SHARP data. S19 was omitted because its time base is quite short still (the star 
was not detected before 2003). Since S24 would only contribute marginally to the potential, it was left out, too.
Finally, we did not select S31, since the nearby sources S59 and S60 were confused with S31 in the earlier NACO data. Not surprising, the final sample
contains the same stars as \cite{eis05} had reported orbits for. 

In order to balance the relative weights of the stars used, we had fitted the five additional stars first alone, leaving
also the potential free (but applying the priors). While the such obtained fits were not of interest per se, they still provided
a smooth, unbiased model for each star. Hence, we used
the resulting reduced $\chi^2$ values to rescale the astrometric and radial velocity errors such that all stars yielded a value of 1. The scaling factors applied 
ranged from 1.20 to 2.33, the latter value being extreme and occurring for S13, which perhaps suffered from confusion in the SHARP data and of which the
data in 2006/2007 was affected by confusion with S2. Our procedure guaranteed that such a star with a high astrometric noise would not 
contribute overly much to the combined $\chi^2$. We obtained the results given in rows 7, 8 and 9 of table~\ref{summ_fit}:
These numbers agree with each other within the uncertainties. The combined fit including the S2 2002 data also agrees with the corresponding S2-only fit.
This is not true for the combined fit excluding the S2 2002 data, which is hardly compatible with the respective S2-only fit. A possible reason is that
the S2 data before 2002 is only relying on the SHARP measurements, which not only have larger formal errors but also is more affected by unrecognized confusion events than the NACO data. 

By fitting the combined data at various, fixed values of $R_0$ we obtain again the scaling of
mass and distance:
\begin{eqnarray}
M_\mathrm{MBH} &=& (3.95 \pm 0.06|_\mathrm{stat} \pm 0.18|_\mathrm{R_0})\nonumber\\
&&\times 10^6\,M_\odot \left( \frac{R_0}{8\,\mathrm{kpc}} \right)^{2.19}\;(\mathrm{incl. 2002})\,,\nonumber \\
M_\mathrm{MBH} &=& (4.01 \pm 0.07|_\mathrm{stat} \pm 0.18|_\mathrm{R_0})\nonumber\\
&&\times 10^6\,M_\odot \left( \frac{R_0}{8\,\mathrm{kpc}} \right)^{2.07}\;(\mathrm{excl. 2002})\,,\nonumber \\
M_\mathrm{MBH} &=& (3.88 \pm 0.10|_\mathrm{stat} \pm 0.41|_\mathrm{R_0})\nonumber\\
&& \times 10^6\,M_\odot \left( \frac{R_0}{8\,\mathrm{kpc}} \right)^{3.07}\;(\mathrm{excl. S2})\,,
\end{eqnarray}

\subsubsection{Other systematic errors for $R_0$}
\label{syserrors}
Beyond what was considered before, the physical model for the potential is another source of uncertainty. For example using a relativistic model instead of
a Keplerian orbit model increased the distance by $\Delta R_0=0.18\,$kpc ($0.09\,$kpc) when including (excluding) the 2002 data. This is consistent with the formal error on $R_0$. Since we do not detect explicitly relativistic effects, we stay with Keplerian orbits and consider the shift of the value as an uncertainty for $R_0$.  
Fitting a Plummer model (as in Section~\ref{post}) instead of a point mass potential increases the distance by a similar value: $0.14\,$kpc ($0.03\,$kpc)
when including (excluding) the 2002 data. The additional degree of freedom in this fit increased the formal uncertainty 
by $0.11\,$kpc added in squares. Finally, we adopted for the uncertainties of the potential
an error of $\Delta R_0 = 0.25\,$kpc.

An additional, systematic error is whether the use of priors (equation~\ref{priorsUsed}) is correct. In order to address this, we repeated the
combined orbit fits without the 2D priors. We obtained the numbers in rows 10 and 11 of table~\ref{summ_fit}.
The influence of the priors on the value of $R_0$ is relatively small (compare rows 7 and 8 with 10 and 11 in table~\ref{summ_fit}). We adopt for this source of uncertainty $\Delta R_0 = 0.10\,$kpc.

Furthermore, rows 7, 8 and 9 of table~\ref{summ_fit} show that the uncertainty of the weights of the 2002 data from S2
in a combined fit alters $R_0$ by $\Delta R_0 = 0.13\,$kpc.  
Deselecting S2 from the fits changes the result by $\Delta R_0 = 0.07\,$kpc. 
Finally, we assign $\Delta R_0 = 0.15\,$kpc for the uncertainties related to the selection of data.

Adding  up the uncertainties yields that the uncertainty of the distance to GC is still rather large with $\Delta_\mathrm{total} R_0 = 0.35\,$kpc. Table~\ref{r0syserr} summarizes the error terms for $R_0$.

\begin{table*}
\caption{Results for the central potential from orbital fitting, from either S2 data only (rows 1 - 6) or a combined fit 
using in addition S1, S8, S12, S13, S14 (rows 7 - 12). In rows 9 and 12, the combined fit was done without S2. 
The third column indicates whether the 2002 data from S2 was used or not; the fourth column
informs about which of the priors from equation~\ref{priorsUsed} have been used.
\label{summ_fit}}
{\scriptsize
\begin{center}
\begin{tabular}{llcc|ccccccc}
&Fit & S2 & priors & $R_0$&$M_\mathrm{MBH} $&
$\alpha $&$\delta $&$v_\alpha $&
$v_\delta $&$v_z  $\\
&&2002&& (kpc) &$(10^6 M_\odot)$ & (mas) & (mas) & ($\mu$as/yr) & ($\mu$as/yr) & (km/s)\\
\hline
\hline
1&S2 only & yes & 2D, $v_z$ & $8.31\pm0.33$& $4.29\pm0.35$&$0.51\pm0.64$&$2.18\pm0.89$&$-5\pm87$&$119\pm78$&$0.8\pm6.2$\\
2&S2 only & no & 2D, $v_z$ & $7.36\pm0.43$& $3.54\pm0.35$& $0.81\pm0.66$& $-0.63\pm1.39$& $-69\pm91$& $103\pm81$& $-0.8\pm6.2$ \\
\hline
3&S2 only & yes & $v_z$ &  $8.48\pm0.38$& $4.45\pm0.41$& $0.37\pm0.73$& $2.33\pm0.94$& $76\pm131$& $231\pm107$& $0.8\pm6.1$\\
4&S2 only & no & $v_z$ &  $7.31\pm0.45$& $3.51\pm0.36$& $0.92\pm0.75$& $-0.84\pm1.43$& $-83\pm137$& $154\pm114$& $-0.9\pm6.3$\\
\hline
5&S2 only & yes & none &  $8.80\pm0.53$& $4.93\pm0.75$& $0.31\pm0.71$& $2.44\pm0.89$& $74\pm$127& $220\pm107$& $29\pm36$\\
6&S2 only & no & none &  $6.63\pm0.91$& $2.85\pm0.74$& $0.96\pm0.75$& -2.00$\pm2.38$& $-111\pm148$& $162\pm115$& $-42\pm44$\\
\hline
\hline
7&comb. & yes & 2D, $v_z$& $8.33\pm0.17$& $4.31\pm0.22$& $0.80\pm0.63$& $2.19\pm0.60$& $-28\pm71$& $100\pm68$ &$0.0\pm5.0$\\
8&comb. & no & 2D, $v_z$& $8.20\pm0.18$& $4.22\pm0.22$& $1.07\pm0.58$& $1.54\pm0.64$& $-32\pm73$& $86\pm71$& $0.0\pm5.1$\\
9&w/o S2 & - & 2D, $v_z$& $8.40\pm0.29$& $4.51\pm0.49$& $1.49\pm0.99$& $2.61\pm1.37$& $-66\pm94$& $-116\pm94$& $-1.3\pm5.1$\\
\hline
10&comb. & yes & $v_z$& $8.38\pm0.16$& $4.36\pm0.21$& $0.73\pm0.65$& $2.10\pm0.61$& $51\pm106$& $211\pm97$& $-0.4\pm5.1$\\
11&comb. & no & $v_z$& $8.22\pm0.20$& $4.25\pm0.26$& $1.22\pm0.81$& $1.59\pm0.83$& $0\pm133$& $164\pm123$& $-0.5\pm6.3$\\
12&w/o S2& - & $v_z$&$8.42\pm0.31$&$4.61\pm0.55$&$6.2\pm2.0$&$6.0 \pm1.9$&$-335\pm294$&$-15\pm281$&$-1.0\pm5.0$
\end{tabular}
\end{center}
}
\end{table*}

\begin{table}
\caption{Systematic errors for the distance to the GC, $R_0$.
\label{r0syserr}}
{\scriptsize
\begin{center}
\begin{tabular}{lc}
\hline
\hline
Error source &$\Delta R_0$(kpc) \\
\hline
Fit error including position and velocity&\\
uncertainty of coordinate system&0.17\\
Assumed potential&0.25\\
Using priors or not&0.10\\
Selection of data&0.15\\
\hline
Total&0.35
\end{tabular}
\end{center}
}
\end{table}
\subsubsection{Final estimate for $R_0$ and mass}
\label{finalR0}
We finally adopt the potential from the combined fit including the S2 2002 data, the difference to the one excluding that data is negligible given the formal fit errors
 (section~\ref{combifit}). This potential will be used in section~\ref{starsWithOrbits} to determine the orbits of the other stars for which we expect to find an orbital solution.
Hence, we find
\begin{equation}
R_0 = 8.33 \pm 0.17|_\mathrm{stat} \pm 0.31|_\mathrm{sys}\,\mathrm{kpc}\,\,.
\label{finalR0eq}
\end{equation}
It should be noted that this value is consistent within the errors with values published earlier \citep{eis03b,eis05}. The improvement of our current work is the more rigorous treatment of the systematic errors. Also it is worth noting that adding more stars did not change the distance much over the 
equivalent S2-only fit. For the mass we adopt
\begin{eqnarray}
M_\mathrm{MBH} &=& (3.95 \pm 0.06|_\mathrm{stat} \pm 0.18|_\mathrm{R_0,\;stat} \pm 0.31|_\mathrm{R_0,\;sys})\nonumber\\
&&\times 10^6\,M_\odot \left( \frac{R_0}{8\,\mathrm{kpc}} \right)^{2.19}\nonumber\\
&=&(4.31 \pm 0.38) \times 10^6\,M_\odot \;\mathrm{for}\, R_0=8.33\,\mathrm{kpc}.
\end{eqnarray}

\subsection{Testing for an extended mass component}
\label{post}
While Newtonian physics seems to describe the S-star system reasonably well, one actually expects to detect deviations from purely Keplerian orbits with accurate enough
astrometric and spectroscopic data. There are two main reasons for this:
\begin{itemize}
\item The relativistic effects as described in Section~\ref{orbitalfitting} lead to deviations \citep{rubilar,wei05,gil06}. 
Note that for S2 the pericenter advances by $0.18^\circ$ per orbital revolution, not far from the precision of the orbit orientation in Table~\ref{orbitsTable}.
\item In addition to the MBH a substantial amount of mass might reside in form of a cluster of dark stellar remnants around the MBH \citep{mor93,mir00,mun05,mouawad,hop06}. This will also lead to a non-Keplerian orbit, with the pericenter precessing in retrograde fashion. 
\end{itemize}
Given our current data base S2 is the only star for which one can hope to find a deviation from a Keplerian orbit. Fitting a relativistic orbit to the S2 data yields a similar  $\chi^2$ (158.5 compared to 158.7 for the Keplerian fit, both with 114 degrees of freedom). Allowing for an extended mass component in addition does not change $\chi^2$ much, typically we found $\chi^2\approx 157.4$ (depending on the details of the model) at the cost of one additional free parameter.

The simplest model for an extended mass component is a constant mass density $\rho(r)$ described by
\begin{equation}
\rho(r) = \rho_0 \,\,.
\end{equation}
More realistic is a  power law model
\begin{equation}
\rho(r) =  \rho_0 \left(\frac{r}{r_0}\right)^\alpha \,\,.
\end{equation}
The power law model is motivated by the findings of \cite{gen03b} who show that the stellar 
number counts display such a density profile, which is also expected 
on theoretical grounds \citep{bah77,you80}. The parameters $\rho_0$ and $\alpha$ are a characteristic density at the given radius and the power law index. We assumed for the following $\alpha=-1.4$, $\alpha=-1.75$ and $\alpha=-2.1$ \citep{hop06}.

Similar investigations \citep{rubilar,mouawad} used a Plummer model:
\begin{equation}
\rho(r) =  \frac{3 \mu M_\mathrm{MBH}}{4 \pi r_\mathrm{core}^3} \left(1+\frac{r^2}{r_\mathrm{core}^2} \right)^{-5/2} \,\,.
\end{equation}
The free parameters of the Plummer model are the core radius $r_\mathrm{core}$ and the mass parameter $\mu$, which corresponds to the ratio of total extended mass versus mass of the central point mass. This model allows a convenient analytical
description of the null hypothesis - no stellar cusp - and roughly
describes the surface light density distribution around Sgr~A* \citep{sco03,sch07}.
We adopt a core radius of $r_\mathrm{core}=15\,$mpc, which matches the observed light profile~\citep{mouawad}.

We fitted the S2 data for all three mass models and included in all cases the relativistic effects. The coordinate system priors were applied (equation~\ref{priorsUsed})
and an additional prior was set on the $R_0 = 8.40 \pm 0.29\,$kpc from the combined fit that excluded S2 completely (row 9, table~\ref{summ_fit}). 
Any such fit can only test for mass inside the S2 orbit; therefore we express the results in terms of mass enclosed between S2's apocenter ($r=0.230''=8.9\,$mpc) and pericenter ($r=0.015''=0.58\,$mpc) relative to the mass of the MBH and call this parameter $\eta$: 
\begin{equation}
\eta \,M_\mathrm{MBH} =4 \pi \int_\mathrm{peri}^\mathrm{apo} dr\, r^2 \int dm \,n(r,m) 
\end{equation}

\begin{table}
\caption{Results from S2 fits including an extended mass component. The parameter $\eta$ describes the ratio of extended mass to the central point mass. The extended mass is accounted for in a spherical shell from the pericenter distance of S2 to the apocenter distance. The table shows the results for various potentials}
\label{rhofits}
{\scriptsize
\begin{center}
\begin{tabular}{l|cc}
Fit&$R_0\,$(kpc)&$\eta$\\
\hline
&\multicolumn{2}{c}{incl. 2002 data of S2}\\
\hline
$\rho=const$&$8.46\pm0.25$&$0.029\pm0.026$\\
power law, $\alpha=-1.4$&$8.49\pm0.26$&$0.020\pm0.017$\\
power law, $\alpha=-1.75$&$8.49\pm0.26$&$0.018\pm0.015$\\
power law, $\alpha=-2.1$&$8.52\pm0.27$&$0.015\pm0.013$\\
Plummer&$8.47\pm0.26$&$0.025\pm0.022$\\
\hline
&\multicolumn{2}{c}{excl. 2002 data of S2}\\
\hline
$\rho=const$&$8.00 \pm 0.33$&$0.018 \pm 0.028 $\\
power law, $\alpha=-1.4$&$8.03\pm0.34$&$0.013\pm0.016$\\
power law, $\alpha=-1.75$&$8.03\pm0.34$&$0.012\pm0.014$\\
power law, $\alpha=-2.1$&$8.05\pm0.35$&$0.012\pm0.014$\\
Plummer&$8.01\pm0.33$&$0.016\pm0.023$\\
\end{tabular}
\end{center}
}
\end{table}

The results are shown in Table~\ref{rhofits} from which
we obtain
\begin{eqnarray}
\nonumber
\eta_\mathrm{S2}&=&0.021\pm0.019|_\mathrm{stat}\pm0.006|_\mathrm{mod}\; \mathrm{(incl.\; 2002)} \nonumber\\
\eta_\mathrm{S2}&=&0.014\pm0.019|_\mathrm{stat}\pm0.003|_\mathrm{mod}\; \mathrm{(excl.\; 2002)}.
\label{etalim}
\end{eqnarray}
The statistical fit error includes the uncertainties due to the coordinate system definition.
The result corresponds to a 1-$\sigma$ upper limit of $\eta\le0.040$ ($0.033$) and a 99\% upper limit of $\eta\le0.066$ ($0.058$) including (excluding) the S2 2002 data,
where the upper limits are defined such that the cumulated probability density function reaches the specified
significance level at the respective value for $\eta$ \citep{fel98}. The (small) uncertainty in $\eta$ due to the model
uncertainty has been excluded for the calculation of the upper limit since Table~\ref{rhofits} shows that it affects rather
the amplitude of $\eta$ than its significance.

So the basic result of this study, improving measurement uncertainties by a factor of six over 
\cite{sch02,ghe05,eis05}, is that a single point mass potential is (still) the best description of the data. Any deviations are smaller than a few percent of the point mass, within the orbits of the central S-star cluster.

\subsection{Stars with orbits}
\label{starsWithOrbits}
\begin{figure*}[htbp]
\begin{center}
\plotone{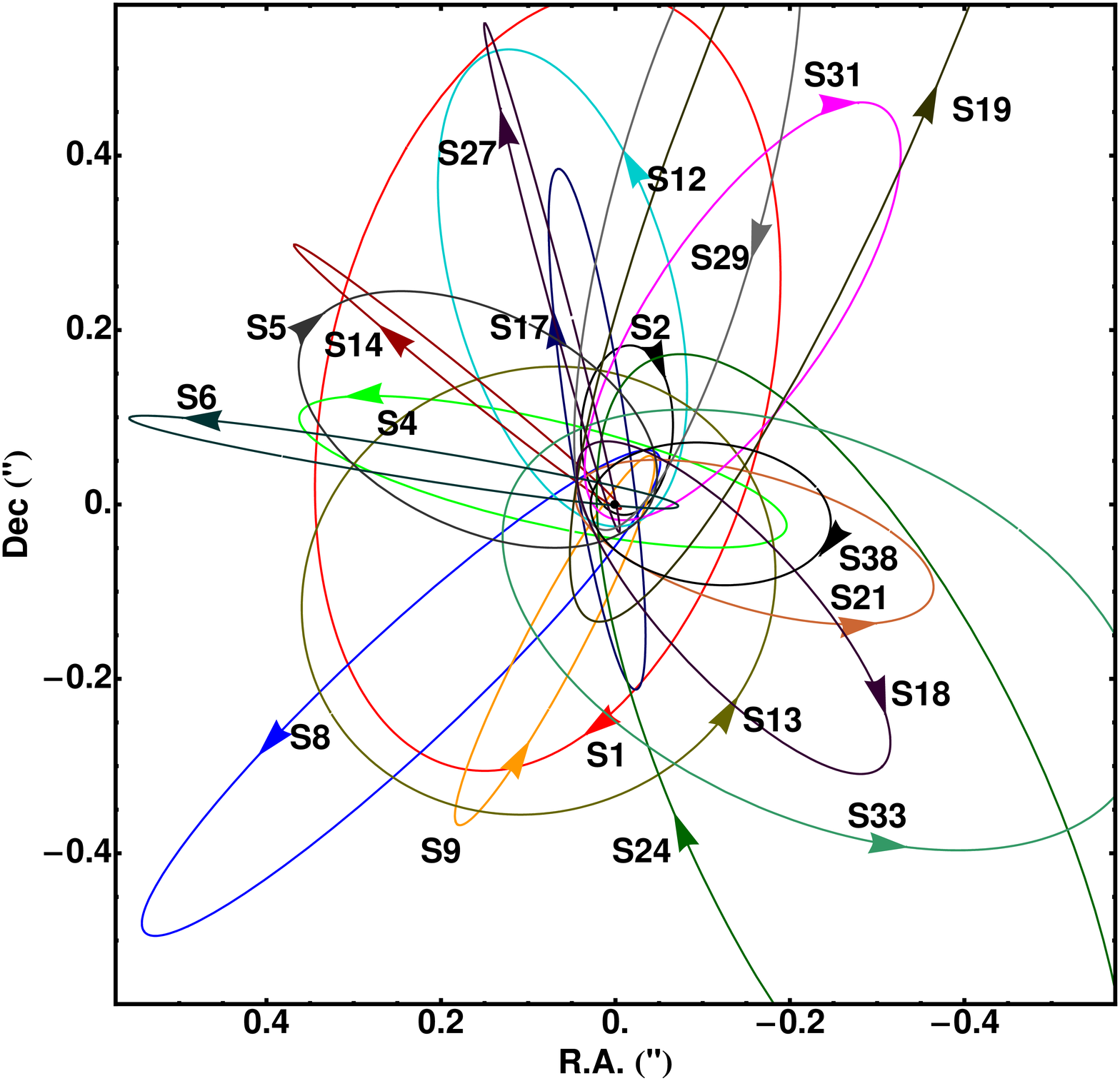}
\caption{The stellar orbits of the stars in the central arcsecond for which we were able to determine orbits. In this illustrative figure, the coordinate system was chosen such
that Sgr~A* is at rest.}
\label{orbits}
\end{center}
\end{figure*}

\begin{table*}
\caption{Orbital parameters of those S-stars, for which we were able to determine orbits. The parameters were determined in the potential as obtained in section~\ref{finalR0},
the errors quoted in this table are the formal fit errors after rescaling them such that the reduced $\chi^2=1$ and including the uncertainties from
the potential. The last three columns give the spectral type ('e' for early-type stars, 'l' for late-type stars), the K-band magnitude and the global rescaling factor
for that star. S111 formally has a negative semi major axis, indicative for a hyperbolic orbit with $e>1$. We also cite the orbital solutions for the stars S96 and S97 which
showed only marginal accelerations, see section~\ref{stellarOrbits}.}
\label{orbitsTable}
{\scriptsize
\begin{center}
\begin{tabular}{lcccccccccc}
Star& $a$['']& $e$& $i\,[^\circ]$& $\Omega\,[^\circ]$& $\omega\,[^\circ]$& $t_P$[yr-2000]& $T$[yr]&Sp&$m_K$&r\\
\hline
S1&$0.508\pm0.028$&$0.496\pm0.028$&$120.82\pm0.46$&$341.61\pm0.51$&$115.3\pm2.5$&$0.95\pm0.27$&$132\pm11$&e&14.7&1.49\\
S2&$0.123\pm0.001$&$0.880\pm0.003$&$135.25\pm0.47$&$225.39\pm0.84$&$63.56\pm0.84$&$2.32\pm0.01$&$15.8\pm0.11$&e&14.0&1.22\\
S4&$0.298\pm0.019$&$0.406\pm0.022$&$77.83\pm0.32$&$258.11\pm0.30$&$316.4\pm2.9$&$-25.6\pm1.0$&$59.5\pm2.6$&e&14.4&1.99\\
S5&$0.250\pm0.042$&$0.842\pm0.017$&$143.7\pm4.7$&$109\pm10$&$236.3\pm8.2$&$-16.4\pm2.5$&$45.7\pm6.9$&e&15.2&1.93\\
S6&$0.436\pm0.153$&$0.886\pm0.026$&$86.44\pm0.59$&$83.46\pm0.69$&$129.5\pm3.1$&$63\pm21$&$105\pm34$&e&15.4&1.45\\
S8&$0.411\pm0.004$&$0.824\pm0.014$&$74.01\pm0.73$&$315.90\pm0.50$&$345.2\pm1.1$&$-16.2\pm0.4$&$96.1\pm1.6$&e&14.5&1.20\\
S9&$0.293\pm0.052$&$0.825\pm0.020$&$81.00\pm0.70$&$147.58\pm0.44$&$225.2\pm2.3$&$-12.2\pm2.1$&$58\pm9.5$&e&15.1&2.23\\
S12&$0.308\pm0.008$&$0.900\pm0.003$&$31.61\pm0.76$&$240.4\pm4.6$&$308.8\pm3.8$&$-4.37\pm0.03$&$62.5\pm2.3$&e&15.5&1.54\\
S13&$0.297\pm0.012$&$0.490\pm0.023$&$25.5\pm1.6$&$73.1\pm4.1$&$248.2\pm5.4$&$4.90\pm0.09$&$59.2\pm3.8$&e&15.8&2.33\\
S14&$0.256\pm0.010$&$0.963\pm0.006$&$99.4\pm1.0$&$227.74\pm0.70$&$339.0\pm1.6$&$0.07\pm0.06$&$47.3\pm2.9$&e&15.7&1.99\\
S17&$0.311\pm0.004$&$0.364\pm0.015$&$96.44\pm0.18$&$188.06\pm0.32$&$31945\pm3.2$&$-8.0\pm0.3$&$63.2\pm2.0$&l&15.3&2.46\\
S18&$0.265\pm0.080$&$0.759\pm0.052$&$116.0\pm2.7$&$215.2\pm3.6$&$151.7\pm2.9$&$-4.0\pm0.9$&$50\pm16$&e&16.7&2.34\\
S19&$0.798\pm0.064$&$0.844\pm0.062$&$73.58\pm0.61$&$342.9\pm1.2$&$153.3\pm3.0$&$5.1\pm0.22$&$260\pm31$&e &16.0&2.31\\
S21&$0.213\pm0.041$&$0.784\pm0.028$&$54.8\pm2.7$&$252.7\pm4.2$&$182.6\pm8.2$&$28.1\pm5.5$&$35.8\pm6.9$&l&16.9&1.55\\
S24&$1.060\pm0.178$&$0.933\pm0.010$&$106.30\pm0.93$&$4.2\pm1.3$&$291.5\pm1.5$&$24.9\pm5.5$&$398\pm73$&l&15.6&1.78\\
S27&$0.454\pm0.078$&$0.952\pm0.006$&$92.91\pm0.73$&$191.90\pm0.92$&$308.2\pm1.8$&$59.7\pm9.9$&$112\pm18$&l&15.6&1.79\\
S29&$0.397\pm0.335$&$0.916\pm0.048$&$122\pm11$&$157.2\pm2.5$&$343.3\pm5.7$&$21\pm18$&$91\pm79$&e&16.7&1.92\\
S31&$0.298\pm0.044$&$0.934\pm0.007$&$153.8\pm5.8$&$103\pm11$&$314\pm10$&$13.8\pm2.2$&$59.4\pm9.2$&e&15.7&1.97\\
S33&$0.410\pm0.088$&$0.731\pm0.039$&$42.9\pm4.5$&$82.9\pm5.9$&$328.1\pm4.5$&$-32.1\pm6.5$&$96\pm21$&e&16.0&2.02\\
S38&$0.139\pm0.041$&$0.802\pm0.041$&$166\pm22$&$286\pm68$&$203\pm68$&$3.0\pm0.2$&$18.9\pm5.8$&l &17.0&2.13\\
S66&$1.210\pm0.126$&$0.178\pm0.039$&$135.4\pm2.6$&$96.8\pm2.9$&$106\pm6.3$&$-218\pm23$&$486\pm41$&e&14.8&1.15\\
S67&$1.095\pm0.102$&$0.368\pm0.041$&$139.9\pm2.3$&$106.0\pm6.1$&$215.2\pm4.8$&$-305\pm16$&$419\pm19$&e&12.1&1.53\\
S71&$1.061\pm0.765$&$0.844\pm0.075$&$76.3\pm3.6$&$34.6\pm1.5$&$331.4\pm7.1$&$-354\pm251$&$399\pm283$&e&16.1&2.44\\
S83&$2.785\pm0.234$&$0.657\pm0.096$&$123.8\pm1.3$&$73.6\pm2.1$&$197.2\pm3.5$&$61\pm25$&$1700\pm205$&e&13.6&1.23\\
S87&$1.260\pm0.161$&$0.423\pm0.036$&$142.7\pm4.4$&$109.9\pm2.9$&$41.5\pm3.7$&$-353\pm38$&$516\pm44$&e&13.6&0.94\\
S111&$-10.5\pm7.1$&$1.105\pm0.094$&$103.1\pm2.0$&$52.8\pm5.4$&$131\pm14$&$-55\pm70$&$-$&l&13.8&0.94\\
\hline
S96&$1.545\pm0.209$&$0.131\pm0.054$&$126.8\pm2.4$&$115.78\pm1.93$&$231.0\pm9.0$&$-376\pm34$&$701\pm81$&e&10.0&1.40\\
S97&$2.186\pm0.844$&$0.302\pm0.308$&$114.6\pm5.0$&$107.72\pm3.15$&$38\pm52$&$175\pm88$&$1180\pm688$&e&10.3&1.15\\
\end{tabular}
\end{center}
}
\end{table*}

Assuming the potential from section~\ref{finalR0} we were able to determine orbits for the stars listed in section~\ref{polyfits}. During these fits, each star was considered
separately and the potential was fixed. 
This yielded a total of 26 measured orbits as expected from section~\ref{polyfits}.
An illustration of the (inner) stellar orbits is shown in Figure~\ref{orbits}, the orbital elements for all 26 stars for which we found orbits are summarized in
Table~\ref{orbitsTable}. For the calculation of the errors quoted, all measurement errors (astrometry and radial velocities) were rescaled such that the
reduced $\chi^2=1$. Furthermore, the uncertainties of the potential were included.

As a double-check, we ran Markov-Chain Monte Carlo (MCMC) simulations \citep{teg04} in order to asses the probability density distribution of the orbital elements in the six dimensional parameter space. Such a chain efficiently samples high-dimensional parameter spaces. The algorithm is simple:
\begin{enumerate}
\item Choose a reasonable starting point in the parameter space and calculate $\chi^2$ for that point.
\item Draw a random jump in the parameter space with the typical jump distance simultaneously for each parameter being
the respective 1-$\sigma$ uncertainty divided by the square root of the number of parameters (hence the
mean jump distance corresponds to a 1-$\sigma$ jump). The uncertainties are obtained from the Hessian matrix at the given point
in parameter space. 
\item Calculate $\chi_n^2$ for the new point.
\item If $\chi_n^2 < \chi^2$ accept the new point, else accept the new point with a probability of $\exp(-(\chi_n^2- \chi^2)/2)$. 
\item Store the new point if it is accepted in the buffer of the chain, otherwise store the old point.
\item Go back to step 2.  
\end{enumerate}
After running this chain for a many iterations, the distribution of points in the buffer of the chain measures the probability density distribution, which thus can be estimated by the chain. The actual implementation needs some extra tricks, e.g. for quicker convergence the parameters should be chosen orthogonal to each other. Interestingly, for
a sufficiently long chain the result does not depend upon the chosen jump distance; that value influences rather how fast the chain samples the parameter space.
 
\begin{figure}[htbp]
\begin{center}
\plotone{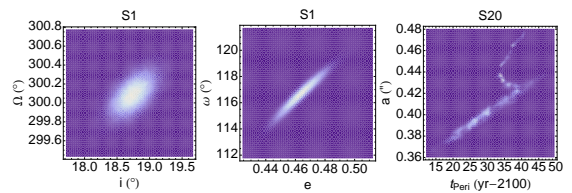}
\caption{Examples from the Markov-Chain Monte Carlo simulations. Each panel shows a 2D cut through the six dimensional phase space of the orbital elements for the respective star. Left: Example of two well constrained and nearly uncorrelated parameters. Middle: Example for two correlated parameters, which are nonetheless well constrained. Right: Example of badly constrained parameters, showing a non-compact configuration in parameter space.}
\label{mcmc}
\end{center}
\end{figure}

For each star we used the MCMC algorithm. Assuming some reasonable potential (e.g. as determined from a preliminary fit to the S2 data) we varied all six orbital elements and checked whether the region in this six-dimensional parameter space which is reached by the chain is compact and reasonably well described by Gaussian functions (see Figure~\ref{mcmc}). 
The advantage of doing so is mainly that, unlike a minimization routine that can be trapped in a local minimum, the MCMC simulations yield a global picture of the probability density distribution.
 
For all 26 stars for which we were able to determine an orbit the probability density distribution was well-behaved, i.e. in all cases the MCMC sampled a compact region in
parameter space, the size of which was consistent with the expectation from the fit errors of the parameters. Examples are shown in Figure~\ref{mcmc}. 
We conclude that the orbital solutions presented in Table~\ref{orbitsTable} are reliable.

Among the stars with orbital solution, six stars are late type (S17, S21, S24, S27, S38 and S111). It is worth noting that for the first time we determine here the orbits of late-type stars in close orbits around Sgr~A*. In particular S17, S21 and S38 have small semi major axes of $a\approx0.25''$. The late-type star S111 is marginally unbound
to the MBH, a result of its large radial velocity ($-740\,$km/s) at $r=1.48''$ which brings its total velocity up to a value $\approx1\sigma$ above the local
escape velocity.

Furthermore we
determined (preliminary) orbits for S96 (IRS16C) and S97 (IRS16SW), showing marginal accelerations ($2.1\sigma$ and $3.9\sigma$ respectively). These
stars are of special interest, since they were proposed to member of a clockwise rotating disk of stars \citep{pau06}.
Similarly, we could not detect an acceleration for S95 (IRS16 NW). This excludes the star from being a member of the counter-clockwise disk \citep{pau06}, since
in that case it should show an acceleration of $\approx150\,\mu$as/yr$^2$, while we can place a safe upper limit of $a<30\,\mu$as/yr$^2$.

\section{Discussion}
\subsection{The distance to the Galactic Center}
Our estimate $R_0= 8.33 \pm 0.17|_\mathrm{stat} \pm 0.31|_\mathrm{sys}\,\mathrm{kpc}$ (equation~\ref{finalR0eq})
is compatible with our earlier work \citep{eis03b,eis05}. While the underlying data base is
partially identical, this work mainly improved the understanding of the systematic uncertainties. In particular, the astrometric data during the pericenter passage of S2
is hard to understand. This is an unfortunate situation, since that data potentially is most constraining for the potential.
During the passage the star sampled a wide range of distances from the MBH, corresponding to
a radially dependent measurement of the gravitational force acting on it. Probably only future measurements of either S2 or other stars passing close
to Sgr~A* will allow one to answer the question, whether the confusion problem close to Sgr~A* is generic or whether 2002 was a unlucky coincidence.

Besides stellar orbits, there are other techniques to determine $R_0$. A classical one is to use the distribution of globular clusters.
\cite{bic06} applied this technique to a sample of 153 globular clusters and obtained $R_0=7.2\pm0.3$. This value is only marginally compatible with our result.
However, the error quoted by \cite{bic06} corresponds to the formal fit error derived
from their figure~4. Therefore, one might suspect that systematic problems owed to the method were not yet included in the error estimate.

The fact that the absolute magnitudes of red clump stars is known and that the red clump can be identified in the luminosity function
obtained from the apparent magnitudes of stars in the galactic bulge was used by \cite{nis06}. These authors
obtain $R_0 = 7.52 \pm 0.10|_\mathrm{stat} \pm 0.35|_\mathrm{sys}\,$kpc, where the statistical error is owed mainly
to the uncertainty of the local red clump starsÕ luminosities and the systematic error terms 
includes uncertainties in the extinction and population corrections, the zero point of photometry, and the fitting of the 
luminosity function of the red clump stars. This result is in agreement with our measurement, given the errors of both results.

The known absolute magnitudes from RR Lyrae stars and Cepheids are the key to the work from \cite{gro08}. Their
result  $R_0 = 7.94\pm0.37|_\mathrm{stat}\pm0.26|_\mathrm{stat}\,$kpc is fully consistent with our result. The statistical error here
is due to the photometric measurement errors, the zero point of photometry and the uncertainty of extinction correction. The systemtatic
error includes the calibration of period-luminosity relations used and the selection effect, which could affect the result since only 39
Cepheids and 37 RR Lyrae stars have been used for this statistical approach.

\subsection{Limits on the binarity of Sgr~A*}
It is interesting to see how our data limits the possible existence of a second, intermediate mass black hole (IMBH) in the GC. Here, we do not aim at a rigorous 
treatment of the problem (which would be beyond the scope of this paper) but limit ourselves to estimates that appear reasonable given our findings.

\begin{figure}[htbp]
\begin{center}
\plotone{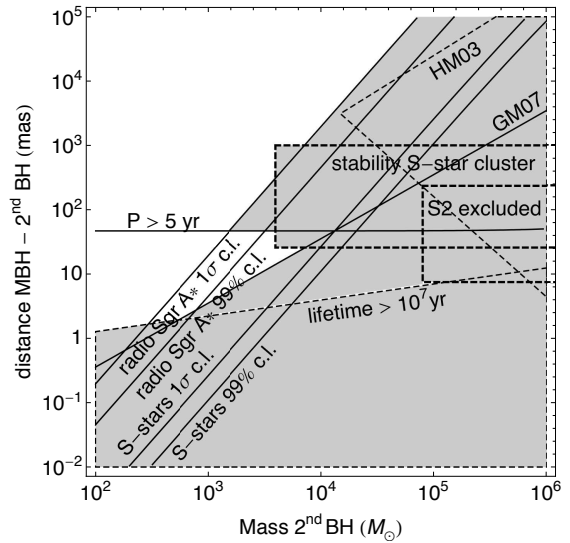}
\caption{Constraints on the binarity of Sgr~A* as function of the mass of the secondary black hole and the distance between the black holes. The shaded areas are excluded due to various arguments. The diagonal lines assume an orbital motion of Sgr~A* around the secondary and
correspond to velocity limits obtained either from the S-stars or the motion of Sgr~A* \citep{rei04}. We estimate that only periods longer than $5\,$yr would lead to an observable effect, thus excluding an area towards higher masses and large distances. Demanding that the lifetime of the binary black hole exceeds $10^7\,$yr 
yields another constraint (from \cite{han03}). These authors also made similar arguments for the motion of the Sgr~A*, the resulting constraint is replicated in this plot (denoted as HM03).
The stability of the S-star cluster puts a further constraint \citep{mik08}, as does the stalling radius found by \cite{gua07}, denoted as GM07.
Finally also the S2 orbit excludes some part of the diagram, since it apparently is Keplerian.}
\label{binlim}
\end{center}
\end{figure}

The first constraint comes from the fact that the center of mass does not move fast. If the central mass were in orbit with an IMBH, the orbital reflex motion of Sgr~A* might show up in our data. The upper limit on the velocity which we obtain from row 7 in table~\ref{summ_fit} corresponds to a line in a phase space plot of IMBH mass versus IMBH-MBH distance (Figure~\ref{binlim}), separating configurations at smaller masses from systems with higher masses. 
From our data, we would not have been able to detect such an orbital motion of the MBH if the orbital period $P$ were too short, namely much shorter than the orbital period of S2. We estimate that configurations with $P>5\,$yr would be discoverable. Taken together, this excludes an area towards higher masses and larger distances.
This constraint assumes implicitly that the stellar cluster rests relative to the MBH since
it was derived in the combined coordinate system. Using the velocity calibration of the maser system would have yielded a slightly weaker constraint.
However, an even stronger constraint comes from the radio measurements of Sgr~A* \citep{rei04}. The limit on the motion of radio Sgr~A* in galactic latitude ($v_b=-0.4\pm0.9\,$km/s) can also be used. Since this velocity limit is much smaller than the upper limit on the MBH motion from the stellar orbits, it is more constraining. Also
for this data it seems reasonable to assume that only systems with  $P>5\,$yr would have been discovered.
Similar arguments constraining the binarity of Sgr~A* have been put forward by \cite{han03}, whose results we also show in Figure~\ref{binlim}.

Secondly, two black holes in close orbits will loose energy via gravitational waves and thus spiral in. Demanding a life time of at least $10^7\,$yr for the IMBH-MBH system
excludes configurations towards smaller distances and higher masses. Dynamical stability can also be demanded for the S-star cluster as such. \cite{mik08} have shown
that an IMBH with a mass of $10^{-3}\,M_\mathrm{MBH}$  in a distance of $1\,$mpc would make the S-stars cluster unstable. It is reasonable to assume that this also holds for larger masses and radii at least as large as the S-star cluster extends ($\approx1''$).

Based on simulations, \cite{gua07} concluded that an IMBH will reach a stalling radius that  is proportional to the mass of the IMBH: $a_\mathrm{stall}=3.5\,\mu\mathrm{as}\times M_\mathrm{IMBH} [M_\odot]$ (for our values of mass and distance). Since one does not expect an IMBH to reside at a much smaller radius, this puts another constraint on the IMBH-MBH binary.

Finally, also the S2 orbit allows us to exclude part of the phase space. Motivated by the findings of section~\ref{post} and equation~\ref{etalim}, we simply assume that no mass larger than $0.02\,M_\mathrm{MBH}$ can be hidden inside the S2 orbit. Actually, also somewhat smaller radii than the pericenter distance $r_p$ of S2 are excluded, since this would still perturb the orbit figure notably. We estimate that down to $0.5 r_p$ no IMBH more massive than $0.02\,M_\mathrm{MBH}$ can reside.

\subsection{Properties of the stellar orbits}
We obtained orbits for 20 early-type stars. This relatively large number - \cite{eis05} had six orbits, \cite{ghe05} seven - allows us to 
assess distributions of orbital parameters and study the properties of the stellar orbits thereby characterizing the 
S-star population. 
\subsubsection{Orientations of orbital planes}
\label{stellarOrbits}
\begin{figure*}[htbp]
\begin{center}
\plotone{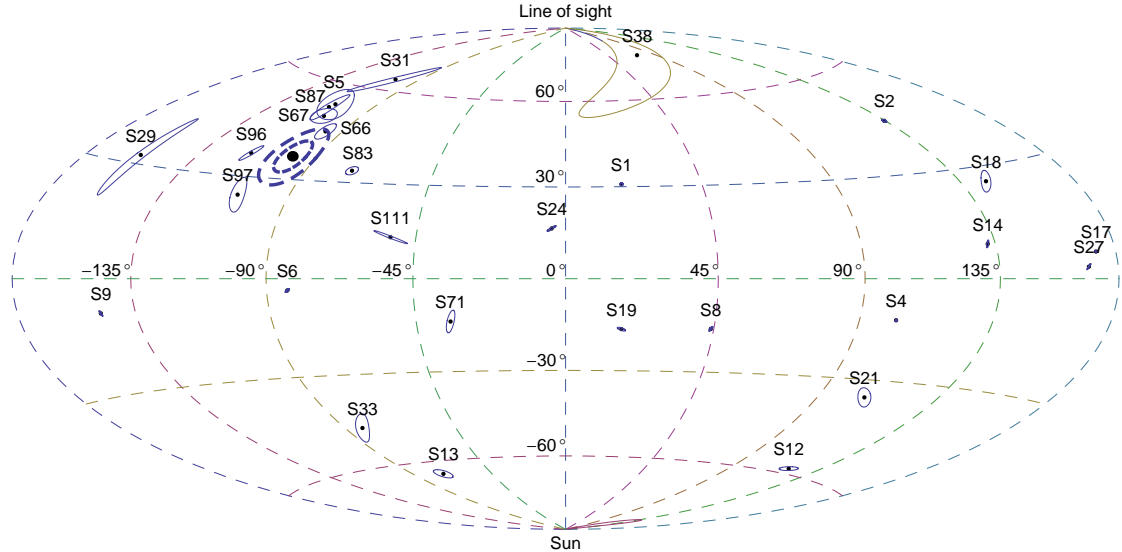}
\caption{Orientation of the orbital planes of those S-stars for which we were able to determine orbits. The orientation of the orbits in space is described by the orbital angular momentum vector, corresponding to a position in this all sky plot, in which the vertical dimension corresponds to the inclination $i$ of the orbit and the horizontal dimension to the longitude of the ascending node $\Omega$. 
A star in a face-on, clockwise orbit relative to the line of sight, for instance, would be located at the top of the graph, while a star with an edge-on seen orbit would
be located on the equator of the plot.
The error ellipses correspond to the statistical $1\,\sigma$ fit errors only, thus the area covered by each is 39\% of the probability density function.
Stars with an ambiguous inclination have been plotted at their more likely position.
The stars S66, S67, S83, S87, S96 and S97 which were suspected to be part of the clockwise stellar disk by \cite{pau06} at $(\Omega=99^\circ,\,i=127^\circ)$ actually are found very close to the position of the disk. 
The latter is marked by the thick black dot and the dashed lines, indicating a disk thickness of $14^\circ\pm4^\circ$, the value found by
\cite{pau06}. The orbits of the other stars are oriented randomly.
}
\label{angmom}
\end{center}
\end{figure*}

Figure~\ref{angmom} illustrates the orientations of the orbital planes for all stars from Table~\ref{orbitsTable}. 
\cite{pau06} suggested that the six stars S66, S67, S83, S87, S96 and
S97 (E17, E15 (S1-3), E16 (S0-15), E21, E20 (IRS16C) and E23 (IRS16SW) in their notation)
are members of the clockwise disk. Our findings explicitly confirm this.
All six stars have an angular distance to the disk between $9^\circ$ and $21^\circ$ with a mean and standard deviation of $15^\circ\pm4^\circ$. This is 
somewhat (a factor of 2) more than the disk thickness of $14^\circ\pm4^\circ$ found by \cite{pau06}. However, statistically the difference
is not very significant and only the inner edge of the disk is sampled here. 
All six disk stars have a semi major axis of $a\approx 1''$ and a small eccentricity ($e\approx0.2 - 0.4$) in agreement with the estimates from \cite{pau06}. 
The orbital plane of S5 is also consistent with the disk given its distance of $18^\circ$. However, the lower brightness ($m_K=15.2$) of the star and the higher
eccentricity ($e>0.8$) of the orbit make it unlikely that S5 is a true disk member. 
The next closest star to the disk beyond the six disk stars and S5 is S31 with an angular distance of $27^\circ$.
We also note that the orbital solutions for S96 and S97 derived from marginal accelerations are consistent with the disk hypothesis. Therefore we are confident in these orbits, too.

We used a Rayleigh test \citep{wil83} to check whether the distribution of orbital angular momenta for the 22 other stars for which we found orbits is compatible with a random distribution. We found a probability of randomness of $p=0.74$; meaning that the non-disk stars do not show a preferred orbit orientation.
Using the projection method from \cite{cue07} we obtained $p=1.0$. The same statement also holds when testing for randomness of the subset of early-type stars.

\subsubsection{Distribution of semi major axes}
\begin{figure}[htbp]
\begin{center}
\plotone{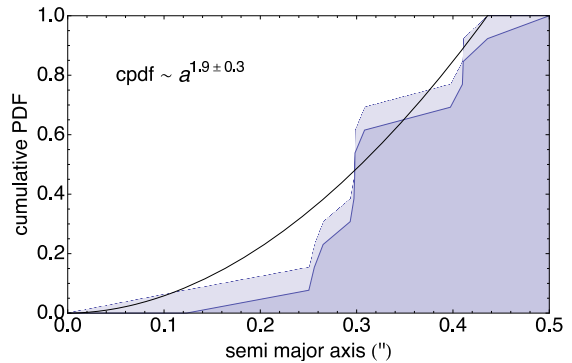}
\caption{The cumulative pdf for the semi major axis of the early-type stars with $a<0.5''$. The two curves 
correspond to the two ways to plot a cumulative pdf, with values ranging either from 0 to (N-1)/N or from 1/N to 1.
The distribution can be represented by $n(a) \sim a^{0.9\pm0.3}$.}
\label{cpdfsma}
\end{center}
\end{figure}

Figure~\ref{cpdfsma} shows the cumulative probability distribution function (pdf) for the semi major axes of stars which have semi major axis smaller than 0.5'', thus excluding
the stars that are identified to be members of the clockwise disk. The statistic is limited still (15 stars make up this sample), but nevertheless the distribution allows us to estimate the functional behavior of the pdf $n(a)$. Due to the small number of data points we did not bin the data but used a log-likelihood fit for $n(a)$.
We found $n(a) \sim a^{0.9\pm0.3}$. This can be converted to a number density profile as a function of radius \citep{ale05}. We obtain $n(r)\sim r^{-1.1\pm0.3}$, consistent with the mass profile in \cite{gen03b} who found $\rho(r)\sim r^{-1.4}$ and with the newer work in \cite{sch07} who found  $\rho(r)\sim r^{-1.2}$  for the innermost region of the cusp.

\subsubsection{Distribution of eccentricities}
\begin{figure}[htbp]
\begin{center}
\plotone{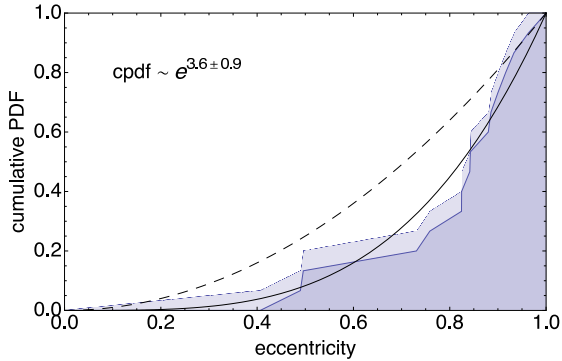}
\caption{The cumulative pdf for the eccentricities of the early-type stars which are not identified as disk members. The two curves 
correspond to the two ways to plot a cumulative pdf, with values ranging either from 0 to (N-1)/N or from 1/N to 1. The distribution is 
only marginally compatible with $n(e) \sim e$ (dashed line), the best fit is $n(e)\sim e^{2.6\pm0.9}$.}
\label{cpdfecc}
\end{center}
\end{figure}

The distribution of eccentricities allows us to estimate the velocity distribution. Figure~\ref{cpdfecc} shows the cumulative pdf for the eccentricities of those 
young (early-type) stars which are not associated with the clockwise stellar disk. Using again a log-likelihood fit, we find $n(e)\sim e^{2.6\pm0.9}$.
The profile still is barely consistent with $n(e) \sim e$, corresponding to an isotropic, thermal velocity distribution \citep{sch03,ale05}.
This would be the expectation for a relaxed stellar system. However, given that the maximal lifespan for B stars ($\lesssim10^8\,$yr) is much shorter than 
the local two body relaxation (TBR) time ($\approx 10^9\,$yr, \cite{ale05}) one does not expect a thermal distribution. In this light, it is interesting to notice that the distribution appears to be a bit steeper (i.e. peaked towards higher eccentricities) than a thermal distribution. This might be a first hint towards the formation scenario for the S-stars.
For example, it is exactly what one expects in the binary capture scenario \citep{per07}, in which the S-stars are initially captured on very 
eccentric orbits ($e\gtrsim0.98$), and then subsequent relaxation gradually smears out the distribution of eccentricities towards a thermal
distribution. From the time scales involved, one expects that the latter is not reached completely, so a high eccentricity bias remains, 
which in turn might fit nicely together with our indication for a steeper than thermal eccentricity distribution. 

This lays out a very interesting perspective for the continuation of the orbital monitoring. Increasing the statistics of the
eccentricity distribution by determining more stellar orbits will allow us to test explicitly whether it truly deviates from a thermal distribution
and thus provides us with a quantitative test for formation scenarios of the S-stars. 

\subsection{Estimates of the extended mass component}
In addition to the population of stars not yet resolved by current instrumentation
a cluster of dark objects  - e.g. stellar mass black holes (SBHs) as proposed in \cite{mor93,mir00,mun05,hop06} -  is plausibly present in the GC. 
As shown in section~\ref{post} the orbital data allows to test for such extended mass components. 
Here we investigate several theoretical and observational constraints on the extended mass distribution and relate these to $\eta$.
We mostly assume that the extended mass distribution is due to SBHs with a mass of $M_\star=m_\star\,M_\odot$ with $m_\star=10$ \citep{tim96}, since this component is likely to make up most of the mass
of a potential dark cluster \citep{ale07}. 
\subsubsection{Stellar number counts}
\cite{gen03b} and \cite{sch07} have inferred a stellar density profile for the GC from completeness-corrected stellar number counts. 
Assuming that the luminous objects trace the total mass, the number
density profile that is determined reliably on the $>0.01\,$pc scale can
be extrapolated to the S2 orbit. We obtain
\begin{equation}
\eta = 3.7 \times 10^{-4}\times \left( \frac{m_\star}{10} \right)\,\,,
\end{equation}
This extrapolation is quite uncertain,
since mass segregation predicts that the SBHs should have a much steeper slope than the
less-massive luminous stars in the central $0.01\,$pc, where the SBHs dominate the total mass \citep{hop06,ale07}. 
Therefore both the mass-to-number ratio and the slope of the
density profile are expected to have a significant radial dependence.
\subsubsection{The drain limit}
The drain limit is a
conservative theoretical upper limit of the number of compact objects
that can exist in steady-state around a MBH. It is 
given by the condition that the number of SBHs that can be packed inside any given radius in steady state has to be smaller than the number of SBHs scattered into the MBH over the age of the Galaxy \citep{ale04}. This can be translated into a theoretical limit
for $\eta$. Close to $m_\star=10$ and using $M_\mathrm{MBH}=4\times10^6 M_\odot$ and $t=10\,$Gyr the relation can be approximated by 
\begin{equation}
\eta  \lesssim 0.0011 \times \left( \frac{m_\star}{10} \right)^{-0.7}\,\,.
\end{equation}
The drain limit could be violated for a non steady-state situation. Indeed,
the existence of the young star disks with a relatively well-defined age of $6\,$My suggests that star formation in the GC is episodic.
However, the amount of mass from SBHs would hardly exceed $10^3\,M_\odot$ even assuming
an optimistic, top-heavy initial mass function, given that the total amount of
mass in the disks is $\approx10^4\,M_\odot$.
\subsubsection{Dynamical modeling of the dark cluster in the GC}
The expected degree of central concentration of SBHs around the
MBH can be estimated by modeling the dynamical evolution of a
system with a present-day mass function similar to that of the GC
\citep{ale05}. Monte Carlo
simulations of the GC using the H\'{e}non method and including also
stellar collisions and tidal disruptions (M. Freitag, priv. comm.; see
also \cite{fre06}) but neglecting star formation yield a rather flat mass density profile of
$10^8(M_\odot/\mathrm{pc}^3)\, (r/0.01\,\mathrm{pc})^{-0.5}$, 
which translates to
$\eta\sim10^{-4}$. Due to the statistical nature of this method the density profile at the very center is not well determined. An
alternative analytic solution for the steady state distribution using a much more idealized formulation of the mass segregation
problem \citep{hop06} yields a similar result of
$\eta\sim5 \times 10^{-4}$. However, in this method the fixed boundary conditions
far from the MBH may artificially maintain a high density in the
center by preventing the expansion of the system.  Nevertheless, the
fact that these two different methods yield similar results also consistent
with the drain limit lends some
credence to this estimate.
\subsubsection{Diffuse X-ray emission of a dark cluster}
A cluster of compact objects will accrete the surrounding gas and thus lead to X-ray emission, which for current X-ray satellites ($\approx1''$) would be barely resolved. Indeed, the X-ray source at the position of Sgr~A* is slightly extended \citep{bag03}. 
We fit the radial profile of Sgr~A* as reported by \cite{bag03} by the superposition of a point source with a Gaussian width of $\sigma_\mathrm{pt}=0.375''$ \citep{bag03} and an extended component with a free width $\sigma_\mathrm{ext}$. We obtain as empirical description for the surface brightness
profile of \cite{bag03}, Figure~6:
\begin{equation}
B(r)\,[\mathrm{cts/arcsec}^2]=73.5 \, e^{-r^2/2 \sigma_\mathrm{pt}^2} + 40.3 \,e^{-r^2/2 \sigma_\mathrm{ext}^2} 
\end{equation}
with $\sigma_\mathrm{ext}=1.05''$. Thus, we obtain for the extended luminosity (assuming the same spectral index of point like and extended component)
$L_\mathrm{X, ext} = 1.95\times10^{33}\,\mathrm{erg/s}$, accounting for $\approx80$\% of the total X-ray luminosity.

The expected X-ray luminosity of a single compact object is given by the mass accretion rate and the radiation efficiency. 
A simple estimate is given by assuming Bondi accretion \citep{bon52}: 
\begin{equation}
\dot{M_B}=4\pi \lambda(G M_\star)^2 n_e \mu \,m_p\, c_s^{-3}\, \approx 10^9\,\mathrm{g/s}\,\,,
\end{equation}
where $\lambda=1/4$, $n_e=26\,\mathrm{cm}^{-3}$ the electron number density, $\mu=0.7$ the mean atomic weight, $m_p$ the proton mass, $c_s=\sqrt{5\, k\, T_e/3\,\mu \,m_p}$ the speed of sound and $T_e=1.3\,$keV  \citep{bag03}.
\cite{pes03} estimate the accretion rate by
\begin{equation}
\dot{M_P}= \pi r_\mathrm{acc}^2 \,\rho\, v\approx 10^9\,\mathrm{g/s}\,\,,
\end{equation}
with the accretion radius $r_\mathrm{acc}=2 G M_\star/v_\mathrm{eff}^2 \approx 3\times10^{11}\,$cm.
Using the density from above and the Keplerian velocity at $r=1''$ one obtains consistently $\dot{M_P}\approx10^9\,$g/s$\,\approx\dot{M_B}$.

The radiation efficiency depends on the type of object considered \citep{hal96}. For neutron stars $10\%$ is assumed \citep{pes03}, since the
accreted material will fall onto a hard surface and the energy released can be radiated away, resulting in a luminosity of $L_\mathrm{NS}\approx10^{29}\,$erg/s.
For SBHs due to the absence of a surface most of the emission will be thermal
bremsstrahlung yielding only $L_\star\approx2\times10^{20}\,$erg/s \citep{hal96}. This shows that $L_\mathrm{X, ext}$ cannot be due to
SBHs, since one would need $10^{13}$ objects to explain the observed luminosity. In the case of neutron stars, one would need $\approx20000$ objects within $r\lesssim1''$ in order to account for the observed luminosity, corresponding to $\eta\approx0.07$. 
However, this number exceeds the estimate
of the segregated cusp model of \cite{hop06} who predict only
$\approx100$ neutron stars there.
\subsubsection{X-ray transients in a dark cluster}
\cite{mun05} report an overabundance of X-ray transients in the inner parsec of the GC compared to the overall distribution of X-ray sources. The sources are classified as  X-ray binaries (XRBs).
These authors suggest a dynamical origin of the XRBs, namely an exchange of type
 $\mathrm{Binary} + \mathrm{SBH}\rightarrow\mathrm{XRB}+ \mathrm{Star}$. The rate density for this reaction is 
\begin{equation}
\gamma_+=n_\star \,n_\mathrm{b}\, \Sigma\, \sigma_1
\end{equation}
where $n_\mathrm{b}$ is the density of binaries, 
\begin{equation}
\Sigma= \pi a^2 + 2\pi a G (M_\mathrm{b}+M_\star)/\sigma_1^2
\end{equation}
the exchange cross section and $\sigma_1=(G M_\mathrm{MBH}/3r)^{1/2}$ the 1D velocity dispersion.
According to \cite{mun05} the number of XRBs is limited by dynamical friction which yields a characteristic life time of
$\tau=10 \,\mathrm{Gyr}\,(M_\star/M_\odot)^{-1} (r/\mathrm{pc})^{1/2}$. Refining this argument, we also take into account the back reaction $\mathrm{XRB}+ \mathrm{Star}\rightarrow\mathrm{Binary} + \mathrm{SBH}$ and assume for simplicity equal exchange cross sections for forward and backward reaction.
Both effects together yield a rate density of
\begin{equation}
\gamma_-=\frac{1}{2} n\,  n_\mathrm{XRB}\,\Sigma\, \sigma_1 + \frac{ n_\mathrm{XRB}}{\tau}, 
\end{equation}
where $n$ is the number density of stars, $n_\mathrm{XRB}$ the number density of 
XRBs.\footnote{The factor $1/2$ takes care of the fact that for the back reaction either the SBH or the ordinary star of the XRB could be replaced by the interaction partner.}
In equilibrium one has $\gamma_+=\gamma_-$, which allows one to solve for $n_\mathrm{XRB}$. After integrating $n_\mathrm{XRB}$ over volume out to $1\,$pc and assuming $a=0.1\,$AU, $n=10^5\,\mathrm{pc}^{-3} (r/\mathrm{pc})^{-2}$, $n_\mathrm{b}=0.1 n$, $M_\mathrm{b}=3M_\odot$, $M_\star=10M_\odot$ and $n_\star = f_\star n$ one obtains the number of XRBs in the central parsec as $N_\mathrm{XRB}=7\times10^4 f_\star$, where $f_\star$ is the relative number of SBHs to ordinary stars.
A certain fraction $f_\mathrm{X}$ of those will shine up as X-ray transients: $f_\mathrm{X} N_\mathrm{XRB}=N_\mathrm{X}$. Calculating $\eta$
from this yields
\begin{eqnarray}
\eta &=& f_\star \frac{M_\star}{M_\mathrm{MBH}} \int_\mathrm{peri}^\mathrm{apo} 4\pi r^2 n(r) dr\\
&=& 3.8\times 10^{-7}\, \frac{N_\mathrm{X}(<1\mathrm{pc})}{f_\mathrm{X}}\,\,,
\end{eqnarray}
relating  the number of X-ray transients in the central parsec with $\eta$.
Using the values $f_\mathrm{X}\lesssim0.01$ and $N_\mathrm{X}=4$ \citep{mun05} we obtain $\eta \approx 1.5\times10^{-4}$.

A more detailed investigation by \cite{dee07} shows that within $r<0.7\,$pc a cusp of SBHs with $\approx20000$ members
is consistent with the number of discrete X-ray sources in the GC. Converting this number for the assumed profile of
$n(r)\sim r^{-7/4}$ into $\eta$ yields $\eta \approx 2.1\times10^{-4}$,
which is very similar to our estimate in the previous paragraph. 
\subsubsection{Further aspects}
There are at least three other aspects of an extended mass component in the GC which are worth mentioning
but beyond the scope of this paper.
\begin{itemize}
\item {\it Star formation in the presence of a dark cluster.} The process of star formation in the GC might be altered significantly by the presence of a substantial dark component. The additional perturbative gravitational forces
due to the SBHs might assist star formation since they increase the inhomogeneities in a star forming gas cloud. On the other hand, close encounters between individual clumps and SBHs might result in disruption of the clumps. 
\item {\it Interaction of the spin of the MBH with the dark cluster.} The spin of the MBH is subject to evolution by several processes. While gas accretion and major mergers can increase the spin, the accretion of SBHs tends to decrease the spin \citep{hug03,gam04}, assuming many random infall events of isotropically distributed SBHs. 
Furthermore there is the general relativistic spin-orbit-coupling between a SBH and the MBH spin, leading to a change of the spin direction of the MBH but not
to a change of its modulus \citep{lod06}.
\item {\it Dark matter.} Dark matter, which is widely accepted in cosmology, might also show up in dynamic measurements in the GC. 
However, \cite{gne04} show that the density of the dark matter at $0.01\,$pc is $\rho_\mathrm{DM}\approx3\times10^5\,M_\odot/\mathrm{pc}^3$, which is negligible compared to the theoretically
predicted stellar density there \citep{hop06}. See also \cite{vas08}.
\end{itemize}
\subsubsection{Conclusions for an extended mass component}
\label{conclusion}
The various estimates for $\eta$ all consistently point towards 
an expected value of $\approx10^{-3}-10^{-4}$, approximately two orders of magnitude smaller than what we can
measure with orbital dynamics today. Nevertheless, some astrophysical insights are possible.

Among the most important scientific questions in the GC is the origin of the S-stars, being a population of apparently young stars 
close to the MBH \citep{ghe03, mar08}. 
One possible origin is that these stars have reached their current orbits by TBR. Then
the S-stars would have an isotropic, thermal velocity distribution,
naturally explaining the observed random distribution of angular momentum vectors (Figure~\ref{angmom}).
The number of stars visible is by far too low to make TBR efficient enough to account for the
present population of S-stars. A hypothetical cluster of SBHs could accelerate the process.
The Chandrasekhar TBR timescale \citep{bin87} is given by
\begin{equation}
t_r \approx \frac{0.34\, \sigma^3}{G^2 \langle M_\star \rangle^2 n_\star \ln \Lambda}\,\,.
\end{equation}
For a power law cusp around a MBH, the velocity dispersion and the density are related to each other. Assuming $\ln \Lambda\approx10$, a power law index of $-3/2$ (which is approximately what is observed) and a population of stars with a single mass one obtains a relaxation time independent of radius
\begin{equation}
t_r \approx 1.8 \times 10^{5} \,\mathrm{yr}\, \,\eta^{-1} \left( \frac{m_\star}{10}\right)^{-1}\,\,,
\label{trelax}
\end{equation}
Thus, if the S-stars formed at the same epoch as the stellar disks $6\times10^6$ years ago \citep{pau06} and reached their present-day orbits by TBR, one needs $\eta \gtrsim 0.033$ for $m_\star=10$ \citep{tim96}. This exceeds the expectations by at least two orders of magnitudes.
If the S-stars were not born in the presently observed disks,
but in older, now-dispersed disks, one can use Equation~\ref{trelax} with the typical age of B stars ($\approx5 \times 10^7\,$yr). For $m_\star=10$ this yields $\eta\gtrsim3.5\times10^{-3}$, which could be marginally compatible with the other estimates for $\eta$.

In order to assess the expected progress we simulated future observations with existing instrumentation and similar sampling. Continuing the orbital monitoring for two more years will lower the statistical error to $\Delta \eta \approx 0.01$, corresponding to $t_r\approx2\times10^7\,$yr. This means we will soon be able to test the hypothesis
that the S-stars formed in the disks and reached their current orbits by TBR. Furthermore there is a chance to rule out any
TBR origin of the S-stars observationally in the near future, namely when $\eta\lesssim3.5\times10^{-3}$ is reached.

\section{Summary}
We continued our long-term study of stellar orbits around the MBH in the Galactic Center. This work is based on our large, high quality data base which is based on high resolution imaging and spectroscopy from the years 1992 to 2008. The main results are
\begin{itemize}
\item The best current coordinate reference
system uses all available IR positions of the SiO maser stars \citep{rei07} for the definition of the origin and assumes that the stellar cluster around Sgr~A* is intrinsically at rest such that it can be used for the calibration of the coordinate system velocity. Having more measurements of the maser sources both in the radio and the IR domain
we eventually will be able to directly tie the coordinate system velocity to radio Sgr~A* with a sufficient precision. Then the intermediate step of 
cross calibration with the stellar cluster can be dropped and the coordinate system definition would be independent from the assumption that the stellar cluster is at rest with respect to Sgr~A*.
\item We obtained orbits for 28 stars. Eleven of those can
contribute to the determination of the gravitational potential, we used up to six. For the first time we were able to determine orbital
parameters for six of the late-type stars in our
sample. Furthermore, we confirm unambiguously the earlier report
\citep{pau06} that six of the stars are members of the clockwise disk.
\item Overall, we improved measurement uncertainties by a factor of six over the most recent set of Galactic Center 
papers \citep{sch02,ghe05,eis05}. A single point mass potential continues to be the best fit to these improved data as well. The main contribution to the error in the mass of Sgr~A* and the distance to the 
Galactic Center are systematic uncertainties. While the value of the mass is driven by the distance estimate, the latter is subject to many systematic uncertainties that amount to $0.31\,$kpc. The statistical error now decreased to $0.17\,$kpc and became smaller than the systematic one. The most fruitful way to overcome current limitations would probably be the observation of another close pericenter passage of an S-star. Our current best values are:
\begin{eqnarray}
M&=&  (3.95 \pm 0.06|_\mathrm{stat} \pm 0.18|_\mathrm{R_0,\;stat} \pm 0.31|_\mathrm{R_0,\;sys})\nonumber\\
&&\times 10^6\,M_\odot \times (R_0/8\,\mathrm{kpc})^{2.19}\nonumber\\
&&=(4.31 \pm 0.36) \times 10^6\,M_\odot \;\mathrm{for}\; R_0=8.33\,\mathrm{kpc}\nonumber\\
R_0&=& 8.33 \pm 0.17|_\mathrm{stat} \pm 0.31|_\mathrm{sys}\,\mathrm{kpc}
\end{eqnarray}
It should be noted that this value is consistent within the errors with values published earlier \citep{eis03b,eis05}. The improvement of our current work is the more rigorous treatment of the systematic errors. Also it is worth noting that adding more stars did not change the distance much over the 
equivalent S2-only fit.
\item We have obtained an upper limit for the mass enclosed within the S2 orbit in units of the mass of the MBH:
\begin{equation}
\eta=0.021\pm 0.019|_\mathrm{stat}\pm 0.006|_\mathrm{model}\,\,.
\end{equation}
which corresponds to a $1\sigma$ upper limit of $\eta\le0.040$. 
\end{itemize}

\clearpage

\appendix
\section{Acknowledgements}
We would like to thank D.T. Jaffe, U Texas, for helpful discussions and U. Bastian, U Heidelberg,
for access to GAIA internal notes. T. A. is supported by Minerva grant 8563, ISF grant 968/06 and a New
Faculty grant by Sir H. Djangoly, CBE, of London, UK. We also thank our referee, whose report helped in
improving the manuscript.
~\\
~\\
~\\
\section{A. List of NACO data sets}
\label{nacocom}
{\scriptsize
\begin{center}
\begin{tabular}{lcccccc|lcccccc}
Date&Band&$\frac{\mathrm{mas}}{\mathrm{pix}}$&DIT &NDIT&\#& \#&
Date&Band&$\frac{\mathrm{mas}}{\mathrm{pix}}$&DIT &NDIT&\#& \#\\
&&&[s]&&frames&S-stars&
&&&[s]&&frames&S-stars\\
\hline
2002.25&Ks&27&0.5&8&10&35& 2006.32&H&13&17&2&36&97\\
2002.34&Ks&27&20&3&20&84&  2006.41&H&13&17&2&48&80\\
2002.39&Ks&13&15&1&11&64&  2006.49&H&13&17&2&48&53\\
2002.409&Ks&13&15&1&86&70& 2006.49&Ks&13&17&2&94&58\\
2002.412&Ks&13&15&1&58&47& 2006.57&H&13&60&1&32&55\\
2002.414&Ks&13&15&1&45&83& 2006.58&Ks&13&2.4&14&38&75\\
2002.58&Ks&13&15&4&60&83&  2006.65&H&13&17&2&44&57\\
2002.66&H&13&15&4&25&86&   2006.726&Ks&13&17.2&2&48&85\\
2002.66&Ks&13&15&4&20&84&  2006.728&Ks&13&17.2&2&40&89\\
2003.21&H&13&20&1&32&84&   2006.75&Ks&13&17.2&2&24&88\\
2003.35&H&13&10&6&30&89&   2006.78&Ks&13&17.2&2&48&93\\
2003.36&Ks&13&5&12&8&82&   2006.80&Ks&13&2.4&14&48&87\\
2003.445&Ks&13&5&3&102&81& 2007.17&Ks&13&12&3&32&100\\
2003.451&H&13&20&1&116&90& 2007.21&H&13&17&2&102&105\\
2003.452&Ks&13&20&1&182&86&2007.21&Ks&13&2.4&14&48&100\\
2003.454&H&13&20&1&150&85& 2007.214&Ks&13&17.2&2&96&100\\
2003.454&Ks&13&10&2&208&81&2007.252&Ks&13&10&3&48&99\\
2003.55&H&13&20&3&72&81&   2007.255&H&13&10&3&96&103\\
2003.676&H&13&20&3&32&81&  2007.255&Ks&13&10&3&63&97\\
2003.678&H&13&2&30&32&65&  2007.46&Ks&13&17.2&2&110&100\\
2003.76&Ks&13&5&12&34&89&  2007.54&H&13&10&3&48&96\\
2004.24&H&13&10&3&41&85&   2007.55&H&13&10&3&96&106\\
2004.33&H&13&15&2&73&92&   2007.69&H&13&17&2&48&104\\
2004.35&Ks&13&10&3&52&70&  2007.69&Ks&13&17.2&2&48&100\\
2004.44&H&13&15&2&48&94&   2007.692&Ks&13&17.2&2&48&100\\
2004.51&Ks&13&30&1&272&86& 2008.15&Ks&13&17.2&2&48&101\\
2004.52&H&13&30&1&48&82&   2008.20&Ks&13&17.2&2&68&106\\
2004.57&H&13&15&2&47&70&   2008.27&Ks&13&17.2&2&96&93\\
2004.57&Ks&13&15&2&92&46&  2008.46&Ks&13&17.2&2&96&101\\
2004.66&Ks&13&15&2&100&89& 2008.47&H&13&17.2&2&65&88\\
2004.73&H&13&25&1&16&92&   2008.60&Ks&13&17.2&2&90&104\\
2005.27&Ks&13&2&15&48&95\\
2005.37&Ks&13&2&15&71&91&\multicolumn{7}{l}{DIT: single detector integration time}\\
2005.47&Ks&13&10&2&77&83&\multicolumn{7}{l}{NDIT: number of single integrations per image file}\\
2005.58&Ks&13&15&4&23&93\\
2005.67&Ks&27&30&1&19&54\\
\end{tabular}
\end{center}
}
\clearpage

\section{B. Polynomial fits to the S-stars data}
\label{polytab}
The following table lists the polynomial fits to the S-stars data (except S2 which is not well described by polynomial fits). 
For stars with a significant (at the 5-$\sigma$ level) astrometric acceleration pointing towards Sgr~A* we report quadratic fits.
For stars with significant $da/dt$ we report the cubic fit. Otherwise
linear fits are given. Similarly, for stars for which detected changes in the
radial velocities, we report linear fits. For stars for which we determined orbits but did not detect changes in $v_\mathrm{vrad}$ we report the
weighted averages. 

{\scriptsize
\begin{center}
\begin{tabular}{l|l}
Name, $m_K$& $\alpha\, [\mathrm{mas}] =$\\
$t_0$ [yr] for $(\alpha,\,\delta)$ & $\delta\, [\mathrm{mas}] =$\\
$t_0$ [yr] for $v_z$ & $v_z\, [\mathrm{km/s}] =$\\ 
\hline
S1, 14.7 & $(-87.2 \pm 0.4) + (19.70 \pm 0.15) \Delta t + (0.665 \pm 0.022) (\Delta t)^2 + (-0.0483 \pm 0.0026) (\Delta t)^3$\\
2000.41 & $(-125.4 \pm 0.5) + (-32.04 \pm 0.16) \Delta t + (1.080 \pm 0.023) (\Delta t)^2 + (0.0398 \pm 0.0032) (\Delta t)^3$\\
2005.77&$(1094.5 \pm 8.4) + (-45.8 \pm 4.2) \Delta t  + (13.6 \pm 2.6) (\Delta t)^2$\\
\hline
S4, 14.4 & $(269.9 \pm 0.2) + (15.95 \pm 0.16) \Delta t + (-0.544 \pm 0.070) (\Delta t)^2 + (-0.0260 \pm 0.0101) (\Delta t)^3$\\
2003.07 & $(124.8 \pm 0.1) + (-0.36 \pm 0.08) \Delta t + (-0.129 \pm 0.023) (\Delta t)^2 + (0.0134 \pm 0.0039) (\Delta t)^3$\\
2006.40&$(-687.9 \pm 13.3) + (-66.7 \pm 9.1) \Delta t$\\
\hline
S5, 15.2 & $(352.0 \pm 0.3) + (-4.93 \pm 0.12) \Delta t + (-0.510 \pm 0.078) (\Delta t)^2$\\
2005.05 & $(200.4 \pm 0.4) + (8.00 \pm 0.15) \Delta t + (-0.341 \pm 0.094) (\Delta t)^2$\\
2006.40 & $129\pm39$\\
\hline
S6, 15.4 & $(484.6 \pm 0.1) + (6.38 \pm 0.04) \Delta t + (-0.128 \pm 0.026) (\Delta t)^2$\\
2005.05 & $(99.1 \pm 0.2) + (0.74 \pm 0.05) \Delta t + (-0.075 \pm 0.031) (\Delta t)^2$\\
2006.40 & $118\pm21$\\
\hline
S7, 15.3 & $(533.2 \pm 0.2) + (-4.03 \pm 0.04) \Delta t$\\
2000.41 & $(-29.8 \pm 0.3) + (-3.08 \pm 0.05) \Delta t$\\
\hline
S8, 14.5 & $(336.1 \pm 0.2) + (15.03 \pm 0.08) \Delta t + (-0.347 \pm 0.009) (\Delta t)^2$\\
2000.41 & $(-212.6 \pm 0.2) + (-14.60 \pm 0.10) \Delta t + (0.228 \pm 0.010) (\Delta t)^2$\\
2005.77&$-(53.6 \pm 7.6) + (-31.2 \pm 5.3) \Delta t$\\
\hline
S9, 15.1 & $(181.1 \pm 0.3) + (1.65 \pm 0.16) \Delta t + (-0.254 \pm 0.021) (\Delta t)^2$\\
2001.46 & $(-335.9 \pm 0.3) + (-8.69 \pm 0.15) \Delta t + (0.634 \pm 0.019) (\Delta t)^2$\\
2006.40&$614\pm27$\\
\hline
S10, 14.1 & $(64.2 \pm 0.1) + (-5.14 \pm 0.03) \Delta t$\\
2001.46 & $(-381.9 \pm 0.1) + (2.96 \pm 0.03) \Delta t$\\
\hline
S11, 14.3 & $(142.4 \pm 0.3) + (8.79 \pm 0.05) \Delta t$\\
2000.41 & $(-548.9 \pm 0.2) + (-5.39 \pm 0.04) \Delta t$\\
\hline
S12, 15.5 & $(-66.4 \pm 0.3) + (4.73 \pm 0.21) \Delta t + (1.066 \pm 0.052) (\Delta t)^2 + (-0.1002 \pm 0.0066) (\Delta t)^3$\\
2002.07 & $(252.0 \pm 0.3) + (29.01 \pm 0.15) \Delta t + (-1.606 \pm 0.028) (\Delta t)^2 + (0.0745 \pm 0.0044) (\Delta t)^3$\\
2005.77 & $318\pm7$\\
\hline
S13, 15.8 & $(-187.3 \pm 1.6) + (15.22 \pm 0.79) \Delta t + (3.954 \pm 0.129) (\Delta t)^2 + (-0.0880 \pm 0.0206) (\Delta t)^3$\\
2001.46 & $(-32.0 \pm 1.0) + (44.79 \pm 0.48) \Delta t + (-0.085 \pm 0.068) (\Delta t)^2 + (-0.3469 \pm 0.0116) (\Delta t)^3$\\
2006.40&$(-7.3 \pm 22.8) + (186.7 \pm 14.7) \Delta t$\\
\hline
S14, 15.7 & $(64.2 \pm 2.4) + (17.48 \pm 0.89) \Delta t + (2.838 \pm 0.085) (\Delta t)^2 + (-0.2773 \pm 0.0142) (\Delta t)^3$\\
2000.41 & $(36.7 \pm 2.2) + (14.13 \pm 0.81) \Delta t + (2.485 \pm 0.076) (\Delta t)^2 + (-0.2473 \pm 0.0129) (\Delta t)^3$\\
2006.40 & $300.3 \pm 25.2$\\ 
\hline
S17, 15.3 & $(10.3 \pm 1.0) + (2.81 \pm 0.47) \Delta t + (-0.373 \pm 0.072) (\Delta t)^2 + (0.1068 \pm 0.0139) (\Delta t)^3$\\
2001.46 & $(-162.1 \pm 1.0) + (20.04 \pm 0.39) \Delta t + (1.408 \pm 0.068) (\Delta t)^2 + (-0.1411 \pm 0.0146) (\Delta t)^3$\\
2005.85&$(594.7 \pm 5.2) + (-84.0 \pm 4.6)\Delta t$ \\
\hline
S18, 16.7 & $(-202.5 \pm 0.5) + (-17.08 \pm 0.18) \Delta t + (0.946 \pm 0.115) (\Delta t)^2$\\
2005.47 & $(-70.2 \pm 0.6) + (-18.22 \pm 0.22) \Delta t + (0.008 \pm 0.146) (\Delta t)^2$\\
2006.40 & $-257\pm53$\\
\hline
S19, 16.0 & $(38.7 \pm 1.1) + (-10.27 \pm 0.31) \Delta t + (-2.461 \pm 0.216) (\Delta t)^2$\\
2005.98 & $(-117.3 \pm 0.9) + (-19.78 \pm 0.26) \Delta t + (6.068 \pm 0.177) (\Delta t)^2$\\
2006.40&$(-2314.9 \pm 36.4) + (10.4 \pm 15.4) \Delta t  + (88.0\pm 13.7) (\Delta t)^2$\\
\hline
S20, 15.7 & $(220.8 \pm 0.6) + (-4.94 \pm 0.31) \Delta t$\\
2005.98 & $(109.5 \pm 0.4) + (-6.29 \pm 0.21) \Delta t$\\
\hline
S21, 16.9 & $(-334.1 \pm 0.2) + (-11.19 \pm 0.06) \Delta t + (1.001 \pm 0.039) (\Delta t)^2$\\
2005.47 & $(-128.7 \pm 0.4) + (4.15 \pm 0.10) \Delta t + (0.598 \pm 0.068) (\Delta t)^2$\\
2006.40&$410\pm12$\\
\hline
S22, 16.6 & $(191.2 \pm 0.2) + (22.81 \pm 0.11) \Delta t$\\
2005.47 & $(-268.4 \pm 0.3) + (-6.89 \pm 0.14) \Delta t$\\
\hline
S23, 17.8 & $(307.4 \pm 1.1) + (-13.81 \pm 0.34) \Delta t + (-0.953 \pm 0.196) (\Delta t)^2$\\
2005.47 & $(-89.1 \pm 0.9) + (-11.17 \pm 0.23) \Delta t + (0.525 \pm 0.153) (\Delta t)^2$\\
\hline
S24, 15.6 & $(-177.1 \pm 0.3) + (6.32 \pm 0.14) \Delta t + (0.065 \pm 0.019) (\Delta t)^2$\\
2001.46 & $(-566.7 \pm 0.3) + (10.40 \pm 0.13) \Delta t + (0.210 \pm 0.019) (\Delta t)^2$\\
2006.40&$(-824.9 \pm 6.5) + (-23.0 \pm 4.8) ( \Delta t)$\\
\hline
S25, 15.2 & $(-95.3 \pm 0.3) + (-2.88 \pm 0.06) \Delta t$\\
2001.46 & $(-426.4 \pm 0.3) + (1.08 \pm 0.05) \Delta t$\\
\hline
S26, 15.1 & $(514.4 \pm 0.2) + (6.39 \pm 0.05) \Delta t$\\
2001.46 & $(440.6 \pm 0.2) + (0.81 \pm 0.05) \Delta t$\\
\hline
S27, 15.6 & $(146.9 \pm 0.3) + (0.66 \pm 0.15) \Delta t + (-0.032 \pm 0.019) (\Delta t)^2$\\
2001.46 & $(523.1 \pm 0.3) + (3.74 \pm 0.14) \Delta t + (-0.112 \pm 0.018) (\Delta t)^2$\\
2005.77&$-114\pm3$\\
\end{tabular}
\end{center}
{\scriptsize
\begin{center}
\begin{tabular}{l|l}
Name, $m_K$& $\alpha\, [\mathrm{mas}] =$\\
$t_0$ [yr] for $(\alpha,\,\delta)$ & $\delta\, [\mathrm{mas}] =$\\
$t_0$ [yr] for $v_z$ & $v_z\, [\mathrm{km/s}] =$\\ 
\hline
S28, 17.1 & $(-19.7 \pm 0.6) + (4.06 \pm 0.22) \Delta t + (0.145 \pm 0.031) (\Delta t)^2$\\
2001.46 & $(427.6 \pm 0.6) + (12.09 \pm 0.27) \Delta t + (-0.648 \pm 0.036) (\Delta t)^2$\\
\hline
S29, 16.7 & $(-206.5 \pm 0.2) + (1.79 \pm 0.16) \Delta t + (0.256 \pm 0.036) (\Delta t)^2$\\
2003.49 & $(519.3 \pm 0.3) + (-14.93 \pm 0.28) \Delta t + (-0.792 \pm 0.064) (\Delta t)^2$\\
2007.94&$-273\pm38$\\
\hline
S30, 14.3 & $(-560.3 \pm 0.1) + (0.92 \pm 0.02) \Delta t$\\
2000.41 & $(384.3 \pm 0.1) + (3.37 \pm 0.02) \Delta t$\\
\hline
S31, 15.7 & $(-303.7 \pm 0.5) + (5.89 \pm 0.23) \Delta t + (0.561 \pm 0.040) (\Delta t)^2 + (0.0269 \pm 0.0068) (\Delta t)^3$\\
2001.39 & $(301.6 \pm 0.5) + (-15.88 \pm 0.24) \Delta t + (-0.347 \pm 0.045) (\Delta t)^2 + (-0.0421 \pm 0.0068) (\Delta t)^3$\\
2006.40&$-366\pm23$\\
\hline
S32, 16.6 & $(-323.3 \pm 0.1) + (-4.17 \pm 0.05) \Delta t$\\
2005.43 & $(-356.2 \pm 0.2) + (0.62 \pm 0.07) \Delta t$\\
\hline
S33, 16.0 & $(-411.7 \pm 0.5) + (-13.54 \pm 0.20) \Delta t + (0.253 \pm 0.020) (\Delta t)^2$\\
2000.41 & $(-396.2 \pm 0.4) + (0.90 \pm 0.17) \Delta t + (0.170 \pm 0.017) (\Delta t)^2$\\
2006.42&$-139\pm33$\\
\hline
S34, 15.5 & $(302.3 \pm 0.3) + (9.58 \pm 0.07) \Delta t$\\
2002.07 & $(-469.7 \pm 0.3) + (3.78 \pm 0.07) \Delta t$\\
\hline
S35, 13.3 & $(540.5 \pm 0.1) + (1.70 \pm 0.02) \Delta t$\\
2000.41 & $(-437.8 \pm 0.1) + (3.16 \pm 0.03) \Delta t$\\
\hline
S36, 16.4 & $(276.5 \pm 0.2) + (-1.15 \pm 0.16) \Delta t$\\
2004.56 & $(246.4 \pm 0.3) + (-0.71 \pm 0.17) \Delta t$\\
\hline
S37, 16.1 & $(331.2 \pm 0.4) + (-6.16 \pm 0.18) \Delta t$\\
2005.47 & $(390.2 \pm 0.3) + (10.17 \pm 0.13) \Delta t$\\
\hline
S38, 17.0 & $(-179.6 \pm 0.6) + (-30.56 \pm 0.61) \Delta t + (4.923 \pm 0.488) (\Delta t)^2$\\
2006.94 & $(55.2 \pm 0.7) + (-10.91 \pm 0.98) \Delta t + (-1.951 \pm 0.754) (\Delta t)^2$\\
2008.26&$-185\pm70$\\
\hline
S39, 16.8 & $(-102.2 \pm 0.7) + (-11.80 \pm 0.16) \Delta t + (1.385 \pm 0.132) (\Delta t)^2$\\
2005.50 & $(268.9 \pm 1.2) + (33.61 \pm 0.29) \Delta t + (-2.386 \pm 0.233) (\Delta t)^2$\\
\hline
S40, 17.2 & $(144.0 \pm 1.5) + (3.96 \pm 0.48) \Delta t + (-3.772 \pm 0.512) (\Delta t)^2$\\
2006.42 & $(33.4 \pm 2.9) + (1.71 \pm 0.98) \Delta t + (-2.621 \pm 1.026) (\Delta t)^2$\\
\hline
S41, 17.5 & $(-221.0 \pm 0.6) + (-0.58 \pm 0.37) \Delta t$\\
2004.94 & $(-299.0 \pm 0.6) + (-1.99 \pm 0.38) \Delta t$\\
\hline
S42, 17.5 & $(-160.1 \pm 0.7) + (-6.18 \pm 0.43) \Delta t$\\
2004.98 & $(-354.3 \pm 1.1) + (18.00 \pm 0.63) \Delta t$\\
\hline
S43, 17.5 & $(-493.2 \pm 0.3) + (5.39 \pm 0.12) \Delta t$\\
2005.47 & $(-134.4 \pm 0.4) + (7.78 \pm 0.16) \Delta t$\\
\hline
S44, 17.5 & $(-92.2 \pm 0.5) + (-9.48 \pm 0.43) \Delta t$\\
2006.52 & $(-246.0 \pm 1.0) + (-10.57 \pm 0.76) \Delta t$\\
\hline
S45, 15.7 & $(193.3 \pm 0.2) + (-6.61 \pm 0.08) \Delta t$\\
2005.47 & $(-515.0 \pm 0.3) + (-4.06 \pm 0.13) \Delta t$\\
\hline
S46, 15.7 & $(246.1 \pm 0.4) + (0.57 \pm 0.12) \Delta t$\\
2001.46 & $(-574.3 \pm 0.4) + (5.57 \pm 0.10) \Delta t$\\
\hline
S47, 16.3 & $(383.6 \pm 0.8) + (-3.82 \pm 0.70) \Delta t$\\
2006.52 & $(245.2 \pm 0.4) + (5.05 \pm 0.38) \Delta t$\\
\hline
S48, 16.6 & $(438.5 \pm 0.5) + (-0.33 \pm 0.14) \Delta t + (-0.442 \pm 0.098) (\Delta t)^2$\\
2005.47 & $(472.1 \pm 0.5) + (12.26 \pm 0.14) \Delta t + (-0.343 \pm 0.098) (\Delta t)^2$\\
\hline
S49, 17.5 & $(585.6 \pm 0.6) + (15.61 \pm 0.34) \Delta t$\\
2005.63 & $(51.6 \pm 0.7) + (1.30 \pm 0.34) \Delta t$\\
\hline
S50, 17.2 & $(-504.7 \pm 0.3) + (-2.84 \pm 0.13) \Delta t$\\
2005.47 & $(-528.8 \pm 0.3) + (9.51 \pm 0.18) \Delta t$\\
\hline
S51, 17.4 & $(-473.5 \pm 0.3) + (7.86 \pm 0.14) \Delta t$\\
2005.47 & $(-299.4 \pm 0.3) + (8.01 \pm 0.15) \Delta t$\\
\hline
S52, 17.1 & $(200.8 \pm 0.5) + (2.68 \pm 0.56) \Delta t$\\
2006.94 & $(286.0 \pm 1.3) + (-3.13 \pm 1.24) \Delta t$\\
\hline
S53, 17.2 & $(323.7 \pm 0.7) + (12.15 \pm 1.04) \Delta t$\\
2007.46 & $(514.2 \pm 0.7) + (9.04 \pm 1.03) \Delta t$\\
\hline
S54, 17.5 & $(135.4 \pm 1.2) + (-1.05 \pm 0.90) \Delta t$\\
2006.59 & $(-60.1 \pm 0.9) + (-26.90 \pm 0.67) \Delta t$\\
\hline
S55, 17.5 & $(93.8 \pm 2.9) + (-23.72 \pm 4.31) \Delta t$\\
2006.78 & $(-95.8 \pm 3.0) + (22.74 \pm 4.18) \Delta t$\\
\hline
S56, 17.0 & $(143.5 \pm 1.2) + (-15.96 \pm 2.81) \Delta t$\\
2007.81 & $(154.6 \pm 0.5) + (0.16 \pm 1.16) \Delta t$\\
\hline
S57, 17.6 & $(393.6 \pm 1.1) + (-9.99 \pm 1.43) \Delta t$\\
2007.46 & $(-147.4 \pm 0.8) + (-4.05 \pm 1.07) \Delta t$\\
\hline
S58, 17.4 & $(-339.6 \pm 0.4) + (5.98 \pm 0.11) \Delta t + (0.189 \pm 0.069) (\Delta t)^2$\\
2005.47 & $(-569.7 \pm 0.5) + (4.03 \pm 0.13) \Delta t + (0.584 \pm 0.085) (\Delta t)^2$\\
\hline
S59, 17.2 & $(-222.8 \pm 0.4) + (5.87 \pm 0.89) \Delta t$\\
2007.89 & $(240.7 \pm 0.8) + (-3.83 \pm 1.71) \Delta t$\\
\hline
S60, 16.3 & $(-277.6 \pm 0.9) + (3.38 \pm 2.19) \Delta t$\\
2007.82 & $(168.0 \pm 0.5) + (-19.17 \pm 1.02) \Delta t$\\
\hline
S61, 17.9 & $(-205.9 \pm 1.4) + (-18.08 \pm 2.17) \Delta t$\\
2007.90 & $(-45.0 \pm 1.4) + (-20.35 \pm 2.12) \Delta t$\\
\hline
S62, 17.8 & $(-59.2 \pm 1.3) + (-10.31 \pm 2.34) \Delta t$\\
2007.90 & $(67.8 \pm 1.7) + (2.28 \pm 3.47) \Delta t$\\
\end{tabular}
\end{center}
{\scriptsize
\begin{center}
\begin{tabular}{l|l}
Name, $m_K$& $\alpha\, [\mathrm{mas}] =$\\
$t_0$ [yr] for $(\alpha,\,\delta)$ & $\delta\, [\mathrm{mas}] =$\\
$t_0$ [yr] for $v_z$ & $v_z\, [\mathrm{km/s}] =$\\ 
\hline
S63, 17.5 & $(196.0 \pm 2.1) + (-12.04 \pm 4.59) \Delta t$\\
2007.93 & $(-134.7 \pm 2.8) + (-9.44 \pm 5.89) \Delta t$\\
\hline
S64, 17.5 & $(-12.3 \pm 1.1) + (-16.42 \pm 0.46) \Delta t$\\
2005.51 & $(238.5 \pm 1.3) + (8.57 \pm 0.57) \Delta t$\\
\hline
S65, 13.7 & $(-777.0 \pm 0.1) + (2.17 \pm 0.02) \Delta t$\\
2000.41 & $(-269.5 \pm 0.1) + (-1.49 \pm 0.02) \Delta t$\\
\hline
S66, 14.8 & $(-47.7 \pm 0.2) + (12.90 \pm 0.10) \Delta t + (0.015 \pm 0.010) (\Delta t)^2$\\
2000.41 & $(-1006.8 \pm 0.2) + (-1.39 \pm 0.08) \Delta t + (0.047 \pm 0.008) (\Delta t)^2$\\
2004.63&$12\pm22$\\
\hline
S67, 12.1 & $(461.4 \pm 0.2) + (-13.69 \pm 0.07) \Delta t + (-0.036 \pm 0.008) (\Delta t)^2$\\
2000.41 & $(872.4 \pm 0.2) + (1.98 \pm 0.08) \Delta t + (-0.058 \pm 0.009) (\Delta t)^2$\\
2003.95&$1\pm33$\\
\hline
S68, 12.9 & $(275.8 \pm 0.4) + (4.90 \pm 0.07) \Delta t$\\
2000.41 & $(764.9 \pm 0.4) + (2.83 \pm 0.08) \Delta t$\\
\hline
S69, 16.8 & $(-16.3 \pm 0.3) + (-0.25 \pm 0.17) \Delta t$\\
2005.47 & $(762.7 \pm 0.5) + (0.69 \pm 0.25) \Delta t$\\
\hline
S70, 16.9 & $(-348.1 \pm 0.1) + (-3.53 \pm 0.05) \Delta t$\\
2005.47 & $(714.6 \pm 0.2) + (-4.09 \pm 0.08) \Delta t$\\
\hline
S71, 16.1 & $(-562.1 \pm 0.2) + (7.94 \pm 0.07) \Delta t + (0.152 \pm 0.047) (\Delta t)^2$\\
2005.47 & $(-762.1 \pm 0.4) + (14.91 \pm 0.09) \Delta t + (0.324 \pm 0.068) (\Delta t)^2$\\
2007.44&$-237\pm85$\\
\hline
S72, 14.3 & $(-668.1 \pm 0.2) + (8.88 \pm 0.03) \Delta t$\\
2000.41 & $(-858.8 \pm 0.2) + (-5.86 \pm 0.03) \Delta t$\\
\hline
S73, 16.1 & $(-310.2 \pm 0.1) + (-9.93 \pm 0.05) \Delta t$\\
2004.03 & $(-995.4 \pm 0.2) + (-8.95 \pm 0.06) \Delta t$\\
\hline
S74, 16.9 & $(-98.1 \pm 0.2) + (-0.78 \pm 0.09) \Delta t$\\
2005.47 & $(-861.4 \pm 0.2) + (4.64 \pm 0.09) \Delta t$\\
\hline
S75, 17.1 & $(-154.5 \pm 0.2) + (6.03 \pm 0.12) \Delta t$\\
2005.47 & $(-727.1 \pm 0.2) + (1.06 \pm 0.12) \Delta t$\\
\hline
S76, 12.8 & $(355.2 \pm 0.1) + (-3.70 \pm 0.02) \Delta t$\\
2000.41 & $(-925.9 \pm 0.1) + (4.34 \pm 0.02) \Delta t$\\
\hline
S77, 15.8 & $(364.2 \pm 0.4) + (10.27 \pm 0.17) \Delta t$\\
2005.50 & $(-824.5 \pm 0.4) + (-6.73 \pm 0.20) \Delta t$\\
\hline
S78, 16.5 & $(453.4 \pm 0.2) + (-17.40 \pm 0.12) \Delta t$\\
2005.47 & $(-665.2 \pm 0.4) + (-7.24 \pm 0.18) \Delta t$\\
\hline
S79, 16.0 & $(646.6 \pm 0.3) + (1.09 \pm 0.16) \Delta t$\\
2005.47 & $(-533.5 \pm 0.3) + (1.31 \pm 0.16) \Delta t$\\
\hline
S80, 16.9 & $(991.7 \pm 0.3) + (-5.38 \pm 0.14) \Delta t$\\
2005.47 & $(-351.6 \pm 0.3) + (4.66 \pm 0.13) \Delta t$\\
\hline
S81, 17.2 & $(769.9 \pm 1.2) + (-2.23 \pm 2.55) \Delta t$\\
2007.68 & $(-481.4 \pm 1.1) + (-6.90 \pm 2.34) \Delta t$\\
\hline
S82, 15.4 & $(54.4 \pm 0.3) + (-7.78 \pm 0.12) \Delta t$\\
2005.47 & $(944.8 \pm 0.4) + (-14.28 \pm 0.20) \Delta t$\\
\hline
S83, 13.6 & $(-927.6 \pm 0.1) + (-4.96 \pm 0.05) \Delta t + (0.086 \pm 0.005) (\Delta t)^2$\\
2000.41 & $(293.2 \pm 0.2) + (-11.50 \pm 0.06) \Delta t + (-0.001 \pm 0.006) (\Delta t)^2$\\
2004.24&$-557\pm26$\\
\hline
S84, 14.4 & $(-1134.8 \pm 0.2) + (4.17 \pm 0.04) \Delta t$\\
2001.46 & $(-28.1 \pm 0.1) + (1.82 \pm 0.03) \Delta t$\\
\hline
S85, 15.6 & $(-904.4 \pm 0.1) + (5.62 \pm 0.04) \Delta t$\\
2005.47 & $(393.1 \pm 0.1) + (-0.01 \pm 0.07) \Delta t$\\
\hline
S86, 15.5 & $(-1035.0 \pm 0.2) + (1.74 \pm 0.11) \Delta t$\\
2004.56 & $(211.4 \pm 0.2) + (-6.21 \pm 0.08) \Delta t$\\
\hline
S87, 13.6 & $(-855.5 \pm 0.1) + (10.97 \pm 0.05) \Delta t + (0.019 \pm 0.005) (\Delta t)^2$\\
2000.41 & $(-996.2 \pm 0.2) + (-3.45 \pm 0.07) \Delta t + (0.038 \pm 0.007) (\Delta t)^2$\\
2005.77&$19\pm11$\\
\hline
S88, 15.8 & $(-1009.5 \pm 0.3) + (-4.05 \pm 0.05) \Delta t$\\
2000.41 & $(-500.5 \pm 0.3) + (-7.98 \pm 0.05) \Delta t$\\
\hline
S89, 15.3 & $(-942.1 \pm 0.2) + (-4.61 \pm 0.03) \Delta t$\\
2000.41 & $(-630.0 \pm 0.2) + (-2.42 \pm 0.04) \Delta t$\\
\hline
S90, 16.1 & $(531.9 \pm 0.3) + (0.83 \pm 0.12) \Delta t$\\
2005.47 & $(-970.1 \pm 0.2) + (0.50 \pm 0.10) \Delta t$\\
\hline
S91, 12.2 & $(778.9 \pm 0.2) + (11.15 \pm 0.03) \Delta t$\\
2000.41 & $(-681.5 \pm 0.2) + (2.82 \pm 0.03) \Delta t$\\
\hline
S92, 13.0 & $(987.3 \pm 0.1) + (5.79 \pm 0.03) \Delta t$\\
2000.41 & $(24.8 \pm 0.3) + (1.23 \pm 0.05) \Delta t$\\
\hline
S93, 15.6 & $(1083.8 \pm 0.3) + (-2.80 \pm 0.12) \Delta t$\\
2005.47 & $(174.2 \pm 0.4) + (-2.52 \pm 0.18) \Delta t$\\
\hline
S94, 16.7 & $(-154.5 \pm 0.5) + (-10.55 \pm 0.24) \Delta t$\\
2005.47 & $(910.3 \pm 0.6) + (2.16 \pm 0.27) \Delta t$\\
\hline
S95, 10.2 & $(22.3 \pm 0.2) + (6.03 \pm 0.04) \Delta t$\\
2000.41 & $(1214.1 \pm 0.2) + (0.75 \pm 0.03) \Delta t$\\
\end{tabular}
\end{center}
{\scriptsize
\begin{center}
\begin{tabular}{l|l}
Name, $m_K$& $\alpha\, [\mathrm{mas}] =$\\
$t_0$ [yr] for $(\alpha,\,\delta)$ & $\delta\, [\mathrm{mas}] =$\\
$t_0$ [yr] for $v_z$ & $v_z\, [\mathrm{km/s}] =$\\ 
\hline
S96, 10.0 & $(1132.3 \pm 0.2) + (-8.68 \pm 0.03) \Delta t$\\
2000.41 & $(482.2 \pm 0.2) + (7.64 \pm 0.04) \Delta t$\\
2000.25&$158\pm5$\\
\hline
S97, 10.3 & $(1040.8 \pm 0.1) + (7.81 \pm 0.02) \Delta t$\\
2000.41 & $(-972.0 \pm 0.2) + (2.46 \pm 0.03) \Delta t$\\
2005.99&$470\pm50$\\
\hline
S98, 15.6 & $(-908.0 \pm 0.1) + (-7.60 \pm 0.03) \Delta t$\\
2001.46 & $(725.6 \pm 0.2) + (2.59 \pm 0.04) \Delta t$\\
\hline
S99, 16.9 & $(-970.1 \pm 0.2) + (-10.32 \pm 0.08) \Delta t$\\
2005.47 & $(824.6 \pm 0.2) + (1.28 \pm 0.09) \Delta t$\\
\hline
S100, 15.4 & $(-977.8 \pm 0.2) + (-0.77 \pm 0.04) \Delta t$\\
2001.46 & $(566.5 \pm 0.3) + (-2.15 \pm 0.06) \Delta t$\\
\hline
S101, 17.4 & $(-857.5 \pm 0.3) + (3.42 \pm 0.16) \Delta t$\\
2005.47 & $(509.6 \pm 0.5) + (7.71 \pm 0.26) \Delta t$\\
\hline
S102, 17.6 & $(-770.4 \pm 0.4) + (-4.88 \pm 0.18) \Delta t$\\
2005.47 & $(462.0 \pm 0.3) + (7.14 \pm 0.15) \Delta t$\\
\hline
S103, 18.3 & $(-779.3 \pm 0.7) + (10.97 \pm 0.43) \Delta t$\\
2005.63 & $(560.7 \pm 0.6) + (-2.73 \pm 0.41) \Delta t$\\
\hline
S104, 17.6 & $(-686.8 \pm 0.7) + (10.23 \pm 0.33) \Delta t$\\
2005.51 & $(496.4 \pm 0.6) + (-1.24 \pm 0.27) \Delta t$\\
\hline
S105, 16.5 & $(-1143.6 \pm 0.2) + (3.37 \pm 0.09) \Delta t$\\
2005.47 & $(513.5 \pm 0.2) + (-7.43 \pm 0.08) \Delta t$\\
\hline
S106, 17.1 & $(-1234.2 \pm 0.2) + (1.27 \pm 0.10) \Delta t$\\
2005.47 & $(266.4 \pm 0.4) + (2.23 \pm 0.18) \Delta t$\\
\hline
S107, 14.8 & $(-1240.3 \pm 0.2) + (-0.53 \pm 0.05) \Delta t$\\
2001.46 & $(-45.1 \pm 0.2) + (5.33 \pm 0.05) \Delta t$\\
\hline
S108, 17.0 & $(-981.1 \pm 0.3) + (3.58 \pm 0.12) \Delta t$\\
2005.47 & $(-797.7 \pm 0.3) + (1.77 \pm 0.14) \Delta t$\\
\hline
S109, 17.3 & $(-887.9 \pm 0.2) + (5.93 \pm 0.11) \Delta t$\\
2005.47 & $(-778.2 \pm 0.4) + (-4.27 \pm 0.18) \Delta t$\\
\hline
S110, 16.9 & $(-778.5 \pm 0.3) + (-3.01 \pm 0.13) \Delta t$\\
2005.47 & $(-724.7 \pm 0.2) + (-0.88 \pm 0.11) \Delta t$\\
\hline
S111, 13.8 & $(-1109.9 \pm 0.2) + (-3.69 \pm 0.06) \Delta t + (0.054 \pm 0.007) (\Delta t)^2$\\
2000.41 & $(-895.5 \pm 0.2) + (-8.15 \pm 0.08) \Delta t + (0.039 \pm 0.008) (\Delta t)^2$\\
2005.77&$-741\pm5$\\
\hline
S112, 17.5 & $(-893.7 \pm 0.2) + (4.30 \pm 0.08) \Delta t$\\
2005.47 & $(953.4 \pm 0.3) + (11.15 \pm 0.11) \Delta t$\\
\hline
\end{tabular}
\end{center}
}

\end{document}